\documentstyle[prl,aps,eqsecnum,epsfig]{revtex}

\begin{document}
\draft

%==========================================================

\def\nid{\noindent}
\def\half{\mbox{\small $\frac{1}{2}$}}
\def\quarter{\mbox{\small $\frac{1}{4}$}}
\def\sfrac#1#2{\mbox{\small $\frac{#1}{#2}$}}
\def\beq{\begin{equation}}
\def\eeq{\end{equation}}
\def\eeql#1{\label{#1} \end{equation}}
\def\bea{\begin{eqnarray}}
\def\eea{\end{eqnarray}}
\def\eeal#1{\label{#1} \end{eqnarray}}
\def\sech{\mathop{\rm sech}\nolimits}
\def\scalefig#1{\def\epsfsize##1##2{#1##1}}
\def\vec{\bbox}
\hyphenation{anti-nodal}
\def\im{\mathop{\rm Im}\nolimits}
\def\re{\mathop{\rm Re}\nolimits}

%==========================================================

\title{SUSY Transformations for Quasinormal and\\
Total-Transmission Modes of Open Systems}

\author{P.~T.~Leung${}^{(1)}$, 
Alec~Maassen van den Brink${}^{(1)}$,
W.~M.~Suen${}^{(1,2)}$, \\
C.~W.~Wong${}^{(1)}$
and K.~Young${}^{(1)}$}

\address{${}^{(1)}$Physics Department,
The Chinese University of Hong Kong, 
Hong Kong, China}

\address{${}^{(2)}$McDonnell Center for Space Sciences, Department of Physics,
Washington University, \\
St.~Louis, Missouri 63130, USA}

\date{\today}

\maketitle

%==========================================================
%==========================================================

\begin{abstract}
Quasinormal modes are the counterparts in open systems of normal modes in conservative systems; defined by
outgoing-wave boundary conditions, they have complex eigenvalues $\omega$.  The conditions are studied for a
system to have a supersymmetric (SUSY) partner with the same 
complex quasinormal-mode spectrum (or the same except for one
eigenvalue). The discussion naturally includes total-transmission modes as well (incoming at one extreme and
outgoing at the other). Several types of SUSY transformations emerge, and each is illustrated with examples,
including the transformation among different P\"oschl--Teller potentials and the well-known identity in
spectrum between the two parity sectors of linearized gravitational waves propagating on a Schwarzschild
background. In contrast to the case
of normal modes, there may be multiple essentially isospectral partners, each missing a different state. The SUSY
transformation preserves orthonormality under a bilinear map which is the analog of the usual inner product
for conservative systems.  SUSY transformations can lead to doubled 
quasinormal and total-transmission
modes;  this phenomenon is analysed and illustrated. The existence or otherwise of SUSY partners is also
relevant to the question of {\em inversion}: are open wave systems uniquely determined by their complex
spectra? 

\end{abstract}


%==========================================================
%==========================================================

\section{Introduction}
\label{sect:intro}

%==========================================================

\subsection{Supersymmetric quantum mechanics}
\label{subsect:susyqm}

In this paper we consider supersymmetry (SUSY)
in the one-dimensional Klein--Gordon equation (KGE)

\beq
\left[ \partial_t^2 - \partial_x^2 + V(x) \right]
\phi(x,t) = 0 \quad ,
\eeql{eq:kgt}

\nid
and especially in the corresponding eigenvalue problem

\beq
H \phi_n(x) = \omega_n^2 \phi_n(x) \quad ,
\eeql{eq:kg}

\nid
where

\beq
H = - \partial_x^2 + V(x) \quad .
\eeql{eq:h1}

\nid
The boundary conditions will be specified later.
In so far as the interest centers on the time-independent
problem (\ref{eq:kg}) and the spectrum, the case of the
Schr\"odinger equation, to which reference is usually made, is
included if the eigenvalues are simply relabeled by
$\omega^2 \mapsto \omega$. 

If there exists another system described by 

\beq
{\tilde H} = -\partial_x^2 + {\tilde V}(x) \quad ,
\eeql{eq:h2}

\nid
such that $H$ and ${\tilde H}$ have the same spectrum
(or the same spectrum apart from one state), 
and moreover if the states in the two systems are related by

\beq
{\tilde \phi}(x) = A \phi(x)
\quad ,
\eeql{eq:map}

\nid
where

\bea
A &=& \partial_x + W(x) \quad, \nonumber \\
-A^{\dagger} &=& \partial_x - W(x) \quad ,
\eeal{eq:opa}

\nid
then the
two systems are said to be SUSY partners~\cite{witten1}.  The term ``super"
originates from boson--fermion symmetry in field theory,
but here simply refers to a relationship between the
states of two different systems, whereas symmetry 
relates states of the same system.
SUSY for quantum mechanics is usually discussed for
{\em normal modes\/} (NMs), i.e., solutions to (\ref{eq:kgt})
and (\ref{eq:kg}) which vanish at spatial infinity and
which are square-integrable~\cite{nm}.  The results 
for NMs are well known~\cite{susy1} and will be covered in the discussion
below.  

In order for (\ref{eq:map}) to preserve the spectrum, one needs

\beq
AH = \tilde{H}A
\quad .
\eeql{eq:ahha}

\nid
Our task is simply to analyse this condition when applied
to open systems.

%==========================================================

\subsection{Quasinormal modes and total-transmission modes}
\label{subsect:qnm}

In open systems, waves are not
confined, but can escape to infinity:
acoustic waves escape from a musical instrument; electromagnetic waves
escape from a laser; and linearized gravitational waves
propagating on a Schwarzschild background escape
to infinity and into the horizon.  These systems are often
described (e.g., in the case of gravitational waves~\cite{chand})
by the KGE (\ref{eq:kgt}),
or (e.g., in the case of optics~\cite{lamb})
by the wave equation 

\beq
\left[ \rho(x) \partial_t^2 - \partial_x^2 \right]
\phi(x,t) = 0 \quad ,
\eeql{eq:wet}

\nid
to which the KGE can be
transformed~\cite{tong1}.  
This transformation has certain subtleties when there
are bound states in the KGE;
see Appendix~\ref{sect:appkgwe}.

We assume for the moment that $V$ (or in the case of the wave equation $\rho-1$) has finite support. This assumption
is natural for describing a system in the finite parts of space, surrounded by a trivial medium such as vacuum, and
it leads to a simple formalism. However, some of the superpartners of finitely supported potentials, as well as the
examples of Sections \ref{sect:pt} and~\ref{sect:bh}, prompt the study of SUSY under the weaker condition
$V(x)\rightarrow0$ as $x\rightarrow\pm\infty$. The analysis then becomes considerably more subtle, and is therefore
relegated to Appendix~\ref{sect:apptail}. 

We shall consider first of all states satisfying the so-called outgoing-wave condition, which is most easily stated
for an eigenfunction of frequency $\omega$: 
\beq
  \phi(x)\propto e^{\pm i\omega x}\quad,\quad x\rightarrow\pm\infty\quad,
\eeql{eq:out1}
or equivalently
\beq
  \frac{\phi'(x)}{\phi(x)}=\pm i\omega\quad,\quad x\rightarrow\pm\infty\quad.
\eeql{eq:out2}
Under these boundary conditions, discrete eigenvalues fall into two classes.  First, there could be bound states or
NMs~\cite{nm2}; because $V$ vanishes at infinity, these must (from the Schr\"odinger point of view) have a negative
energy $\omega^2$, and hence $\omega$ is purely imaginary.  According to the condition (\ref{eq:out1}), we put
$\im\omega>0$. Second, there could be quasinormal modes (QNMs) with complex eigenvalues $\omega^2$.  Because these
waves (in contrast to the bound states) are genuinely outgoing, probability (or energy) in the finite parts of space
decreases, so $\im\omega<0$\cite{bounds}. We here assume that the loss is only due to the escape of the waves, i.e.,
only through the boundary conditions.  In particular, the potential $V$ is real.  (Absorption may be described by a
complex $V$, but causality then requires dispersion; the necessary generalization of (\ref{eq:kgt})\cite{abs} will
not be discussed here.)

The complex QNM frequencies are often directly observable: for example
the central frequency and width of an optical line observed from 
a laser cavity, or the rates of repetition and decay of
a gravitational-wave signal that may within the next 
decade be detected by instruments such as LIGO\cite{ligo}. 
By way of orientation, Figure 1a shows a square well
(solid line, ignore broken line),
and Figure 1b shows, on the complex $\omega$-plane,
the distribution of NM and QNM frequencies
(crosses and circle, ignore triangle).
Several features may be noticed.
(a) Bound states or NMs~\cite{nm} are shown
in the upper half-plane; in this example there
are three.
(b)  QNMs have $\im\omega<0$, and
are in the lower half-plane.  Provided $\re\omega\ne0$,
they occur in pairs:
$\omega^{\vphantom{*}}_{-n} = - \omega^*_{n}$, as is readily
shown by conjugating the defining equation and boundary conditions.
(c)  There may be QNMs, not paired, which have $\re\omega=0$;
in this example there is one.
These {\em zero modes\/}\cite{zeromode} will
be of particular importance below. 

Even though the QNM eigenfunctions are not square-integrable
and do not form a Hilbert space (at least not in the conventional
sense), they are nevertheless useful for analyzing outgoing waves.
As stated above, QNM frequencies are directly observable,
and as illustrated by the example in Figure 1,
the spectrum is typically much richer than that
for NMs.  In fact, the spectrum is so rich that,
under some broad conditions, namely that $V(x)$ vanishes
identically outside an interval $[-a,a]$ and has 
a singularity at $x=\pm a$, the QNMs together with any possible NMs
are {\em complete\/} in at least part of spacetime\cite{comp}, 
so that the wave signal observed
at a point $x$ can be represented as
\beq
  \phi(x,t) = \sum_n a_n \phi_n(x) e^{-i\omega_n t} \quad ,
\eeql{eq:comp}
where the sum is over all QNMs (including zero modes)
and NMs\cite{sumnm}.  
In cases where the QNMs are not complete, it may still be possible
to characterize the remainder, which could be, for example,
a power law in $t$\cite{pricetail,tail}. 
Moreover, when the QNMs are complete, one can set up a formalism that
almost completely parallels the case of NMs\cite{tong1,tong2}:
under a bilinear map which is the analog of the usual
inner product (but which is linear
in both vectors rather than linear in one and
conjugate linear in the other), the Hamiltonian turns out to
symmetric (the analog of self-adjoint).  
This allows much of the well-known mathematical formalism 
developed for NMs to be transcribed.  The mathematical
structure can also be written elegantly using a biorthogonal
formalism\cite{bior}.  
One can even second-quantize using these QNMs 
as a basis (e.g., to discuss photon creation
and annihilation in an optical cavity)\cite{quant}.
These developments have been reviewed in Ref.\cite{rmp}.
An important point of this paper is that much of the mathematical structure
is preserved under SUSY (Section~\ref{sect:orth}).

It is straightforward and indeed natural to include 
total-transmission modes (TTMs) in the discussion as well; these
are defined by the incoming-wave condition at one extreme
and the outgoing-wave condition at the other.
The TTMs propagating from the left (TTM$_{\rm L}$) and those
propagating from the right (TTM$_{\rm R}$) satisfy
\beq
 \phi(x)\propto e^{\pm i\omega x}\quad,\quad|x|\rightarrow\infty\quad ,
\eeql{eq:ttl}
with the $+$ ($-$) sign for a TTM$_{\rm L}$ (TTM$_{\rm R}$).

Given the striking parallel between NMs and QNMs, 
and especially since 
the QNM spectrum is often richer than
the NM spectrum, it is logical to
ask whether the entire concept of SUSY can be generalized to
QNMs: namely, given a real potential $V$, can one find a
real partner potential $\tilde{V}$ such that the outgoing waves
for the two have the same set of complex eigenfrequencies (or
the same set apart from one state) and are related by (\ref{eq:map})?  
The same question can be asked of TTMs.
This paper sets out to answer this
question, which turns out {\em not\/} to be more difficult,
even though one has to match a set of {\em complex\/} rather than
real frequencies.  Indeed, when NMs and QNMs are studied together, 
the relationship between the spectra of $H$ and $\tilde{H}$ becomes 
clearer.

Relations between such pairs of open systems have been
discussed in terms of the
transmission amplitude $T(\omega)$ and the reflection
amplitude $R(\omega)$ etc., for solutions at definite
frequencies.  However, an {\em operator\/}
SUSY transformation such as (\ref{eq:map}) is 
more powerful:
given a $\phi(x,t)$ that solves the dynamics of $H$,
one can immediately generate the corresponding solution
$\tilde{\phi}(x,t) = A \phi(x,t)$ which solves the
dynamics of ${\tilde H}$, {\em without\/} having to first
project $\phi$ onto frequency eigenfunctions.

SUSY is closely related to the question of {\em inversion\/}: given the complete set of
complex QNM frequencies, can $V(x)$ be determined uniquely?  Obviously, if 
isospectral SUSY partners exist, then unique inversion 
would not be possible.
The converse need not be the case,
since there could be isospectral systems not related
by a SUSY transformation such as (\ref{eq:map}).  
It may be well to recall the classic results for NMs~\cite{invnm}: 
for a finite interval, say $[0, a]$, if the (real) spectrum is
given for the {\em two\/} dynamical systems defined respectively
by (a) $\phi(0) = \phi(a) = 0$ (the nodal system),
and (b) $\phi'(0) = \phi(a) = 0$ (the antinodal system), then the potential can
be uniquely determined.  
(Equivalently, one can extend the problem to $[-a,a]$ and
seek a {\em symmetric\/} potential $V(x)$ given the spectrum for
(a) the even-parity sector and (b) the odd-parity sector.)
Effective numerical algorithms
have recently been given~\cite{rundell,invnmnum}.  
Since the QNMs of open systems are complex and
carry twice the amount of information, it is tempting to speculate that
the frequencies for {\em one\/} set of boundary conditions 
(e.g., outgoing waves at both ends of a finite interval) may be enough 
for unique inversion.
There is some numerical evidence to support this 
conjecture, at least in some cases\cite{numinv,numinv2}.

The possibility of inversion from {\em one\/} spectrum is intriguing, since
physically all the information one has is the time-dependent
signal and hence in principle the set of complex frequencies under 
{\em one\/} set of boundary conditions\cite{freq}.  
For example, without wishing to belittle
the technical difficulties involved, one may imagine that 
the signal received by LIGO could yield the complex eigenvalues and from 
these one could determine $V(x)$, which describes the background curvature.
This could then, in principle, provide a novel astrophysical tool ---
using gravitational waves to probe the intervening spacetime between the
source and the observer.  It has been suggested that the QNM spectrum may
be a definitive signature of black holes~\cite{bhsign}.
Moreover, a toy model has been given for the corresponding 
forward problem of perturbing a black hole: if the hole is ``dirty", i.e.,
perturbed by surrounding matter,
how would this be revealed in the shifts of the complex frequencies\cite{dirt}?
The inverse of this perturbative problem would 
in principle allow the nature
of the astrophysical environment around a black hole to be revealed through
the shifts in the QNM spectrum.
 
With these motivations, in this paper we 
study SUSY for QNMs and TTMs.  
The formalism is given
in Section~\ref{sect:form}, and is in fact almost the
same as that for NMs\cite{suk}.
SUSY transformations can
be classified into three discrete types plus one continuous type.
Examples for square wells and barriers are given in Section~\ref{sect:ex}.
It is found that the familiar transformations
not only establish an equivalence between the
NM spectra of partner potentials, but preserve the
QNM spectrum as well.  Moreover, by considering NMs
and QNMs together, a clearer picture emerges.
The reflection and transmission
amplitudes for SUSY partners are related in a simple way, as is shown in Section~\ref{sect:reftran}.
Section~\ref{sect:pt} deals with the P\"oschl--Teller (PT) 
potential, which has a number of
interesting features: 
(a) for a range of parameters, the PT potential has
an infinite number of zero modes, and many of these allow 
corresponding SUSY transformations, leading to infinitely many 
essentially isospectral partners;
(b)~two of these transformations lead to partner
potentials also of the PT type, but with different amplitudes; 
(c) thus the partner can again be transformed to
yet other PT systems, {\em ad infinitum\/}; and
(d) a subset of PT potentials are SUSY-equivalent to the
free field, and therefore have total transmission
at all positive energies.
In Section~\ref{sect:bh}, 
the Regge--Wheeler (RW) and Zerilli potentials,
which respectively describe the axial and polar sectors of linearized
perturbations of a Schwarzschild spacetime,
are shown to be SUSY partners.    
This result is not novel, but is here cast 
into the general framework of SUSY.
The orthonormality of QNMs under SUSY
is presented in Section~\ref{sect:orth},
in which SUSY is
generalized to the two-component formalism appropriate to
open systems.
SUSY can also lead to {\em doubled\/}
QNMs and TTMs, in the sense that two such modes occur
at the same frequency.  This phenomenon, which has no
counterpart in conservative systems, is analysed and
illustrated in Section~\ref{sect:double}.
Finally, a discussion is given in Section \ref{sect:disc}.

%==========================================================
%==========================================================

\section{Formalism}
\label{sect:form}

It is readily shown, without reference to any boundary conditions
and hence with equal validity for NMs, QNMs and TTMs, that $H$ and ${\tilde H}$
have the same spectrum (with the possible exception of one state, to be
discussed below), provided (\ref{eq:ahha}) holds.
It is readily shown that the two potentials can be written as
\bea
  V(x) &=& W(x)^2 - W'(x) + \Omega^2 \quad , \nonumber \\
  {\tilde V}(x) &=& W(x)^2 + W'(x) + \Omega^2 \quad ,
\eeal{eq:vw}
with $W(x)$ (called the SUSY potential) as in (\ref{eq:opa})
and for some constant $\Omega^2$.  Moreover, the Hamiltonians
can be represented as
\bea
H&=& A^{\dagger}A + \Omega^2
\quad , \nonumber \\
{\tilde H} &=& AA^{\dagger} + \Omega^2
\quad .
\eeal{eq:hop}
It follows that if $\phi(x)$ is an eigenfunction of $H$,
then the partner function
${\tilde \phi}(x) = A \phi(x)$
(provided it does not vanish)
is an eigenfunction of ${\tilde H}$ with the same eigenvalue.
The two partner systems can be put into one linear space
by introducing Pauli spinors, with $H$ and ${\tilde H}$
associated with $1 \pm \sigma_z$ and $A, A^{\dagger}$
associated with $\sigma_{\pm}$.
This transcription of the usual formalism remains valid even
when the eigenvalues are complex.  We defer
all discussions of normalization to Section~\ref{sect:orth}.

Upon reversing the sign of $W$,
(a) $V$ and ${\tilde V}$ are interchanged (see (\ref{eq:vw})), and
(b) $A$ and $-A^{\dagger}$ are interchanged
(see (\ref{eq:opa}));
thus the inverse mapping from
${\tilde H}$ back to $H$ is (up to a sign) achieved by $A^{\dagger}$.

We may regard (\ref{eq:vw}) as a Riccati equation for $W$ in terms
of the given $V$.  As $V(x)$ is assumed to vanish for 
$x \rightarrow \pm \infty$, we have $W^2=-\Omega^2$
as $x \rightarrow \pm \infty$.  The first-order Riccati equation 
can satisfy two boundary
conditions only at special values of $\Omega^2$;
this condition becomes familiar if we define the
{\em SUSY generator\/} $\Phi(x)$ by
\beq
  W(x) = - \frac{\Phi'(x)}{\Phi(x)} \quad .
\eeql{eq:wphi}
Then (\ref{eq:vw}) implies that
$\Phi$ must be an eigenfunction of $H$ with eigenvalue $\Omega^2$:
\beq
  H \Phi(x) = \Omega^2 \Phi(x) \quad .
\eeql{eq:phieq}
The various SUSY transformations are then related,
in a one-to-one manner, to solutions of (\ref{eq:phieq}).
Before proceeding
further, we should note that even for the QNM case with complex
frequencies, since ${\tilde V}$ is required to be
real, $W$ and hence $\Phi$ and $\Omega^2$ also have to be real
(in the case of $\Phi$ up to an irrelevant overall phase)\cite{complexW}.  

The task is therefore to analyze the boundary conditions.  First, suppose $\Omega^2>0$, so that $\Omega$ is real. 
Then outside the support of $V$, $\Phi$ is oscillatory:  either complex (e.g.\ $e^{i\Omega x}$), inadmissible since
it leads to a complex $W $; or real (e.g.\ $\sin \Omega x$), inadmissible since its nodes lead to singularities in
$W$.  Thus, $\Omega^2\le0$, and we denote $K \equiv |\Omega|$. The somewhat special marginal case $\Omega=0$ is dealt
with in Appendix~\ref{Omega0}. 

At each spatial extreme ($x \rightarrow \pm \infty$), the solution
can be either (a) purely decreasing ($\Phi \propto e^{-K|x|}$), denoted
as~D; (b) purely increasing ($\Phi \propto e^{K|x|}$), denoted
as I; or (c)~a mixture of the two terms, i.e.,
\beq
  \Phi(x)=c e^{K|x|} + d e^{-K|x|} \quad ,
\eeql{eq:mix}
with $c, d \neq 0$, to be denoted as M.  Considering both extremes,
the possible solutions for $\Phi$ can be classified into 
the following types, in obvious notation:
\bea
  \mbox{Type 1 } &=& \mbox{DD} \quad , \nonumber \\
  \mbox{Type 2 } &=& \mbox{II} \quad , \nonumber \\
  \mbox{Type 3 } &=& \mbox{DI, ID} \quad , \nonumber \\
  \mbox{Type 4 } &=& \mbox{MM, MD, MI, DM, IM} \quad .
\eeal{eq:type}
The first three types can only occur at discrete values
of $\Omega$, and can be characterized as follows.
A solution decreasing at both ends is an NM (DD = NM);
we associate these with waves going as $e^{i\Omega |x|}$,
with $\Omega = iK$.  A solution which is purely increasing
at both ends is a QNM (II = QNM); we again associate these
with waves going as $e^{i\Omega |x|}$, but now
with $\Omega = -iK$.  
A DI solution goes as $e^{Kx}$ at both extremes ($K>0$).
We can assoiciate this with either
(a) the time-dependent solution $e^{i\Omega x} e^{-i\Omega t}$
with $\Omega = -iK$, namely a TTM$_{\rm L}$, or
(b) the time-dependent solution $e^{-i\Omega' x} e^{-i\Omega' t}$
with $\Omega' = iK$, namely a TTM$_{\rm R}$.
Purely as a matter of convention, we shall only discuss states
in the lower half-plane (in this case the TTM$_{\rm L}$),
always noting that there is a TTM propagating in the opposite
direction at the opposite value of the frequency which is not shown.  
Thus, with this convention,
DI = TTM$_{\rm L}$; similarly ID = TTM$_{\rm R}$.
This accords with the convention in discussions of
black-hole modes.
We shall in the rest of this section
pay particular attention to the disappearance and appearance
of the four classes of solutions (NM, QNM, TTM$_{\rm L}$, TTM$_{\rm R}$) under SUSY.

In contrast, Type 4 solutions are allowed for continuous $\Omega$.
There is an additional difference.
For Types 1 to~3, the logarithmic derivative outside $I$ is
exactly $\pm K$, so $W' = 0$.  Then, since $V$ vanishes
outside $I$, so does ${\tilde V}$.  However, for Type 4,
the logarithmic derivative is given by 
(e.g., for $x \rightarrow +\infty$)
\beq
  W = - K + \frac{2dK}{c} e^{-2Kx} + \cdots \quad ,
\eeql{eq:wasymp}
so that $\tilde{V}$ would have an exponential tail, taking us outside the simplified theory presented in this section.

We should check immediately that outgoing/incoming boundary 
conditions are preserved under SUSY.
To do so,
let $\phi$ be, for example, an outgoing wave, $\phi(x) = C e^{i\omega x}$ 
as $x \rightarrow \infty$; then operating with $A$, we find
\bea
  {\tilde \phi} (x) &=& \left[ \partial_x + W(x) \right] \phi(x)\nonumber \\
  &=& (i \omega + W_{+} ) C e^{i\omega x} \quad ,
\eeal{eq:checkbc}
where we have introduced the constants
\beq
W(x{\rightarrow}{\pm}{\infty}) = W_{\pm} \quad ,
\eeql{eq:wpm}
which, depending on which type is being discussed, would
equal $\pm K$.
Thus $\phi$ and ${\tilde \phi}$ always satisfy the same type
of boundary conditions.

Two properties of the generator
$\Phi$ ought to be mentioned.  First, it is
annihilated by the operator~$A$: $A \Phi = 0$; this follows
trivially from (\ref{eq:opa}) and (\ref{eq:wphi}).  Second, 
from (\ref{eq:vw}) we see that the partner systems 
are related by reversing the sign of $W$, so in view
of (\ref{eq:wphi}), the corresponding generator 
for the inverse transformation from ${\tilde H}$ to $H$
is ${\tilde \Phi} = \Phi^{-1}$; this function is guaranteed
to be an eigenfunction of ${\tilde H}$ also with the eigenvalue
$\Omega^2$.  
(Despite the notation, ${\tilde \Phi}$ is not
the SUSY partner of $\Phi$: ${\tilde \Phi} \ne A \Phi$.)
Because of the reciprocal relation,
the boundary conditions for ${\tilde \Phi}$
are obvious: under $\Phi \mapsto {\tilde \Phi}$,
$\mbox{D} \mapsto \mbox{I}$,
$\mbox{I} \mapsto \mbox{D}$, and
$\mbox{M} \mapsto \mbox{D}$.

Thus, under SUSY, one state $\Phi$ disappears and 
an extra state ${\tilde \Phi}$
appears, with boundary conditions (e.g., DD, II) readily deduced
from the above description.  In view of the association
of these boundary conditions with the four types of 
solutions (e.g., DD = NM, II = QNM), we can summarize the 
changes in the number of these solutions as in Table 1. In this table, the entries $\pm1$ in the ``NM" column refer
to transitions $0\leftrightarrow1$ only, because there can be no doubled bound states (cf.\ at the end of the
Introduction) in 1-d systems; see also Appendix~\ref{susy2waves}.

In particular, if
$\Phi$ is itself an NM or QNM (Type 1 or~2), 
then in the partner system one state at $\Omega = \pm iK$ disappears and another
state at $-\Omega = \mp iK$ appears instead.
In these cases, the systems, differing by one state whose
frequency is reversed, are said to be 
{\em essentially isospectral\/}, and the mapping $A$ between
them constitutes a {\em good\/}
SUSY transformation~\cite{susy1}.  Where $\Phi$ is not
an NM or QNM of $H$, and ${\tilde \Phi}$ is not an NM or QNM of ${\tilde H}$
(Type~3), then
the two systems 
are {\em strictly isospectral\/}
(for both the real NM and the complex QNM spectra), 
corresponding to {\em broken\/} 
SUSY~\cite{fn2}.

Some of these properties are of course well known; but the
more systematic perspective is possible 
only if attention is paid to the QNMs (and TTMs)
as well.  These general
remarks will next be illustrated with examples.

%==========================================================
%==========================================================

\section{Square wells and barriers}
\label{sect:ex}

In this section we illustrate
the different types of SUSY
transformations using square wells and barriers
for $V$; this has the advantage that all different
cases can be solved analytically.  
All examples are presented graphically
with a uniform format.  The top diagrams show the original
potential $V$ (solid line) and the partner potential ${\tilde V}$
(broken line).  The bottom diagrams show the complex $\omega$-plane.
The NMs and QNMs common to both systems are shown with crosses.  The mode corresponding to $\Phi$ (${\tilde \Phi}$)
is shown with a circle (triangle);
these are present in one system but not in the other.
In each case, we have verified the lowest few QNMs of ${\tilde H}$
numerically.

%==========================================================
\subsection{Type 1: $\Phi$ is a normal mode}
\label{subsect:exnm}

The Type 1 solution, denoted as DD, corresponds to $\Phi$ 
itself being a bound state or NM.  This occurs only if $V$ is
sufficiently attractive. No matter how many bound states
there are, only the ground state is acceptable as the generator $\Phi$, since the others would have nodes leading to
a singular $W$ --- exactly as in the familiar 
discussion of SUSY between NM spectra.  In this case, the two spectra
(both NMs and QNMs) are identical except that:
(a) ${\tilde H}$ has one fewer NM, since the ground state
in $H$, namely $\phi_0 = \Phi$, is annihilated by~$A$;
(b)~${\tilde H}$ has one extra QNM, since the corresponding
function ${\tilde \Phi} = \Phi^{-1}$ is a solution with
eigenvalue $\Omega^2$ but which is increasing (i.e., is a
purely outgoing wave) for $x \rightarrow \pm \infty$.
The systems are essentially isospectral.

As an example, take $V$ to be a square well (Figure 1):

\beq
V(x) = V_0 \, \theta(a{-}|x|)
\quad ,
\eeql{eq:exsqwell1}

\nid
where $\theta$ is the unit step function,
and with parameters say $V_0 = -20$, $a = 1$.  
The even wavefunctions, for example, are given by

\beq
\phi(x) = A \cosh \alpha x \quad,
\eeql{eq:exsqwell2}

\nid
for $|x| < a$, where 

\beq
- \alpha^2 + V_0 = \omega^2 \quad .
\eeql{eq:exsqwell3}

\nid
The logarithmic derivative at $x=a$ must match onto 
an outgoing wave:

\beq
\alpha \tanh \alpha a = i\omega \quad ,
\eeql{eq:exsqwell4}

\nid
where the case of NMs is included if we take 
$\re\omega = 0$, $\im\omega > 0$.
The odd sector can be treated similarly.

For this choice of parameters, there
are 3 bound states, at $\omega = 
4.28i, 3.68i, 2.47i$.  Taking the lowest of these
as the auxiliary function $\Phi$ (so that 
$\Omega = +iK$
with $K = 4.28 $), we find for $|x| < a$

\bea
\Phi(x) &=& A \cos qx \quad, \nonumber \\
W(x) &=& q \tan qx \quad, \nonumber \\
{\tilde V}(x) &=& 
V_0 + 2q^2 \sec^2 qx 
\quad ,
\eeal{eq:exsqwell5}

\nid
where $q = \sqrt{-V_0 -K^2}$ is real; with the present parameters $q = 1.28$. Figure 1a shows the two partner
potentials and Figure 1b compares the spectra.  The NM in $H$ but not ${\tilde H}$ is shown by the
circle at $\omega = iK$; the QNM in ${\tilde H}$ but not $H$ is shown by the triangle at $\omega = -iK$. 

%==========================================================

\subsection{Type 2: $\Phi$ is a quasinormal mode}
\label{subsect:exqnm}

The Type 2 solution, denoted as II, corresponds to $\Phi$ 
itself being a QNM.
Since $\Omega^2$ has to be real, the frequency
is purely imaginary and $\Phi$ is a zero mode.
We note two features.  
First, not all potentials will have a zero mode.
Second, in distinct contrast to the NM case, there may be
{\em several} nodeless QNM zero modes, 
all of which are candidates for the generator $\Phi$.

To illustrate the first property, consider the example of a square
barrier, i.e., (\ref{eq:exsqwell1}) with $V_0 > 0$.
Again we take $a=1$, and allow $V_0$ to vary.
The formulas
are similar to (\ref{eq:exsqwell1})--(\ref{eq:exsqwell5}) 
with the sign of $V_0$ reversed, and will
not be shown.  In the even sector there are two zero modes
for small values of $V_0$; 
as $V_0$ increases, these move closer together and
merge at $V_0 = V_{\rm c} = 0.291$\cite{jordan}; for higher barriers there are no
zero modes.  There are no zero modes in the odd sector, and
in any event an odd state 
is not eligible as a generator
(because there is a node).
To be definite, we take $V_0 = 0.16$,
in which case the zero modes are at $\omega = -0.181i, -2.500i$, with $\alpha = 0.242,\,2.506$ in (\ref{eq:exsqwell2}).
These wavefunctions, being $\cosh \alpha x$ within $[-a,a]$
and a real exponential for $|x| > a$, clearly have
no nodes.

This takes us to the second property alluded to above, 
namely, that the existence of more than one nodeless QNM
is by no means exceptional; some general
remarks are given in Appendix \ref{sect:appnode}.
Multiple nodeless QNMs allow
different SUSY transformations.  Continuing with the example,
we show separately the case where the state at 
$\Omega = \omega_1 = -0.181i$ is chosen
to be $\Phi$ (Figure 2) and the case where the state at 
$\omega_2 = -2.500i$
is chosen (Figure~3).  The two partner potentials ${\tilde V}$
are different.  In each case, one QNM 
has disappeared and an extra NM has emerged.
The systems are again essentially isospectral.

%==========================================================

\subsection{Type 3: $\Phi$ is a total-transmission mode}
\label{subsect:exzr}

To be definite, we take the Type 3 solution to be DI: $\Phi$~is
decreasing as $x \rightarrow -\infty$ and increasing
as $x \rightarrow \infty$, i.e., a TTM${}_{\rm L}$ according
to the convention adopted.  
A simple example can be provided by a multi-step
square barrier:

\beq
V(x) = \left\{
\begin{array}{ll}
V_1 & \quad \mbox{if $|x| \le b$} \\
V_0  & \quad \mbox{if $b < |x| \le a$} \\
0 & \quad \mbox{if $|x| > a$}
\end{array}  
\right. \quad .
\eeql{eq:multistep1}

\nid
We take $b = 0.1, a = 1.0, 
V_1 = -10.0, V_0 = 1.0$.  A TTM${}_{\rm L}$ 
is found for $\Omega = -0.990i$, and it can be used to generate a strictly isospectral SUSY partner:
all the NMs and QNMs are preserved.
Figure 4a shows the partner
potential, and
Figure 4b shows the QNM distribution.
Figure 5a shows the generator,
while Figures 5b and 5c respectively show the
NM ${\tilde \phi}_1$ and the QNM ${\tilde \phi}_2$
in the partner system ${\tilde V}$, with eigenvalues
$\omega_1 = 0.498i$ and $\omega_2 = -1.570i$.
(Because in this example $V$ is symmetric,
there is a TTM${}_{\rm R}$ at the same $\Omega$.
However, that mode would generate a different ${\tilde V}$,
in fact the parity-partner of the one considered here.)

%==========================================================

\subsection{Type 4: Continuous transformations}
\label{subsect:excont}

As an example, 
consider the same square barrier as in Section~\ref{subsect:exqnm},
i.e., $V_{0}=0.16$, $a=1.0$.
There are many sub-categories for Type 4 transformations 
(cf.\ (\ref{eq:type})).  We choose symmetric boundary conditions, i.e.,
(\ref{eq:mix}) with the same values of $c$ and $d$
for both $x \rightarrow \pm \infty$.  
Within the range $|x| \leq a$, 
${\tilde V}$ is again given by (\ref{eq:exsqwell5}).
However, for $|x|>a$, there is an exponential tail:
\beq
  \tilde{V}(|x|{>}a) = 2 K^2 \left[ 1 - 
  \left(  \frac{c e^{K |x|} - d e^{-K |x|}}
  {c e^{K |x|} + d e^{-K |x|}} \right)^{\! \! 2} \, 
  \right] \quad .
\eeql{eq:exptail}
Turning to the mode structure, it turns out that $\tilde{\Phi}$ is ``universal" in that it is an NM, a QNM, and a
TTM$_{\rm L/R}$ at the same time, because outgoing and incoming waves coincide in the partner at $\omega=-iK$. (This
somewhat surprising phenomenon is explained in Appendix~\ref{sect:apptail}.) As in other cases, modes at eigenvalues
$\omega^2\neq-K^2$ are not affected by SUSY. Figure 6 shows an example, where we have chosen $K=3.0$, implying $d/c =
-0.829$. 

%==========================================================
%==========================================================

\section{Reflection and transmission amplitudes}
\label{sect:reftran}

The relation between the reflection and transmission
amplitudes of SUSY partners is known in general~\cite{susy1},
and has also been used in the specific context of
gravitational waves propagating on a black-hole background~\cite{chand}.
Here we record the results for completeness.

The reflection and transmission amplitudes  
(for a wave incoming from the left) are defined by
\beq
  \phi(x)=\left\{\begin{array}{l}
    e^{i\omega x} + R_{\rm L}(\omega) e^{-i\omega x}  \\
    T_{\rm L}(\omega) e^{i\omega x} 
  \end{array} \right. \quad ,
\eeql{eq:rt1}
for $x \rightarrow -\infty$ and $x \rightarrow \infty$
respectively. Now applying $A = \partial_x +W$ and recognizing that
$W(x)$ for large $x$ equals finite constants (\ref{eq:wpm}), the corresponding
wavefunction in the partner system is
\beq
  \tilde{\phi}(x)=\left\{\!\!\begin{array}{l}
    (i\omega {+} W_{-}) e^{i\omega x} + 
    (-i\omega {+} W_{-}) R_{\rm L}(\omega) e^{-i\omega x} \\
    (i\omega {+} W_{+}) T_{\rm L}(\omega) e^{i\omega x} 
  \end{array} \right. 
\eeql{eq:rt2}
respectively for $x \rightarrow \mp \infty$.
Normalizing the incoming wave on the left to unity,
we find the corresponding amplitudes in the SUSY partner to be
\bea
  {\tilde R}_{\rm L} (\omega) &=& 
  \frac{-i\omega + W_- }{i\omega + W_-} R_{\rm L}(\omega)
  \quad , \nonumber \\
  {\tilde T}_{\rm L} (\omega) &=& 
  \frac{i\omega + W_+ }{i\omega + W_-} T_{\rm L}(\omega)\quad .
\eeal{eq:rt3}

The constants $W_{+}$ and $W_{-}$ are real and always
equal in magnitude.  For real $\omega$ it then follows that
$| {\tilde R_{\rm L}(\omega)} |^2 = | R_{\rm L}(\omega) |^2$,
$| {\tilde T_{\rm L}(\omega)} |^2 = | T_{\rm L}(\omega) |^2$.
In the case of Type~3 SUSY transformations,
$W_-=W_+$ and one has the stronger condition
$\tilde{T}_{\rm L}(\omega) =  T_{\rm L}(\omega)$.

Similar relations can be obtained for the amplitudes  
$R_{\rm R}$ and $T_{\rm R}$ defined for a wave incoming
from the right; it suffices to change
$W_{\pm} \rightarrow W_{\mp}$ and $\omega \rightarrow
-\omega$ in the prefactors above.  In fact, one has the relations

\bea
R_{\rm R} (\omega) &=& 
- \frac{ T_{\rm L}(\omega) R_{\rm L}(-\omega)}
{T_{\rm L}(-\omega)}
\quad , \nonumber \\
T_{\rm R} (\omega) &=&
{T_{\rm L}(\omega)}
\quad .
\eeal{eq:lr}

Although the transmission and reflection amplitudes for
pairs of related open systems are useful, they relate to
states of definite frequency, and are less general than
the operator SUSY transformation.

%==========================================================
%==========================================================

\section{P\"oschl--Teller potentials}
\label{sect:pt}
%==========================================================

\subsection{General properties}
\label{subsect:ptgen}

We next illustrate with two examples related to the PT
potential~\cite{pt}:
\beq
  V(x) = {\cal V} \sech^2 (x/b) \quad ,
\eeql{eq:pt}
which has an exponential tail:
$V(x) \sim {\cal V} e^{-2|x|/b}$. This means that the theory of Section~\ref{sect:form} is not strictly sufficient
(cf.\ Appendix~\ref{sect:apptail}), but for simplicity we shall gloss over the complications as much as possible, only pointing them out where appropriate.
The PT potential admits analytic solutions, and because of its
exponential tail is sometimes used as a proxy to illustrate
some properties of the effective potentials that
arise in the study of gravitational waves; the latter
also have exponential tails for $x \rightarrow -\infty$, where
$x$ is the tortoise coordinate and $-\infty$ corresponds to the event
horizon. (See Section~\ref{sect:bh} for details.)  The PT potential
is especially interesting from the point of SUSY, in 
at least three ways.

(a)  There is a class of SUSY transformations (Types 1 and~2) which
map one PT potential to another of the same width
(${\cal V} \mapsto {\tilde {\cal V}}$, $b\mapsto b$).

(b)  There are Type 4 transformations which map the free
field (${\cal V} = 0$) to a PT potential with $b^2{\cal V} = -2$.

(c)  By again transforming this potential
as in (a), one obtains a string of other PT potentials with
$b^2{\cal V} = -\ell(\ell{+}1)$, with $\ell$ being an integer.

These results are interesting because (a) implies that
PT potentials are shape-invariant under SUSY, while
(b) and (c) show that certain PT potentials are SUSY-equivalent
to a free field, and therefore have total transmission for
{\em all\/} positive energies. Attractive PT potentials have a finite number of bound states,
while repulsive PT potentials obviously have none; thus the existing SUSY literature, which focuses on NMs,
has not given a systematic account of these properties.

To pave the way for discussing SUSY, we first
summarize the known results for the PT potential. 
For any $\omega$, the
wavefunction that is outgoing at $x \rightarrow -\infty$ reads  

\bea
 \phi(\omega,x) &=&\left[ \xi (1-\xi) \right]^{-i\omega b/2}  \nonumber \\
 &&{}\times{}_2F_1(\half{+}q{-}i\omega b,\half{-}q{-}i\omega b;1{-}i\omega b;\xi)
 \quad,
\eeal{eq:expt01}

where $_2F_1$ is the hypergeometric function, and
the variable $\xi$ is related to $x$ by
\beq
  \xi = \frac{1}{1 + e^{-2x/b}} \quad .
\eeql{eq:expt03}
The parameter
\beq
  q = \sqrt{\quarter -b^2{\cal V}} 
\eeql{eq:expt02}
will be especially important below.
For generic $q$, these solutions are outgoing at the
following values of~$\omega$:
\beq
  \omega = \omega^{\pm}_n (q)\equiv-\frac{i}{b}(n+\half\pm q) \quad , 
\eeql{eq:expt04}
for $n = 0, 1, 2, \ldots$, which thus are 
the (Q)NM frequencies of the PT potential. 
The states have parity $(-1)^n$. However, if $q=\half+\ell$ ($\ell=0,1,2,\ldots$), the modes are restricted to
$i\omega b=-\ell,-\ell+1,\ldots,\ell$: apparently, for half-integer $q$ the two strings of modes (\ref{eq:expt04})
annihilate each other where they coincide. (Moreover, for these $q$ the remaining modes are of a special type, coined
``universal" in Section~\ref{subsect:excont}; see further Appendix~\ref{sect:apptail}.)

From (\ref{eq:expt02}), it is clear that the
value $b^2{\cal V}$ is important, and in fact the situation
can be classified as follows.

(a) If $\quarter <b^2{\cal V}$ (so that $q$ is imaginary),
the QNMs are paired with $\omega_n^{-} = - (\omega_n^{+})^*$.  This case
will not be discussed here, because there are no real SUSY generators.

(b) If $0 <b^2{\cal V} < \quarter$ (so that
$q$ is real and the potential is repulsive),
these QNMs are all zero modes 
lying on the imaginary axis.  This case is interesting
because many of these QNMs can be used as the auxiliary
function $\Phi$.

(c) If $b^2{\cal V} = 0$, one obtains the free field, whose SUSY partners will be
discussed in Section~\ref{subsect:free}.

(d)  If $b^2{\cal V} < 0$, one obtains an attractive
PT potential, and it will be seen that these could
be SUSY partners of repulsive PT potentials.

(e)  If $b^2{\cal V} = -\ell(\ell+1)$, the attractive PT potential
is SUSY-equivalent to a free field.

These cases will be discussed in detail below.
In fact, some of them could be anticipated from
the eigenvalue formula (\ref{eq:expt04}), which has the
important property:
\beq
  \omega_{\tilde n}^{\pm} ({\tilde q}) =\omega_{n}^{\pm} (q)  
\eeql{eq:match1}
for integers $n$ and ${\tilde n}$ iff
\beq
  {\tilde q} = q \pm(n-\tilde{n}) \quad .
\eeql{eq:match2}
Likewise
\beq
  \omega_{\tilde n}^{\mp} ({\tilde q}) =\omega_{n}^{\pm} (q)  
\eeql{eq:match3}
for integers $n$ and ${\tilde n}$ iff
\beq
 {\tilde q} = -q\pm(\tilde{n}-n)\quad .
\eeql{eq:match4}
Secondly, although (\ref{eq:expt02}) implicitly defines
$q \ge 0$ (where it is real), negative values can be admittted
as well, and
\beq
 \omega_n^{+} (q) \equiv \omega_n^{-} (-q) \quad ;
\eeql{eq:match5}
these are just different
labelings of the same state.
So by allowing negative values of $q$,
(\ref{eq:match3})--(\ref{eq:match4}) could be subsumed under
(\ref{eq:match1})--(\ref{eq:match2}).
The equivalence in spectrum implied by
(\ref{eq:match1}) and (\ref{eq:match3})
(which is not strict isospectrality because $n$ and ${\tilde n}$ have
to be non-negative) already suggests a SUSY relationship,
which will be presented in the following.

%==========================================================

\subsection{Results for truncated potential}
\label{subsect:pttrun}

Because of the subtleties when the potential has a tail, we first consider a slightly altered problem --- the PT
potential truncated at $x = |a|$: 
\beq
  V(x) = {\cal V} \sech^2 (x/b) \, \theta (a{-}|x|)\quad .
\eeql{eq:expttrun1}
We take ${\cal V} = 3/16$, $b=1$, and $a = 2$
(solid line in Figure~7a).  
The QNMs can be found by requiring the logarithmic derivative at $x=\pm a$ to be~$\pm i\omega$; the resulting frequencies are shown in 
Figure 7b (crosses and circle only). The lowest QNMs are approximately the same as in
the untruncated case, though the higher QNMs are altered beyond 
recognition~\cite{pottail}, while many new QNMs appear
because of the reflections at $x = \pm a$.  The important point is that
there are still two zero modes, in this instance at 
$\omega = -0.224i, -1.301i$. Either of these can be used as $\Phi$; to
be definite, we take the first.  The SUSY transformation is thus of Type 2.
The partner potential is shown in Figure 7a (broken line).  We note two
features: (a) ${\tilde V}$ vanishes outside $[-a,a]$, and (b) ${\tilde V}$ in this
case is attractive. The overall spectrum is shown in Figure 7b.  Notice, as
before, the appearance of an extra NM accompanying the disappearance 
of the QNM.

%==========================================================

\subsection{Results for full potential}
\label{subsect:ptfull}

We now return to the full PT potential, the eigenvalues for
which are given by (\ref{eq:expt04}).  
For $b^2{\cal V} < \quarter$, all the QNMs $\omega_n^{\pm}$ are zero modes.
For states with odd $n$, the node at $x=0$ prevents
their use as the generator~$\Phi$. If ${\cal V}>0$,
all the even-$n$ eigenfunctions are nodeless (Appendix~\ref{sect:appnode}) and 
therefore eligible as the generator,
which we denote
as $\Phi_n^{\pm}$.  This then leads to an infinite number
of SUSY potentials $W_n^{\pm}$, and correspondingly an infinite
number of partner potentials ${\tilde V}_n^{\pm}$.
Since we start with a QNM, these refer to Type~2 SUSY. For attractive potentials, the construction goes through at least for $n=0$: $\Phi_0^-$ is the ground state and hence certainly nodeless, and $\Phi_0^+$ is readily checked to have the same property.
Some arithmetic leads to the following explicit expression:

%\hrule

\beq
{\tilde V}_n^{\pm} (x) = \left[ 
{\cal V} + A^{\pm}_n 
- 2 B^{\pm}_n C^{\pm}_n G_{2}(x)
+ 2(B^{\pm}_n)^2 G_{1}(x)^2 \sech^2 (x/b)
+ 4B^{\pm}_n G_{1}(x) \tanh (x/b) \right]
\sech^2 (x/b) \quad ,
\eeql{eq:expt11}

%\hrule

\nid
where
\bea
%  F&&_{n,p}^{\pm}(x) = \nonumber \\
%  &&{}_2F_1(\half{+}q{-}i\omega_n^{\pm}b{+}p\, ,
%  \half{-}q{-}i \omega_n^{\pm}b{+}p\,;
%  1{-}i\omega_n^{\pm}b{+}p\,;\xi) \quad ,\nonumber
  F_{n,p}^{\pm}(x) &=& {}_2F_1(\half{+}q{-}i\omega_n^{\pm}b{+}p\, ,
  \half{-}q{-}i \omega_n^{\pm}b{+}p\,;
  1{-}i\omega_n^{\pm}b{+}p\,;\xi) \quad ,\nonumber \\
%\eea
%\vspace{-8mm}
%\bea
G_{n,p}^{\pm}(x) &=& F_{n,p}^{\pm}(x) / F_{0,p}^{\pm}(x)\quad ,\nonumber \\
A_n^{\pm} &=& (-2n-1 \mp 2q)b^{-2} \quad , \nonumber \\
B_n^{\pm} &=& \frac{ n ( n \mp 2q)b^{-2}} { -2n + 1 \pm 2q} \quad , \nonumber \\
C_n^{\pm} &=& \frac{ (n-1) (n-1 \mp 2q) }{ 3-2n \pm 2q} \quad .
\eeal{eq:expt13}

Figure 8 shows the PT potential $V$ and some of 
its SUSY partners ${\tilde V}_n^{\pm}$ obtained in this manner.

%==========================================================
\subsection{The case $n=0$}
\label{subsect:tree}

We omit a detailed discussion of all
the SUSY partners, but those generated with the $n=0$
states as the generator deserve special attention.
From (\ref{eq:expt13}), we see that $B_0^{\pm} = 0$,
and consequently ${\tilde V}_0^{\pm}(x)$ are again of the
PT form, with the same width $b$ but different amplitudes, given by
\beq
  b^2\tilde{\cal V}=b^2{\cal V}-1\mp 2q=b^2{\cal V}-1\mp\sqrt{1-4b^2{\cal V}}
\quad .
\eeql{eq:tree2}
This can be expressed more succinctly in terms of $q$:
\beq
  {\tilde q} = | 1\pm q | \quad.
\eeql{eq:relateq}
The equivalence in spectrum can now be readily verified from
(\ref{eq:match2}) and (\ref{eq:match4}). In fact, we can use the freedom in the labeling convention provided by (\ref{eq:match5}) to rewrite (\ref{eq:relateq}) as $\tilde{q}=q\pm1$.

Since the partner ${\tilde V}(x)$ is another PT potential, we can apply
these transformations again and again.  The result is a chain:
\bea
  \cdots &\leftrightarrow& -m{+}q \leftrightarrow \cdots \leftrightarrow -1{+}q 
  \leftrightarrow q \nonumber \\
  &\leftrightarrow& 1{+}q \leftrightarrow \cdots \leftrightarrow m{+}q
  \leftrightarrow \cdots \quad.
\eeal{eq:chain2}
This shows that the repeated application of SUSY in this
manner forms a group
isomorphic to the integers.  The PT potentials at both
extremes of (\ref{eq:chain2}) become increasingly attractive,
with more and more NMs.

The PT potential is 
self-replicating in the sense that 
(some of) its SUSY partners are the same potential
with a different amplitude.  In Appendix~\ref{sect:apprep} we show
that conversely this condition leads uniquely to the PT potential.

%==========================================================

\subsection{SUSY partners of the free field}
\label{subsect:free}

The free field $V(x) =0$
is a special case of the
PT potential, corresponding to $q = \half$.  Apply a Type 4 transformation
by using the generator
\beq
  \Phi(x) = \cosh Kx \quad .
\eeql{eq:free01}
A simple calculation then gives
\beq
  {\tilde V}(x) = -2K^2 \sech^2 Kx \quad ,
\eeql{eq:free02}
which is of the form (\ref{eq:pt}). We see that an
attractive PT potential is a SUSY partner
of the free field if
\beq
  b^2{\cal V} = -2 \quad
\eeql{eq:free04}
or equivalently $q = \sfrac{3}{2}$.  This potential has one NM,
which is just ${\tilde \Phi} = \Phi^{-1}$.

For half-integer $q$ the QNM $\omega_0^+$ is missing, see below (\ref{eq:expt04}). However, its limit as
$q\rightarrow\half+\ell$, being the even solution (cf.\ Section~\ref{subsect:excont}) at eigenvalue
$b^2\Omega^2=-(\ell+1)^2$, still generates a SUSY transform which can be verified to increase $q$ by one.
Successively applying these (Type~4) transformations one obtains increasingly attractive PT potentials, described by
$b^2{\cal V} = -2, -6, -12, \ldots , -\ell(\ell{+}1), \ldots$, $q = \sfrac{3}{2}, \sfrac{5}{2}, \sfrac{7}{2} ,
\ldots, \ell {+} \half, \ldots$, with the number of NMs being respectively $1, 2, 3, \ldots, \ell, \ldots$. 

These PT potentials are interesting because,
being SUSY-equivalent to the free field, they obviously
have total transmission for {\em all\/} complex frequencies.
This property has recently been discussed in the context
of electron transmission through semiconductor devices~\cite{bytong}, 
but here it falls naturally into the SUSY framework.

%==========================================================

\subsection{Coalescence of QNMs}
\label{subsect:coal}

The eigenvalue formula (\ref{eq:expt02}) implies that
QNMs coalesce in two ways as the system parameters
are tuned; these are quite distinct
scenarios, and it is well to emphasize their difference.

First, as $q \rightarrow 0$, the QNMs
$\omega_n^{+}(q)$ and $\omega_n^{-}(q)$ merge.
At the point of coalescence, they become a double pole;
this is related to an interesting Jordan-block
structure in the evolution operator~\cite{jordan}.
Beyond this point, the two QNMs separate in the
real direction (i.e., $q$ becomes imaginary).
The transition is exactly the same as a harmonic
oscillator going through critical damping.
This case will be further discussed in Section~\ref{sect:double}.
The same occurs for $q \rightarrow n$.

In contrast, as $q \rightarrow \sfrac{1}{2}$, the QNMs
$\omega_n^{+}(q)$ and $\omega_{n+1}^{-}(q)$ merge.
However, when they coalesce, the two poles disappear.
With all poles eliminated, the dynamics become essentially
trivial, with total transmission.
The same occurs for $q \rightarrow n + \half$.

%==========================================================
%==========================================================

\section{Regge--Wheeler and Zerilli potentials}
\label{sect:bh}

An application of particular interest concerns linear\-ized gravitational waves propagating on a Schwarzschild
background.  In each angular-momentum sector $l\ge2$ and for each spin $s$ (for gravitational waves, $s = 2$), these
waves can be described by a scalar field $\phi$ of the radial variable, satisfying a KGE with the potential $V$
describing the nontrivial background~\cite{chand}. However, the axial sector is described by the RW equation, which
is (\ref{eq:kgt}) with the potential
\beq
  V(x) = 2 \left( 1 - \frac{2m}{x_{*}} \right) 
  \frac{x_{*}(n{+}1) - 3m}{x_{*}^{3}}\quad ,
\eeql{eq:v-rw}
while the polar sector is described by the Zerilli equation, which is (\ref{eq:kgt}) with the potential
\bea
  &&\tilde{V}(x) = \left( 1 - \frac{2m}{x_{*}} \right) \times 
%\nonumber \\ && 
  \frac
  {2n^{2} (n{+}1) x_{*}^{3} + 6n^{2} m x_{*}^{2} + 18 n m^{2} x_{*} 
  + 18 m^{3}}
  {x_{*}^{3} (n x_{*} + 3m )^2}
  \quad .
\eeal{eq:v-z}
Here $m$ is the black-hole mass, and $x$ is the tortoise coordinate, related to the circumferential radius $x_{*}$ by
\beq
  x = x_{*} + 2 m \log \left( \frac{x_{*}}{2m} - 1 \right)\quad .
\eeql{eq:tort}
Note that $x_{*} \in (2m , +\infty)$ maps to $x \in ( -\infty , +\infty)$.  The constant $n$ is related to the
angular momentum $l$ by
\beq
  n = \frac{1}{2} (l{-}1) (l{+}2)\quad.
\eeql{eq:q}
It has been noticed~\cite{chanddet} that the two equations have the same QNM spectrum. For example, Figure 9 shows
the distribution of QNM frequencies for $l=s=2$ for both equations. This can be understood because the QNMs of the
two systems are related by
\beq
  \tilde{\phi} = \left[ \frac{n(n{+}1)}{3m} + 
  \frac{3m(x_{*}-2m)}{x_{*}^{2}(nx_{*}+3m)}\right]\phi+\frac{d}{dx}\phi\quad,
\eeql{eq:bhsusy}
which has been referred to as ``intertwining"\cite{price}.
Eq.~(\ref{eq:bhsusy}) is exactly a SUSY transformation, if we identify
\beq
  W(x) = \frac{n(n{+}1)}{3m} + 
  \frac{3m(x_{*}-2m)}{x_{*}^{2}(nx_{*}+3m)} \quad .
\eeql{eq:wbh}

The corresponding generator $\Phi$ must be an eigenfunction of the RW equation with eigenvalue $\Omega^2$, where
\beq
  \Omega= -i\frac{n(n{+}1)}{3m}\quad .
\eeql{eq:spmode}
This is exactly the so-called algebraically special frequency\cite{spmode,onozawa,and}. With the same eigenvalue,
${\tilde \Phi} = \Phi^{-1}$ should be an eigenfunction of the Zerilli equation.  Eq.~(\ref{eq:wbh}) shows that
$W_+=i\Omega$; hence, $\Phi$ is incoming from the right. Since $\Omega$ is not special for $x\rightarrow\infty$ (cf.\
Appendix~\ref{sect:apptail}; this can easily be verified using the Leaver series solution~\cite{leaver}), this
already establishes that $\Phi$ is not a QNM --- as has sometimes been suggested\cite{onozawa}. However,
$W_-=i\Omega$ does not prove conclusively that it is outgoing to the left (cf.\ Appendix~\ref{tail-pre}), as implied
by a conflicting suggestion that $\Phi$ may be a TTM$_{\rm R}$\cite{and}. In fact it can be shown that
$\Phi(x{\rightarrow}{-\infty})$ is {\em not\/} outgoing, so that $\Phi$ is not a TTM$_{\rm R}$ and (\ref{eq:bhsusy})
is a Type~4 transform, being of category (b2) on the left in the terminology of Appendix~\ref{susy1wave}. However,
these and other aspects will be reported separately. 

%==========================================================
%==========================================================

\section{Orthonormality}
\label{sect:orth}

%==========================================================

\subsection{Orthonormality for NMs}
\label{subsect:orthnm}

In the familiar discussion of SUSY for NMs, orthonormality
can be preserved.  There are two issues:
(a) orthogonality is preserved because the transformed
NMs are eigenvectors of the self-adjoint operator 
${\tilde H}$; and  
(b) normalization is preserved if the transformation
is changed to

\beq
\phi_n \mapsto {\underline {\tilde \phi}}{}_n
= N_n {\tilde \phi}_n  = N_n A \phi_n
\quad ,
\eeql{eq:ortho01}

\nid
for each eigenstate $n$ other than the ground state $\Phi$,
with 

\beq
N_n^{-2} = \frac{
\langle {\tilde \phi}_n | {\tilde \phi}_n \rangle }
{ \langle \phi_n | \phi_n \rangle }
\quad .
\eeql{eq:ortho02}

\nid
It is readily shown that $N_n^{-2} = \omega_n^2 - \Omega^2$,
a result that can also be read off as a special case
of the derivation below for QNMs.  Since
$\omega_n^2 - \Omega^2$ is the eigenvalue of the operator
$A^{\dagger} A$ (see (\ref{eq:hop})), 
the normalized SUSY transformation (\ref{eq:ortho01})
can also be written in operator form

\beq
\phi \mapsto {\underline {\tilde \phi}}
= A \left( A^{\dagger} A \right)^{-1/2} \phi
\quad ,
\eeql{eq:ortho03}

\nid
valid for any state $\phi$, not just
frequency eigenfunctions.
In this and similar formulas below, it will be
understood that the domain is restricted to
the subspace orthogonal to $\Phi$,
on which therefore $A^{\dagger}A$ is non-zero.
This makes it formally easy to
verify the preservation of inner products
in the mapping from this subspace:

\bea
\langle {\underline {\tilde \psi}} | {\underline {\tilde \phi}} \rangle
&=&
\langle \psi | (A^{\dagger}A)^{-1/2} \, A^{\dagger} A \, 
\, (A^{\dagger}A)^{-1/2} |  \phi \rangle
\nonumber \\
&=& \langle \psi | \phi \rangle 
\quad .
\eeal{eq:ortho04}

\nid
However, when operating on a
general wavefunction $\phi$, the factor $(A^{\dagger}A)^{-1/2}$
can only be evaluated by projecting $\phi$ onto the eigenfunctions,
and scaling each component by $N_n$.  Thus, in practice,
the significant result is the evaluation
of the factor $N_n$.
We now generalize these concepts to QNMs.

%==========================================================

\subsection{Normalization and inner product for QNMs}
\label{subsect:qnmnorm}

It is necessary to first digress and briefly review the
concepts of orthogonality and normalization for QNMs.
The central issue is that with the outgoing-wave
boundary condition, $H$ is {\em not\/} self-adjoint
in the usual sense, and different QNMs are
{\em not\/} orthogonal under the usual inner product.  Likewise,
the norm $\int |\phi_n|^2 dx$ is divergent
for a QNM, since the wavefunction grows exponentially
at spatial infinity.

An appropriate normalizing factor for QNMs was first introduced
by Zeldovich~\cite{zel}, and later generalized 
and applied to other
situations~\cite{zelgen}, including models of
linearized waves propagating on a background that is
a Schwarzschild black hole plus a perturbation~\cite{dirt}:

\bea
( \phi_n , \phi_n ) &=&
 2 \omega_n \int_{-a}^{a} \phi_n(x)^2 \, dx
\nonumber \\
&&{} + i \left[ \phi_n(-a)^2 + \phi_n(a)^2 \right]
\quad .
\eeal{eq:ortho05}

\nid
This expression goes as $\phi_n^2$ rather than
$|\phi_n|^2$, and is in general not real.
The upper limit of the integral and the first surface term
can be shifted to any $b_1 > a$ without affecting the
value of (\ref{eq:ortho05}); likewise, 
the lower limit and the second surface term
can be shifted to any $b_2 < -a$.  This 
expression is the correct
normalizing factor in the sense that, e.g., under a
perturbation $V \mapsto V + \Delta V$, the complex eigenvalues
of the QNMs change by

\beq
\Delta (\omega_n^2) = 
\frac{ \int \phi_n(x)^2 \Delta V(x) \,dx }
{ (\phi_n , \phi_n ) }
\quad .
\eeql{eq:ortho06}

\nid
Since one no longer has positivity, there is the possibility
that $(\phi_n , \phi_n) =0$.
This exceptional case can be separately
taken care of~\cite{jordan}, and some interesting
aspects are dealt with in Section~\ref{sect:double}. 

To go beyond the normalizing factor and
discuss the analog of orthogonality, one has to first
regard each state as a two-component vector 
$\bbox{\psi} = ( \psi^1 , \psi^2 )^{\rm T} \equiv 
( \psi, \partial_t \psi )^{\rm T}$,
which is most easily motivated by noticing that the
dynamics requires two sets of initial data.
In terms of the two-component vector, 
one can define a bilinear map~\cite{tong1,tong2,rmp}

\bea
( \bbox{\psi} , \bbox{\phi} )
&=& i \left\{ \int_{-a}^{a} \left[ \psi^1(x) \phi^2(x) +
\psi^2(x) \phi^1(x) \right] dx \right.
\nonumber \\
&&{}+\left. \vphantom{\int_{-a}^{a}}
\left[ \psi^1(-a) \phi^1(-a) + \psi^1(a) \phi^1(a) \right] \right\}
\quad .
\eeal{eq:ortho13}

\nid
This will be seen to take the place of the usual
inner product.  Note that for an eigenfunction,
$( \bbox{\phi}_n , \bbox{\phi}_n )$ agrees with (\ref{eq:ortho05}).
The dynamics can be written in the first-order form
$i\partial_t \bbox{\phi} = {\cal H} \bbox{\phi}$,
with the $2 \times 2$ evolution operator

\beq
{\cal H} = i \pmatrix{ 0 & 1 \cr \partial_x^2 - V & 0 }
\quad .
\eeql{eq:ortho14}

\nid
The important property is that ${\cal H}$ is
symmetric under the bilinear map:

\beq
( {\cal H} \bbox{\psi} , \bbox{\phi} )
= ( \bbox{\psi} , {\cal H} \bbox{\phi} )
\quad ,
\eeql{eq:ortho07}

\nid
in the proof of which the surface terms generated
in the integration by parts exactly cancel against
those in (\ref{eq:ortho13}).
The relation
(\ref{eq:ortho07}) is the analog of self-adjointness, and
leads to the usual proof that
for two eigenvectors,

\beq
( \bbox{\phi}_m , \bbox{\phi}_n ) = 0
\eeql{eq:ortho08}

\nid
whenever $\omega_m \neq \omega_n$.  Provided that $( \bbox{\phi}_n , \bbox{\phi}_n ) \neq 0$~\cite{jordan},
one can normalize these eigenfunctions in the usual way,
i.e., by requiring (\ref{eq:ortho08}) to be 
$2 \omega_m \delta_{mn}$
in general.  We shall henceforth call this
property orthonormality (and (\ref{eq:ortho08}) alone
as orthogonality), with the understanding that reference
is made to the bilinear map (\ref{eq:ortho13}) rather than
the usual inner product.  It also follows trivially that, provided
this orthonormal system is complete
(which is the case under fairly broad assumptions, see
Section~\ref{sect:intro}),
then the time evolution of a wavefunction $\bbox{\psi}(t)$
is given by

\beq
\bbox{\psi}(t) = \sum_n a_n \bbox{\phi}_n \, e^{-i\omega_n t}
\quad ,
\eeql{eq:ortho09}

\nid
where the $a_n$ are obtained by projecting the
initial data:

\beq
a_n = \frac{ (  \bbox{\phi}_n , \bbox{\psi}(t{=}0) ) }
{ ( \bbox{\phi}_n , \bbox{\phi}_n ) }
\quad .
\eeql{eq:ortho10}

\nid
The preservation of 
orthonormality under SUSY should therefore be sought in terms of
the bilinear map (\ref{eq:ortho13}).

%==========================================================

\subsection{Normalization factor for QNMs under SUSY transformation}
\label{subsect:orthqnm}

We first present a derivation
of orthonormality that does not explicitly
require the two-component formalism.
With orthogonality in the sense of the
bilinear map already guaranteed by (\ref{eq:ortho08}),
it remains only to compute the normalizing factor

\bea
({\tilde \phi}_n , {\tilde \phi}_n )
&=& 2\omega_n  \int_{-a}^{a} 
\left[ (\partial_x + W) \phi_n \right]^2 dx
\nonumber \\
&&{} + i \left[ {\tilde \phi}_n(-a)^2 + {\tilde \phi}_n(a)^2 \right]
\quad.
\eeal{eq:ortho11}

\nid
Integrate by parts to convert 
$(\partial_x \phi)^2$ to $-(\partial_x^2 \phi)\phi$
plus a surface term,
express the second derivative in terms of $V - \omega_n^2$
by means of the eigenvalue equation, and write the potential
as $V = W^2 - W'+ \Omega^2$.  Then, apart from 
a term proportional to $\omega_n^2 - \Omega^2$,
the integrand becomes a total
derivative $\partial_x(W\phi^2)$.  
Using $W(\pm a)^2 = -\Omega^2$
and $\partial_x \phi_n (\pm a) = \pm i \omega_n \phi_n(\pm a)$
then leads to

\beq
\frac{ ({\tilde \phi}_n , {\tilde \phi}_n ) }
{(\phi_n , \phi_n ) }
= \omega_n^2 - \Omega^2   
\quad .
\eeql{eq:ortho12}

\nid
Incidentally, the case of NMs on $[-a,a]$ (which corresponds to
nodal conditions at the ends of the interval)
can be recovered by simply dropping all surface terms.

Since the ratio (\ref{eq:ortho12}) takes the same form as for NMs, and is the eigenvalue of $A^{\dagger}A$, we can
again write the normalized SUSY transformation 
for each eigenfunction as (\ref{eq:ortho03}).

%==========================================================

\subsection{SUSY transformation for two-component form}
\label{subsect:twocomp}

We have so far deliberately avoided the explicit two-component
form of the SUSY transformation, since, for an 
eigenfunction $\phi_n$, one can always
express $\partial_t \phi_n$ in terms
of $\phi_n$.  However, in order 
to perform SUSY transformations on general outgoing
wavefunctions (e.g., when given a time-dependent state,
to find its SUSY partner at all times), the second
component must be considered.

Since the SUSY transformation must commute with time-evolution
and $\phi^2 = \partial_t \phi^1$, it is clear that both
components must be transformed in the same way.
Thus, the (unnormalized) SUSY transformation on the two-component vector is
${\cal A} = \mbox{diag } ( A , A )$, which satisfies

\beq
( {\cal A} \bbox{\psi} , \bbox{\phi} )
= ( \bbox{\psi} , {\cal A}^{\dagger} \bbox{\phi} )
\quad ,
\eeql{eq:ortho17}

\nid
where ${\cal A}^{\dagger} \equiv \mbox{diag } ( A^{\dagger} , A^{\dagger} )$.
In deriving (\ref{eq:ortho17}), one has to integrate 
by parts.  In the integrand, 
${\cal A}$ turns into ${\cal A}^{\dagger}$ because
$A$ contains a single derivative $\partial_x$,
which reverses sign upon integration by parts.
The surface terms are seen to work out by using,
e.g., $\phi^2(a) = \partial_t \phi^1(a)
= -\partial_x \phi^1(a)$, and the known values of $W_{\pm}$.

Note that ${\cal A}^{\dagger} {\cal A} = ( H - \Omega^2 ) \openone$,
${\cal A} {\cal A}^{\dagger}  = ( {\tilde H} - \Omega^2 ) \openone$,
i.e., these products do not relate to the two-component
${\cal H}$.

With (\ref{eq:ortho17}), it is 
straightforward to show that the normalized SUSY transformation

\beq
\bbox{\phi} \mapsto \underline{\tilde {\bbox{\phi}}} =
{\cal A} \, ( {\cal A}^{\dagger} {\cal A} )^{-1/2} \bbox{\phi}
\quad ,
\eeql{eq:ortho18}

\nid
defined on the subspace orthogonal to ${\bbox {\Phi}}$,
preserves the bilinear map, in a manner that exactly parallels
(\ref{eq:ortho04}).  In the exceptional case of SUSYs that generate
a doubled mode (see Section~\ref{sect:double}), the subspace
has to the exclude {\em two\/} states on which
$H - \Omega^2$ vanishes.

These properties show that two apparently unrelated concepts,
namely SUSY and the linear-space structure for QNMs in open systems
(e.g., the replacement of inner products by the bilinear
map), turn out to be consistent.  This should not be
surprising: the bilinear map (\ref{eq:ortho13})
has an intrinsic geometric meaning for all outgoing states,
not just QNMs~\cite{factor}.

%==========================================================
%==========================================================

\section{Doubled modes}
\label{sect:double}

%==========================================================

\subsection{Examples}
\label{sect:doubleex}

In this section, we deal with the possibility
that doubled modes may be produced by a SUSY transformation.
Consider a system $H$ with an NM $\Phi(x)$ at 
$\Omega = iK$ and  
a QNM $\Psi(x)$ at $-\Omega= -iK$.
Then both ${\tilde \Phi}(x) = \Phi(x)^{-1}$
and ${\tilde \Psi}(x) = A \Psi(x)$
are QNMs of ${\tilde H}$ at $-\Omega$.
Since there can only be {\em one\/} QNM at
any frequency\cite{jordan},
${\tilde \Psi}(x) \propto {\tilde \Phi}(x)$.
The proportionality constant can be evaluated by

\bea
\frac{\tilde{\Psi}}{\tilde{\Phi}}
&=& \frac{\tilde{\Psi}(-a)} {\tilde{\Phi}(-a)}
=2i \Omega\Psi(-a) \Phi(-a)
\nonumber\\
&=& \frac{\tilde{\Psi}(a)} {\tilde{\Phi}(a)}
= -2i\Omega\Psi(a)\Phi(a)
\quad.
\eeal{eq:doub01}

\nid
Incidentally, the agreement of these two expressions can also be 
seen without invoking SUSY.  First one notes that $\Phi$ and $\Psi$,
being eigenfunctions of ${\cal H}$
with distinct eigenvalues, are orthogonal.  In the expression
for the bilinear map, the integral term vanishes because
the frequencies are opposite,
leaving only the surface terms.  Thus one finds
$0=(\Psi,\Phi)=i[\Psi(-a)\Phi(-a)+\Psi(a)\Phi(a)]$.

Secondly, consider a perturbation
of $H$ which splits the eigenvalues (originally
both $\Omega^2$) of $\Phi$ and $\Psi$,
in which case their counterparts 
${\tilde \Phi}(x) = \Phi(x)^{-1}$ and $A \Psi(x)$ must
also split into two states.  Thus the one state represented
by $A \Psi \propto \Phi^{-1}$ must be the limit of two states that
coalesce when the splitting is switched off, and in some
sense to be made precise below must count as a {\em doubled\/}
state.  

This situation has already been encountered.  Consider
a PT potential with $q = 1$.
Its modes include (see (\ref{eq:expt04}))
the NM $\Phi$ at $\omega_0^{-} = i/2$ and 
the QNM $\Psi$ at $\omega_1^{-} = -i/2$,
both with eigenvalue $\Omega^2 = -1/4$.
Use $\Phi$ to generate
the SUSY partner with ${\tilde q} =0$.  Its QNMs 
${\tilde \Phi}$ at ${\tilde \omega}_0^{+}$ and 
${\tilde \Psi} = A \Psi$ at ${\tilde \omega}_0^{-}$ 
both occur at the same eigenvalue $-1/4$.
(Other pairs also coalesce, but this particular pair
is special in that one of the two states is the reciprocal
of the generator.)
This pair of states form a doubled mode, and we have
verified that $\tilde{\Psi}\propto\tilde{\Phi}$ (cf.\ (\ref{eq:doub01})) indeed holds in this example.

As another example,  
consider a system $H$ with a TTM${}_{\rm L}$ $\Phi(x)$ at an eigenvalue
$\Omega^2$.
The partner system ${\tilde H}$
must have a TTM${}_{\rm R}$ ${\tilde \Phi}(x) = \Phi(x)^{-1}$ at the
same eigenvalue.  But if $H$ should have a TTM${}_{\rm R}$ 
$\Psi(x)$ also
with the eigenvalue $\Omega^2$, then ${\tilde \Psi}(x) = A \Psi(x)$
is a TTM${}_{\rm R}$ of ${\tilde H}$ with the eigenvalue $\Omega^2$
as well.  As before, we expect these states to be proportional, and a calculation analogous to (\ref{eq:doub01}) shows that in
this case

\bea
  \frac{\tilde{\Psi}}{\tilde{\Phi}}&=&-2i\Omega\Psi(-a)\Phi(-a)\nonumber\\
  &=&-2i\Omega\Psi(a)\Phi(a)\quad.
\eeal{eq:doub02}

Again, this situation has been encountered.  Consider
the example in Figures 4 and~5.  Since $V$
is symmetric, it is guaranteed that
for every TTM${}_{\rm L}$ $\Phi(x)$ there will also be a TTM${}_{\rm R}$ 
$\Psi(x) = \Phi(-x)$
at the same eigenvalue.  With this substitution for $\Psi$, the condition (\ref{eq:doub02}) is readily verified for any symmetric potential.

%==========================================================

\subsection{Double zeros of Wronskians}
\label{sect:wron}

In this subsection we make precise what is meant by a doubled mode
and in the next subsection we describe the SUSY transformation
of such doubled modes.
We shall do so for QNMs, the case of TTMs differing only by
some signs.  In the
original system $H$, define outgoing-wave solutions of the wave equation $f(\omega,x)$
and $g(\omega,x)$, satisfying the boundary conditions
\bea
  f(\omega,x{<}{-}a) &=& e^{-i\omega x} \quad , \nonumber \\
  g(\omega,x{>}a) &=& e^{i\omega x} \quad ,
\eeal{eq:doub03}
where $\,' = \partial_x$.
(For TTMs, $e^{i\omega x}$ on the r.h.s.\ also for $f$.)
A QNM is a solution that satisfies {\em both\/} the left
boundary condition (as on $f$) {\em and\/} the right boundary
condition (as on $g$); in other words, at a QNM
frequency, $f$ and $g$ are linearly dependent.
Thus, the zeros of the Wronskian
\bea
  J_{\rm q}(\omega)&\equiv&J(f,g;\omega)\nonumber\\
  &=& f'(\omega,x) g(\omega,x) - f(\omega,x) g'(\omega,x)
\eeal{eq:doub04}
identify the QNMs.  In fact, it is easy to show that the Wronskian
is related to the transmission amplitude introduced earlier by:

\beq
J_{\rm q}(\omega) = - \frac{2i\omega}{T_{\rm L} (\omega)}
\quad .
\eeql{eq:wrontrans}

Now consider the analogous construction in the partner
system ${\tilde H}$, obtained by using an NM $\Phi$
of $H$ as the generator.  By our convention,
$\Phi$ is associated with a frequency $\Omega$ on the
positive imaginary axis, and $W(\pm a) = - \Phi'(\pm a) / \Phi(\pm a)
= \mp i \Omega$.  The SUSY transformation gives

\bea
{\tilde f}(\omega,x) &=& (\partial_x + W) f(\omega,x)
\nonumber \\
{\tilde g}(\omega,x) &=& (\partial_x + W) g(\omega,x)
\quad ,
\eeal{eq:doub05}

\nid
leading to the Wronskian

\bea
{\tilde J}_{\rm u}(\omega) 
&=& 
\left\{\left[ (\partial_x + W) f \right]' 
\left[ (\partial_x + W) g \right] \right.
\nonumber \\
&& \left. - \,\,
\left[ (\partial_x + W) f \right]
\left[ (\partial_x + W) g \right]' \,  \right\}
\quad .
\eeal{eq:doub07}

\nid
This Wronskian is however unnormalized (as indicated by the
subscript), since only $C{\tilde f}$ and $D{\tilde g}$ satisfy the
normalization conventions at $-a$ and $+a$ respectively,
where 

\beq
C = -D = \frac{i}{\omega - \Omega}
\quad .
\eeql{eq:doub06}

\nid
These normalizing factors have no zeros
or poles in the lower half-plane, which is the only
region of interest for QNMs.  Thus the normalized
Wronskian in the partner system is

\bea
{\tilde J}_{\rm q}(\omega) &=&
CD {\tilde J}_{\rm u}(\omega)
\nonumber \\
&=& (\omega - \Omega)^{-2} {\tilde J}_{\rm u}(\omega)
\quad.
\eeal{eq:doub06a}

When (\ref{eq:doub07}) is written out, some terms cancel by using
$J_{\rm q}' = 0$, the second derivatives can be eliminated
by the defining equation, and $V$ is expressed in terms
of $W$ and~$\Omega^2$.  Some arithmetic then leads to
\bea
  {\tilde J}_{\rm u}(\omega) &=&
  (\omega^2-\Omega^2)\,J_{\rm q}(\omega)\quad,\label{eq:doub08a} \\
  {\tilde J}_{\rm q}(\omega)
  &=& \frac{\omega + \Omega} {\omega - \Omega} \, J_{\rm q}(\omega)\quad ,
\eeal{eq:doub08}
which is also readily derived from (\ref{eq:rt3}), since
(\ref{eq:wrontrans}) shows that $J_{\rm q}$ transforms as
$T_{\rm L}^{-1}$ under SUSY.

Eq.~(\ref{eq:doub08}) neatly summarizes the properties
of Type~1 SUSY transformations (cf.\ Table 1):
the QNMs of ${\tilde H}$ and $H$ are the same except that the former
has an extra QNM at $\omega = - \Omega$.  In fact, this
formula can be used in the upper half-plane as well,
where the denominator in (\ref{eq:doub08})
indicates that the NM at $\omega = \Omega$ is absent in
${\tilde H}$.  Moreover, (\ref{eq:ortho12}) on the change
in normalization under SUSY emerges
as a simple consequence of (\ref{eq:doub08a}), 
since the bilinear map is related to the Wronskian by~\cite{rmp}
\beq
  (f(\omega_n),g(\omega_n))=
  -\left[ \frac{\partial J_{\rm q}(\omega)} {\partial \omega} \right]_{\omega_n}
  \quad .
\eeql{eq:fgJ}

Now specialize to the
case where $H$ happens to have a QNM with eigenvalue $\Omega^2$;
this means $J_{\rm q}(\omega)$ has a zero at $\omega = -\Omega$.
By (\ref{eq:doub08}), ${\tilde J}_{\rm q}$ has a
{\em double\/} zero at this point.  This then gives a
precise definition of a doubled mode.  It is clear that
upon perturbation, such a double zero would generically
split up into two simple zeros, i.e., two QNMs.
The existence and properties of double (and higher) zeros
in the Wronskian for QNMs relate to the issue of Jordan blocks (JBs),
which has been extensively discussed recently~\cite{jordan}, and to which
we will return in Section~\ref{sect:susyjb}.

With TTMs, the situation is entirely analogous, and the
double zero refers to a Wronskian defined with boundary
conditions appropriate to the type of TTMs under discussion.
To be specific, consider a SUSY transformation of Type 3a,
which deletes a TTM${}_{\rm L}$ and produces a TTM${}_{\rm R}$
(by convention we only consider TTMs 
in the lower half-plane).
We define Wronskians using the boundary conditions appropriate
to the latter, and in exactly the same way arrive at
(\ref{eq:doub08}).  To interpret the prefactor, we note that
a TTM${}_{\rm R}$ at $\pm \Omega$ is the same as a TTM${}_{\rm L}$
at $\mp \Omega$; thus (\ref{eq:doub08}) in this case simply states
that a TTM${}_{\rm L}$ in the lower 
(i.e., a TTM${}_{\rm R}$ in the upper) half-plane is deleted
while a TTM${}_{\rm R}$ 
in the lower 
(i.e. a TTM${}_{\rm L}$ in the upper) half-plane is produced.

SUSY transformations of Wronskians are discussed in more detail and generality in Appendix~\ref{susy2waves}, in the
context of potentials with tails. 

The coalescence of two modes is similar to degeneracy
(though one of the degrees of freedom is not an eigenstate).
Usually, a degeneracy is associated with some symmetry.
It is therefore interesting to note that in our
example of doubled TTMs, the TTM${}_{\rm L}$ and TTM${}_{\rm R}$
are at the same frequency because of parity; under
SUSY, they then turn into a doubled TTM${}_{\rm R}$ in the
partner system (even though the latter has no parity
invariance).  

%==========================================================

\subsection{SUSY partner of a Jordan block}
\label{sect:susyjb}

Returning to the situation described at the 
beginning of Section~\ref{sect:doubleex}, 
let us examine in some detail the relationship between the 
doubled QNM at $-\Omega$ of $\tilde{H}$ and its SUSY pre-image.  

To conform with the notation of Ref.~\cite{jordan},
denote the state $\Psi$ as $\Psi_j$, where $j$ is a
QNM index.  Now, using the normalization convention
(\ref{eq:doub03}), one has 
$\Psi_j(x)=\Psi_j(-a)f(-\Omega,x)e^{i\Omega a}$, implying 
$\tilde{\Psi}_j(x)=\Psi_j(-a)\tilde{f}(-\Omega,x)e^{i\Omega a}$.
(Analogous formulas involving $\Phi$ and $\tilde{\Phi}$ 
follow by the proportionality (\ref{eq:doub01}) 
and will not be given.) 
Since $\tilde{J}_{\rm u}(\omega)$ has a double zero at $\omega=-\Omega$, 
$\tilde{f}(\omega,x)$ satisfies the outgoing condition at $x=a$ 
not only {\em at\/} $\omega = -\Omega$, but also to first order away from
the pole. This makes it plausible that, in the QNM expansion, 
$\partial_\omega\tilde{f}(\omega,x)|_{-\Omega}$ 
takes the place of the ``missing" eigenfunction when 
$\tilde{\Phi}$ and $\tilde{\Psi}_j$ coincide, 
which has been confirmed in detail\cite{jordan}. One thus defines,
for an arbitrary $\alpha$, a pair of functions ${\tilde \Psi}_{j,n}$,
where the QNM index $j$ labels the JB and $n=0,1$ is an intra-block index:
\bea
  \tilde{\Psi}_{j,0}(x) &=& \tilde{\Psi}_j(x)\nonumber \\
  \tilde{\Psi}_{j,1}(x) &=&
  \Psi(-a) \partial_\omega {\tilde{f}(\omega,x) |}_{-\Omega}e^{i\Omega a}
  + \alpha \tilde{\Psi}_j(x)\quad,
\eeal{eq:Psi01}
where the second function satisfies
\beq
  \bigl( {\tilde H}- \Omega^2 \bigr) 
  {\tilde \Psi}_{j,1} = -2 \Omega {\tilde \Psi}_{j,0}\quad.
\eeql{eq:WEJB}
Using this, one verifies that a time-dependent outgoing 
solution is ${\tilde \Psi}_{j,1}(t) \equiv
( {\tilde \Psi}_{j,1}-it {\tilde \Psi}_{j,0})
e^{i\Omega t}$, showing that the 
associated momentum reads
${\tilde \Psi}_{j,1}^2 = 
-i(-\Omega {\tilde \Psi}_{j,1}+ {\tilde \Psi}_{j,0})$. 
One of the issues to be investigated is the behaviour of the term 
$te^{i\Omega t}$ in $\tilde{\Psi}_{j,1}$ under (inverse) SUSY, 
since such a time dependence cannot occur for $H$.

Next consider the bilinear map for these functions; because
the second components are nontrivial, we explicitly
use the two-component notation.
For a double pole one always has 
$( \tilde{\vec{\Psi}}_{j,0}, \tilde{\vec{\Psi}}_{j,0})=0$, 
cf.~(\ref{eq:fgJ}). 
The undetermined constant $\alpha$ in (\ref{eq:Psi01}) 
can be used to also achieve 
$( \tilde{\vec{\Psi}}_{j,1}, \tilde{\vec{\Psi}}_{j,1})=0$; 
this is useful in wavefunction expansions, 
but for our purposes $\alpha$ can be left arbitrary.

The above merely applies standard JB theory to the double pole 
in the ${\tilde H}$-system. 
Turning now to SUSY transformations, 
one can trivially write 
\bea
  {\tilde \Psi}_{j,1}(x) &=& 
  \Psi(-a) \times \partial_\omega{[(\partial_x+W)f(\omega,x) ]}_{-\Omega}
  e^{i\Omega a}\nonumber \\
  &=&(\partial_x+W)\Psi(-a)\partial_\omega {f(\omega,x)|}_{-\Omega}e^{i\Omega a}
  \quad.
\eeal{eq:doub21}
Thus, $\Psi(-a)\partial_\omega {f(\omega,x)|}_{-\Omega}e^{i\Omega a}$ 
is the SUSY pre-image of ${\tilde \Psi}_{j,1}(x)$. 
However, the former is {\em not\/} outgoing, since
in the $H$-system the Wronskian only has a first-order zero 
at $-\Omega$, and consequently $f(\omega,x)$ satisfies the outgoing-wave
condition at $x=+a$ only {\em at\/} $-\Omega$, and not to first
order away from the eigenvalue.

The normalization of a JB is determined by one overall factor, 
which equals the bilinear map between the two basis states~\cite{jordan}. 
In the present case it is evaluated as
\bea
  \frac{ ( \tilde{\vec{\Psi}}_{j,1}, \tilde{\vec{\Psi}}_{j,0}) }
  {(\vec{\Psi}_j,\vec{\Psi}_j)}&=&
  \left[ \frac{-\sfrac{1}{2}\partial_\omega^2\tilde{J}_{\rm u}(\omega) }
  {-\partial_\omega J_{\rm q}(\omega)} \right]_{-\Omega}\nonumber \\
  &=& -2\Omega\quad ,
\eeal{eq:normjb}
where in the numerator of the first line we have used the result analogous to 
(\ref{eq:fgJ}) for a double pole~\cite{jordan}. 

Let us finally consider the reverse transform generated by 
$-A^{\dagger}$\cite{missing}. It is easy to establish
the following three properties.
(a) $A^{\dagger}{\tilde \Psi}_{j,0} \propto A^{\dagger}{\tilde \Phi}=0$. 
(b) Hence in $A^{\dagger} {\tilde \Psi}_{j,1}(t)$, 
the term $\propto te^{i\Omega t}$ is annihilated.
(c) The remaining term in $A^{\dagger} {\tilde \Psi}_{j,1}$
is $c \Psi_j$.  
The last property is readily seen by observing that
\bea
  (H-\Omega^2) (A^{\dagger}{\tilde \Psi}_{j,1})&=&
  A^{\dagger}({\tilde H}-\Omega^2) {\tilde \Psi}_{j,1}\nonumber \\
  &=& A^{\dagger}(-2\Omega {\tilde \Psi}_{j,0} ) \nonumber \\
  &=& 0\quad ,
\eeal{eq:jb02}
so that $A^{\dagger}{\tilde \Psi}_{j,1}$ is an eigenfunction
of $H$ with eigenvalue~$\Omega^2$.
We have also noted that the 
SUSY pre-image of ${\tilde \Psi}_{j,1}$ 
(i.e., $\partial_\omega{f|}_{-\Omega}$) 
is not an outgoing solution at all.  These remarks
completely resolve the puzzle related to the prefactor $t$
in the time evolution in the ${\tilde H}$-system.

It is instructive to compare various calculations of the
constant of proportionality $c$ above. 
Defining $\bar{\Psi}_j=A^{\dagger}\tilde{\Psi}_{j,1}$ 
(to avoid double stacked tildes), 
in the two-component formalism one can calculate 

\bea
(\bar{\vec{\Psi}}_j,\bar{\vec{\Psi}}_j)
&=& (\tilde{\vec{\Psi}}_{j,1},{\cal AA}^{\dagger}\tilde{\vec{\Psi}}_{j,1})
\nonumber \\
&=& (\tilde{\vec{\Psi}}_{j,1},[\tilde{H}-\Omega^2]
\openone\tilde{\vec{\Psi}}_{j,1})
\nonumber \\
&=& -2\Omega(\tilde{\vec{\Psi}}_{j,1},\tilde{\vec{\Psi}}_{j,0})
\nonumber \\
&=& 4\Omega^2(\vec{\Psi}_j,\vec{\Psi}_j)
\quad.
\eeal{eq:calratio}

\nid
One can get the same result from (\ref{eq:doub08a}), 
for which one has to realize that in 
$\bar{J}_{\rm u}(\omega)=(\omega^2-(-\Omega)^2)\tilde{J}_{\rm q}(\omega)$, 
one overall factor $(\omega+\Omega)^2$ is associated with
$A^{\dagger}\tilde{f}(-\Omega,x)=0$ and should be removed. 
Thus, one should evaluate

\bea
\frac{(\bar{\vec{\Psi}}_j,\bar{\vec{\Psi}}_j)}
{(\tilde{\vec{\Psi}}_{j,1},\tilde{\vec{\Psi}}_{j,0})}&=&
\left\{
\frac{-\partial_\omega[(\omega+\Omega)^{-2}\bar{J}_{\rm u}(\omega)]}
{-\sfrac{1}{2}\partial_\omega^2 \tilde{J}_{\rm q}(\omega)}
\right\}_{\!-\Omega}
\nonumber\\
&=&-2\Omega\quad.
\eeal{eq:calratio2}

\nid
For simple poles, one also knows the sign left undetermined by the 
bilinear map: $\bar{\phi}_n \equiv A^{\dagger}\tilde{\phi}_n
= A^{\dagger}A\phi_n=(\omega_n^2-\Omega^2)\phi_n$. 
For JBs, the corresponding calculation is 

\bea
\bar{\Psi}_j &=& A^{\dagger}A\Psi(-a)\partial_\omega{f|}_{-\Omega}e^{i\Omega a}
\nonumber \\
&=& \Psi(-a)\partial_\omega{[A^{\dagger}A f]}_{-\Omega}e^{i\Omega a}
\nonumber \\
&=& \Psi(-a)\partial_\omega{[ (H-\Omega^2) f]}_{-\Omega}e^{i\Omega a}
\nonumber \\
&=& \Psi(-a)\partial_\omega{[ (\omega^2-\Omega^2) f]}_{-\Omega}e^{i\Omega a}
\nonumber \\
&=& -2\Omega \Psi_j
\quad .
\eeal{eq:calcratio3}

Incidentally, if the eigenvalues $\omega_j^2$ of the JB and
$\Omega^2$ of the SUSY auxiliary function are not equal,
the transformation leaves the JB structure unmodified.
An example of this situation has already been encountered
in the PT potential with $q = 1$, for the
QNMs with $\omega_j^2 \neq -\sfrac{1}{4}$.
We omit the details, which are straightforward, as is
the generalization to higher-order multiple
poles.

%==========================================================
%==========================================================

\section{Discussion}
\label{sect:disc}

In this paper, we have extended the usual discussion
of SUSY as a relation between NMs of partner systems
to include the QNMs as well; TTMs also
come in naturally.  By viewing all these together,
a more complete picture emerges, most conveniently
summarized by Table~1 or say (\ref{eq:doub08}).  For example, in the usual
discussion for NMs only, essentially isospectral systems
are described as being identical except for one NM
present in one system but absent in the other;
now we see that (under Types 1 and~2 transformations)
when an NM appears (disappears), a corresponding QNM
disappears (appears).

QNMs differ from NMs in two regards.  First, they have
complex frequencies; nevertheless, even with 
twice as many constraints, matching the spectra
is not any more difficult.  Second, QNMs need not have
an increasing number of nodes, and it is often possible
to find several nodeless QNMs which generate
distinct SUSY transformations --- whereas the analogous
operation for NMs would be restricted to the nodeless ground state.

The general formalism also provides a framework in which
a number of well-known results find a convenient 
and unified expression.  These include (Types 1 and 2)
SUSY transformations
relating different PT potentials of the same width,
as well as (Type 4) SUSY transformations that take
the free field into a nontrivial PT potential.
The latter property provides one way of understanding the
total-transmission property (at all positive energies)
of a class of PT potentials.  The transformation between
the RW and Zerilli potentials can also
be placed into this framework.

This wider perspective is
gained only because attention is paid to the Klein--Gordon
rather than the Schr\"odinger equation, since the
concept of outgoing waves has no meaning in an equation
that is first-order in time. 

SUSY also preserves the orthonormality
of QNMs in the sense of the bilinear map --- supporting
the view that both the extension of SUSY and the generalization
of the inner product are sensible and useful.

Two further important properties are also preserved. (a) If $V$ has a singularity say at $x=\pm a$ (e.g., a step),
then ${\tilde V}$ will have the same type of singularity (but with the opposite sign).  This is illustrated by many
of the figures, and can be seen from (\ref{eq:vw}) by noticing that the most singular part is $W'$. (b)~If $V$ has no
tail, then for transformations of Types 1--3, ${\tilde V}$ likewise has no tail. These two properties are precisely
the necessary conditions for the QNMs (plus any possible NMs) to be complete~\cite{rmp}.  Thus, SUSY maps a complete
basis to a complete basis, if for Types 1 and~2 $\Phi \mapsto {\tilde \Phi}$ is included as well. 

Finally, this work partially answers the
question of QNM {\em inversion\/}: does one set of
QNM (again plus NM) frequencies (roughly speaking carrying twice the
information of one set of NM frequencies in closed systems) uniquely
determine $V$?  In general the answer is negative,
for there can be strictly isospectral potentials:
see Types 3a and~3b in Table~1, and also
Figure 5.  However, if we consider a half-line problem $x>0$
(say the radial variable in a 3-d system),
imposing a nodal condition at $x=0$ and the outgoing-wave
condition at $x > a$, can 
one set of QNMs uniquely determine $V$?
SUSY transformations (\ref{eq:map}) do not directly rule out
this scenario --- for which there is some numerical 
evidence~\cite{numinv2} --- since these one-sided systems
do not feature TTMs generating
strictly isopectral partners.  Moreover, by
(\ref{eq:vw}) and (\ref{eq:wphi}) the nodal condition maps
a regular $V$ to ${\tilde V} \sim 2/x^2$ for $x \rightarrow 0^+$
(generalizing the result that SUSY increases the
angular momentum by one unit in the hydrogen atom).
It would therefore be interesting to see if an enlarged
class of transformations can address this question.

%==========================================================
%==========================================================

\acknowledgments
The work is supported in part
by a grant (CUHK 4006/98P) from the Hong Kong Research Grants 
Council.  We thank Y.~T.~Liu,
C.~P.~Sun, B.~Y.~Tong and Jianzu Zhang for discussions.

%==========================================================

\appendix

%==========================================================
%==========================================================

\section{Transformation between the wave
and Klein--Gordon equations}

\label{sect:appkgwe}

In several previous papers\cite{comp}, 
we have already commented on the relation between the 
KGE studied in the main text and 
the wave equation (WE) (\ref{wave-eq}) below. However, 
this relation exhibits some subtleties in the presence 
of bound states, or when the domain is the full line. Since the main text has shown both of these 
possibilities to be relevant to SUSY, 
in this appendix the transformation between the two equations 
will be investigated in detail.

Consider the WE

\beq
  [\rho(z)\partial_t^2-\partial_z^2]\psi(z,t)=0\quad.
\eeql{wave-eq}

\nid
In mechanics and optics, $\rho$ is the mass density and the refractive index squared respectively, and thus is
positive. Violation of $\rho>0$ would also make the time evolution singular and would take one outside the class of
hyperbolic equations; we the refore exclude this. Then, one can define $x$ by $dx/dz=n(z)\equiv\sqrt{\rho(z)}$, 
$\phi\equiv\sqrt{n}\psi$, and

\beq
  V=\frac{1}{2n^3}\partial_z^2n-\frac{3}{4n^4}(\partial_zn)^2\quad,
\eeql{Vn}

\nid
upon which $\phi(x,t)$ is seen to satisfy the KGE (\ref{eq:kgt}). 
The transformation determines $x$ up to an immaterial additive constant, 
which can for instance be fixed by letting $z=0$ map to $x=0$. 
Clearly, if $\rho(z)\rightarrow1$ 
for $|z|\rightarrow\infty$ then $V(x)\rightarrow0$ 
for $|x|\rightarrow\infty$ (excluding pathologies).

The WE $\mapsto$ KGE transform thus is 
well-behaved at infinity, but singularities in $\rho$ require caution. 
While a kink in $\rho$ leads to an admissible $\delta$-peak in $V$, 
already a step in $\rho$ is too bad: 
$\int V\,dx=\int Vn\,dz=n'/2n^2+\int(n')^2/4n^3\,dz$, 
with the second term diverging if $n$ tends to a step 
through a sequence of smooth functions. 
The difficulty stems from the fact that both the WE and the KGE 
become distributionally undefined if $\rho$ (or $V$) is more 
singular than a $\delta$-function. The latter is the marginal case, 
in which the $\delta$-function multiplies a wave function which 
is not differentiable but still continuous.

Consider the two-component formalism. 
Since the WE $\mapsto$ KGE transform leaves $\partial_t$ invariant, 
one has 
$\phi^2=\dot{\phi}^1=\sqrt{n}\dot{\psi}^1=n^{-3/2}\psi^2$. 
For the bilinear map this leads to
\bea
(\vec{\phi}_1,&&\vec{\phi}_2{)}_{\rm KGE}\nonumber \\
=&&
  i\biggl[\int_\alpha^\beta\!\!(\phi_1^1\phi_2^2+\phi_1^2\phi_2^1)\,dx
  +\phi_1^1(\alpha)\phi_2^1(\alpha)+\phi_1^1(\beta)\phi_2^1(\beta)\biggr]  
  \nonumber\\
=&&
  i\biggl[\int_{-a}^a\!\!(\sqrt{n}\psi_1^1n^{-3/2}\psi_2^2
     +n^{-3/2}\psi_1^2\sqrt{n}\psi_2^1)n\,dz
  \nonumber \\
&&  
  \phantom{i\biggl[} {}+\psi_1^1(-a)\psi_2^1(-a)
  +\psi_1^1(a)\phi_2^1(a)\biggr]\nonumber\\
=&&
  {(\vec{\psi}_1,\vec{\psi}_2)}_{\rm WE}\quad,
\eeal{eq:eqinner}

\nid
where $\alpha$ and $\beta$ are the images of $-a$ and $a$ 
under the $z\mapsto x$ map respectively, and where in 
the surface 
terms we have used the fact that $n$ must be continuous for the transformation to exist. 
Thus, the WE $\mapsto$ KGE map leaves the linear structure for 
open systems invariant.

When investigating the inverse transformation, 
the form (\ref{Vn}) is not immediately useful since there is no 
variable $z$ to start with. Simple rewriting yields $Hq=0$, 
$dz/dx=q(x)^{-2}$, $\phi=q\psi$ with $\rho=n^2=q^4$. 
The boundary condition to be used is $\lim_{x\rightarrow\infty}q=1$. 
As long as $q>0$, one sees that $n>0$ and hence $\rho>0$ as well. 
However, at a zero of $q$, $dz/dx$ suffers a non-integrable 
singularity and the transformation breaks down. 
For $V>0$ (in field theory, in the absence of broken symmetry $V=m^2>0$) 
this never happens, while if $V(x)<0$ for some $x$, 
the transformation always exists locally but not necessarily globally.

More precisely, the transformation exists iff $H$ has no bound state. 
(For the KGE, the main text shows that these lead to 
outgoing solutions which grow exponentially in time, 
evidently having no WE counterparts.) Namely, 
the breakdown of the transformation means that the bounded 
$\omega=0$ wave function has a node, implying that there is an 
integrable eigenstate with lower $\omega^2$.

For a KGE defined on the positive half-line (with a nodal condition on the fields at the origin), naturally the case
in which $q(x)$ has a single node at $x=0$ is marginal, and $0<x<\infty$ can just be mapped onto the entire $z$-axis. 
For problems defined on the full line, another subtlety shows up for $x\rightarrow-\infty$, where even in the absence
of nodes in $q$ the transformation is regular only if $H$ has a mode at $\omega=0$ (cf.\ Appendix~\ref{Omega0}),
i.e., if $q$ tends to a constant for $x\rightarrow-\infty$ as well. For all other $V$ (which includes all nontrivial
nonnegative potentials), $q$ grows linearly for $x\rightarrow-\infty$, and the $x$-axis maps to a semi-infinite
$z$-interval with a singularity in $\rho$ at the lower boundary.  Of course, by supplementing the differential
equation for $q$ with different boundary or initial conditions, in this latter case one can also map the $x$-axis
onto the negative-$z$ semi-axis, or onto a bounded $z$-interval. 

%==========================================================
%==========================================================

\section{Potentials with tails}
\label{sect:apptail}

\subsection{Preliminaries}
\label{tail-pre}

Determining outgoing waves when the potential has a tail is delicate. Namely, whereas the boundary condition
(\ref{eq:out2}) is readily implemented at finite $x=\pm a$ whenever these enclose the support of $V$, for potentials
with tails the condition can only be imposed asymptotically. However, if $\im\omega<0$ then
$\lim_{x\rightarrow\infty}\phi'(x)/\phi(x)=i\omega$ for {\em any\/} solution at eigenvalue $\omega^2$ except the
(unique up to a prefactor) small one, so that this criterion does not identify the outgoing wave. 

If $V$ vanishes at infinity faster than any exponential, one can impose the boundary condition at some $x=L$ to find
an approximate outgoing wave, which will converge to the exact one as $L\rightarrow\infty$. In practice, for any
required accuracy one thu s can simply truncate $V$, so that this case is still essentially the same as the one with
no tail. 

If however $V(x)\sim e^{-\lambda x}$, the above will only work for $\im\omega>-\lambda/2$, thus identifying only QNMs
within a band below the real axis. By using the $m$th-order Born approximation for the asymptotic region, one can
expand this band to $\i m\omega>-(m{+}1)\lambda/2$\cite{pottail}. 

For more heavily damped QNMs and for potentials which vanish sub-exponentially (i.e., $\lambda=0$ in the above),
sometimes (semi-)analytic solutions enable one to continue the outgoing wave from the upper into the lower-half
$\omega$-plane, cf.\ Section~\ ref{onewave}. Fortunately, this is the case for the PT, RW, and Zerilli potentials. In
general, however, one needs sophisticated numerical techniques such as phase integrals, which involve integrating the
KGE along carefully chosen paths in the $x$-plane\ cite{and}.

\subsection{One outgoing wave}
\label{onewave}

For general potentials the calculation of outgoing waves thus can present difficulties; in fact already their
definition needs some care. Therefore define solutions of the KGE $f(\omega,x)$ and $g(\omega,x)$ such that for
$\im\omega>0$, $f(\omega,x)\sim 1 \cdot e^{-i\omega x}$ for $x\rightarrow-\infty$ and $g(\omega,x)\sim1\cdot
e^{i\omega x}$ for $x\rightarrow\infty$. This makes $f$ and $g$ analytic in this half-plane, and they can be
analytically continued to the whole $\omega$-plane, with a cut on the negative imaginary axis. Functions $\propto f$
and $\propto g$ are said to be outgoing to the left and right, respectively, at frequency $\omega$; a function is
incoming at $\omega$ iff it is outgoing at $-\omega$. 

As a consequence, one will have $g(-\omega^*,x)=g(\omega,x)^*$ and $g(\omega,x{\rightarrow}\infty)\sim1\cdot
e^{i\omega x}$ for generic points in the whole $\omega$-plane. It can happen that the latter relation does not hold
at isolated points $\omega'$ w here $g$ diverges; we will call such frequencies {\em special points\/} and write
locally $g(\omega)=(\omega-\omega')^{-n}\chi(\omega)$, with $\chi(\omega)$ finite near $\omega=\omega'$ and
$\chi(\omega')\neq0$, and with an integer $n\ge1$, to be called the {\em index\/} of the special point. Near the
(possible) branch point $\omega=0$ the behaviour of $g$ can be more complex, and it also can happen that
$\lim_{\omega\rightarrow0}g(\omega)=0$. 

From the Born series for $g$ one sees that, if $V(x{\rightarrow}\infty)\sim e^{-\lambda x}$, frequencies
$\omega'=-ik\lambda/2$ ($k=1,2,\dots$) {\em can\/} be special points: depending on higher-order contributions to~$V$,
the divergence in $g$ could cancel. Examples of such ``miraculous" cancellations include the PT potential, where the
special points generically present in $f(\omega,x)$ given by (\ref{eq:expt01}) at $\omega=-in/b$ ($n=1,2,\ldots$) are
absent if $q=\frac{1}{2}+\ell$ ($\ell=0,1,\ldots$) and $n\ge\ell+1$, as follows from standard properties of the
hypergeometric function\cite{miracle2}; for $\ell=0$ this becomes the free field, which obviously has no special
points at all. Without proof we mention that such a ``miracle" also occurs at the algebraically special frequency
(\ref{eq:spmode}) of the RW equation for any angular momentum. Special points need not lie on the imaginary axis, as
is clear by considering $V(x)\sim e^{-\lambda x}\cos(\mu x)$; however, only if $\re\omega'=0$ can ${\omega'}^2$ be
the eigenvalue of the SUSY generator, so this case needs to be studied most carefully. An example in which the
special point index $n>1$ is provided by $V(x)\sim xe^{-\lambda x}$ with $n=2$, with obvious generalizations. 

Physically, the divergence in $g$ at a special point is irrelevant: it merely comes from the normalization imposed on
$g$. Thus, this divergence cancels against the Wronskian in the Green's function, the proper outgoing wave near
$\omega=\omega'$ in fact being $\chi(\omega)$. Since $\chi(\omega)\sim(\omega-\omega')^ne^{i\omega x}$, the large
part of $\chi$ vanishes at the special point. Because $\chi(\omega)$ is a solution of the KGE with eigenvalue
$\omega^2$, this can only happen if the outgoing wave $\chi(\omega')$, said to be {\em special outgoing}, at the same
time is incoming as well: $\chi(\omega')\propto g(-\omega')$. While this might be counterintuitive, it merely means
that at these frequencies the exponential tail scatters so strongly that the true outgoing wave is completely
different from the one in free space. All this can be verified beyond the Born approximation for the exactly solvable
cases $V(x)=e^{-\lambda x}$\cite{tail} and the PT potential of Section~\ref{sect:pt}. 

On the negative imaginary axis, the analytic continuation in general yields two different branches, 
$g_{\rm l}(\omega)=\lim_{\epsilon\downarrow0}g(\omega-\epsilon)$ and 
$g_{\rm r}(\omega)=\lim_{\epsilon\downarrow0}g(\omega+\epsilon)=g_{\rm l}(\omega)^*$. 

The type of discontinuity this induces is indicated for each $\omega'=-i|\omega'|$ by a second index $m$ so that
$\Delta g(\omega)\equiv g_{\rm r}(\omega)-g_{\rm l}(\omega)\sim(\omega-\omega')^m$. The case in which there is no cut
at all can be thought of as $m=\infty$. Note that because of their normalization, the ``large" parts of $g_{\rm l/r}$
cancel in $\Delta g$, which, also being a solution of the KGE, must thus be small if it does not vanish: $\Delta
g(\omega)\propto g(-\omega)$, with a purely imaginary (frequency-dependent) constant of proportionality. At a special
point, $\chi_{\rm l}$ and $\chi_{\rm r}$ must be proportional, their prefactors having opposite phases, and the cut
index is then defined by $\Delta\chi(\omega)\sim(\omega-\omega')^m$. 

The above treatment might seem to put undue emphasis on exceptions. However, these cases will actually arise by
performing SUSY transforms on a generic potential, as will become clear in the following. The statements in this
subsection have obvious counterparts for $f$. 

\subsection{SUSY transformation of one outgoing wave}
\label{susy1wave}

We first consider how the outgoing wave at one end is transformed under SUSY. The crucial feature to be derived is:
{\em even in the presence of tails, SUSY carries outgoing waves into outgoing waves unless it annihilates them}. The
latter exception can only occur if the eigenvalues $\omega^2$ of the outgoing wave and $\Omega^2$ of the SUSY
generator coincide, for $Ag=0$ iff $g\propto\Phi$. The proof is in fact simple. By (\ref{eq:ahha}), SUSY carries
solutions of the wave equation into solutions for the partner system. For $\im\omega>0$, the operator $A$ of
(\ref{eq:opa}) carries exponentially decreasing waves into exponentially decreasing waves since $W$ is bounded. Since
the frequency independent $A$ maps analytic functions to analytic functions, clearly it commutes with the analytic
continuation to the rest of the $\omega$-plane, completing the proof. 

Normalization of the transformed wave is straightforward, with the result that the outgoing wave in the partner
system reads
\beq
  \tilde{g}(\omega)=\frac{Ag(\omega)}{i\omega+W_+}
\eeql{tildeg}
for generic $\omega$. In particular, this means that special points, branch cuts etc.\ are all conserved under SUSY
if $\omega^2\neq\Omega^2$. For $\omega^2=\Omega^2$, one has to study the limiting behaviour of (\ref{tildeg}) more
carefully. Writing $\Omega=\pm iK$ ($K>0$), one has the following cases.

\medskip(a1) {\em$W_+=K$ and $-iK$ is not a special point}. In this case $\Phi\propto g(iK)$, so that the zero in the
denominator of (\ref{tildeg}) cancels, as it has to because $\tilde{g}(\omega)$ is necessarily regular for
$\im\omega>0$. Hence, $\tilde{g}(iK)=\lim_{\omega\rightarrow iK}\tilde{g}(\omega)=-iAg_1(iK)$, with $g_k(iK)\equiv
{k!}^{-1}\partial_\omega^kg(\omega)|_{iK}$, cf.\ the notation of (\ref{eq:Psi01}). The expression for $\tilde{g}(iK)$
is nonzero because $g_1(iK)\not\propto g(iK)$, as follows from $(H+K^2)g_1(iK)=2iKg(iK)\neq0$.

Let the original system have a discontinuity index $m$ at frequency $-iK$. Frequency differentiation of the KGE
yields $(H+K^2)g_k(-iK)=-2iKg_{k-1}(-iK)+g_{k-2}(-iK)$ on both branches, so that $(H+K^2)\Delta g_k(-iK)$ vanishes
for $k=m$, but is nonzero for $k=m+1$. Thus, $\Delta g_m(-iK)\propto g(iK)$ (since $\Delta g_m(-iK)$ is always
exponentially small, cf.\ at the end of Section~\ref{onewave}) and hence $\Delta\tilde{g}_k(-iK)=0$ for $0\le k\le m$
(for $k<m$ this follows simply from $\Delta g_k(-iK)=0$). However, $\Delta\tilde{g}_{m+1}(-iK)\neq0$ so that
$\tilde{m}=m+1$. In particular, $\tilde{m}\ge1$ so that at $\omega=-iK$, the cut discontinuity vanishes in the
partner. 

It is trivial that $\tilde{\Phi}$ is exponentially large, but is it actually outgoing? The question is answered by a
calculation of the Wronskian
\beq
  J\Bigl(g^{-1}(iK),Ag(-iK)\Bigr)=0\quad,
\eeql{Jg-1g}
for which one has to substitute (\ref{eq:opa}) for $A$ with $W=-g'(iK)/g(iK)$ and use the KGE for $g''$. Thus, up to
proportionality $\tilde{\Phi}$ equals the unique outgoing wave $\tilde{g}(-iK)=Ag_{\rm l/r}(-iK)/2K$. 

\medskip(a2) {\em$W_+=K$ and $-iK$ is a special point with index $n=1$}. The situation at $\omega=iK$ is identical to
case (a1). Near $\omega=-iK$, $g(\omega)=\chi(\omega)/(\omega+iK)$ with $\chi(-iK)\propto g(iK)$ (cf.\
Section~\ref{onewave}), so that $A \chi(-iK)=0$ and the special point is lifted in the partner system, with
$\tilde{g}(-iK)=A\chi_1(-iK)/2K\neq0$, where the r.h.s.\ in general will be different on the two branches. In more
detail, a calculation as for case (a1) shows that now $\tilde{m}=m$.

The counterpart to (\ref{Jg-1g}) is evaluated to be
\bea
  J\Bigl(g^{-1}(iK),A\chi_1(-iK)\Bigr)
  &=&-2iK\chi(-iK)/g(iK)\nonumber\\
  &\neq&0\quad,
\eeal{Jg-1chi}
so that in this case $\tilde{\Phi}$ is mixed in the partner system\cite{Jzero}.

\medskip(a3) {\em$W_+=K$ and $-iK$ is a special point with index $n\ge2$}. The situation at $\omega=iK$ is identical
to case (a1). At $\omega=-iK$ one calculates $\tilde{n}=n-1$ with $\tilde{\chi}(-iK)=A\chi_1(-iK)/2K$ and
$\tilde{m}=m$. The function $\tilde{\Phi}$ is exponentially increasing, hence certainly not (special) outgoing. All
these properties are seen to be analogous to case (a2), except that now the frequency $-iK$ remains special in the
partner system. 

\medskip(b1) {\em$W_+=-K$, $-iK$ is not a special point and $\Phi\propto g_{\rm l}(-iK)=g_{\rm r}(-iK)$ is outgoing}.
Near $\omega=iK$ the transformation is completely regular and $\tilde{g}(iK)=-Ag(iK)/2K\propto\tilde{\Phi}$, since
$\tilde{\Phi}$ is an exponentially decreasing solution with eigenvalue $-K^2$.

Near $\omega=-iK$, $\tilde{g}(\omega)=Ag(\omega)/(i\omega-K)\rightarrow-iAg_1(-iK)=\tilde{g}(-iK)$. This generalizes
the situation of (\ref{eq:jb02}), where $-iK$ is taken to be a double QNM. Setting $\Delta
g(\omega)\sim(\omega+iK)^m$ ($m\ge1$) leads to $\Delta\tilde{g}(\omega)=-i(\omega+iK)^{m-1}A\Delta g_m(-iK)+{\cal
O}[(\omega+iK)^m]$ where $A\Delta g_m(-iK)$ is readily checked to be nonzero, so that $\tilde{m}=m-1$. 

\medskip(b2) {\em$W_+=-K$, $-iK$ is not a special point, and $\Phi$ is mixed}. Cf.\ case (b1): an exponentially
increasing $\Phi$ will always be mixed if there is a discontinuity across the cut, for in that case $g_{\rm
l/r}(-iK)$ are not proportional to a real function. The situation at $\omega=iK$ is identical to case (b1). 

Near $\omega=-iK$, the zero in the denominator of (\ref{tildeg}) is not compensated, so $-iK$ becomes a special point
with index $\tilde{n}=1$ in the partner system and $\tilde{\chi}(-iK)=-iAg(-iK)$. Note how the large parts of $\Phi$
and $g(-iK)$ coincide, leading to the annihilation of this part in $Ag(-iK)$ and leaving only an exponentially
decreasing function. One verifies that $\tilde{m}=m$ for the discontinuity index. 

Note that, if $V$ for instance has finite support, $\tilde{V}$ has a tail $\sim e^{-2Kx}$ as in (\ref{eq:wasymp}),
while $-iK$ is its only special point. Thus, at $i\omega/K=2,3,\ldots$ one has a cancellation of the kind called
``miraculous" in Appendix~\ref{onewave}, but which apparently is not exceptional in the context of SUSY. 

\medskip(b3) {\em$W_+=-K$ and $-iK$ is a special point}. The situation at $\omega=iK$ is identical to case (b1). Near
$\omega=-iK$, the pole in (\ref{tildeg}) leads to $\tilde{n}=n+1$ for the special point index with
$\tilde{\chi}(-iK)=-iA\chi(-iK)$, while $\tilde{m}=m$. 

\medskip One sees that, for each $i=1,2,3$, the partner of a system in the (a$i$) category falls into case (b$i$) and
vice versa. Thus, for each $i$, (a$i$) and (b$i$) transforms are each other's inverse, and inspection of the above
cases confirms that their effects are exactly opposite.

Again, the situation for $x\rightarrow-\infty$ is wholly analogous.

\subsection{Two outgoing waves}
\label{twowaves}

We now combine the left and right boundary conditions to study modes. For this purpose, we introduce the two
Wronskians
\bea
  J_{\rm q}(\omega)&\equiv&f'(\omega)g(\omega)-f(\omega)g'(\omega)\quad,
  \label{Jq}\\
  J_{\rm t}(\omega)&\equiv&f'(-\omega)g(\omega)-f(-\omega)g'(\omega)\quad,
\eeal{Jt}
which keep track of (quasi)normal modes and total-transmission modes respectively. That is, for ordinary points of
$f$ and $g$ an $M$th-order zero in $J_{\rm q}(\omega{=}\omega_0)$ signifies an $M$th-order QNM if $\im\omega_0<0$,
and an $M$th-order NM if $\im\omega_0\ge0$. In the latter case, this of course means that the only possible nonzero
value is $M=1$, in which case $\re\omega_0=0$. Similarly, an $M$th-order zero in $J_{\rm t}(\omega{=}\omega_0)$
signifies an $M$th-order TTM$_{\rm L}$ at $\omega_0$, or equivalently an $M$th-order TTM$_{\rm R}$ at $-\omega_0$.

While QNMs can coexist with NMs, or TTM$_{\rm L}$s with TTM$_{\rm R}$s, at the same eigenvalue $\omega^2$ (cf.\
Section~\ref{sect:double}), (Q)NMs and TTMs clearly exclude each other unless there are special points, at which the
various notions coincide. 

Of course, if $\omega_0$ is a special point of index $n$ for $f$ and $n'$ for $g$, then (Q)NMs are given by zeros of
$(\omega-\omega_0)^{n+n'}J_{\rm q}(\omega)$ and TTM$_{\rm L}$s by those of
$(\omega-\omega_0)^{n'}(\omega+\omega_0)^nJ_{\rm t}(\omega)$. 

For example, let $g(\omega)$ have a special point with index~$1$ at $\omega=\omega_0$. Then at the eigenvalue
$\omega_0^2$ one can have: ($i$) no mode, or ($ii$) an NM and a TTM$_{\rm L}$ of order $1+k$ ($k=0,1,\ldots$) or
($iii$) a QNM of order $1+\ell$ and a TTM$_{\rm R}$ of order $1+k$ ($k,\ell=0,1,\ldots$), with at most one of
$k,\ell$ nonzero. As a second example, if $\omega_0$ is an $n=2$ special point for $f$, then there can at most be a
first-order TTM$_{\rm R}$ at that frequency, since now the notions of TTM$_{\rm R}$ and NM coincide up to second
order and higher-order NMs cannot occur. 

If at least one of the functions $f$ and $g$ has a cut on the negative imaginary axis, then $J_{\rm q}$ will have
such a cut as well (similar to the cut in the Green's function). On the other hand, clearly if there is a branch cut
in $f$ ($g$) then $J_{\rm t}$ will have a cut on the positive (negative) imaginary axis. It follows from the
symmetries satisfied by $f$ and $g$ individually, that if $\phi_{\rm l}$ is a zero-mode at frequency $\omega_0$ on
the left branch of $J_{\rm q}$, then $\phi_{\rm r\vphantom{l}}^{\vphantom{*}}=\phi_{\rm l}^*$ is a zero-mode (of the
same order $M$) on the right branch. As a consequence, if $f_{\rm l}(\omega_0)=f_{\rm r}(\omega_0)$ (for instance if
$f$ has no branch cut at all) then $g(\omega_0)$ should be proportional to a real function; hence, if $\omega_0$ is
an ordinary point for $g$ then $\Delta g$ vanishes there ($m_{\rm R}\ge1$). For TTMs the situation is simpler, since
only one cut at a time is involved: if, say, there is a TTM$_{\rm L}$ at $\omega_0$, then $\max(m_{\rm R},n_{\rm
R})\ge1$. Generalizations to higher-order modes can be derived, but are not given here. 

\subsection{SUSY transformation of two outgoing waves}
\label{susy2waves}

With the six possibilities of Section~\ref{susy1wave} for both \linebreak$x\rightarrow-\infty$ and
$x\rightarrow\infty$, and with the rich variety of possible mode structures in the original system outlined in
Section~\ref{twowaves}, a classification of the possible SUSY transformations may seem daunting even before
considering the subtleties of $\Omega=0$. However, most of the information is contained in the transformation
properties of the Wronskians (\ref{Jq}) and (\ref{Jt}). These are readily derived by substituting (\ref{tildeg}) and
its counterpart for $\tilde{f}$ into (\ref{Jq}) for $\tilde{J}_{\rm q}$ and (\ref{Jt}) for $\tilde{J}_{\rm t}$, using
(\ref{eq:opa}) for $A$ and the KGE for $f''$ and $g''$. One obtains
\bea
  \tilde{J}_{\rm q}(\omega)&=&\frac{(\omega+iK)(\omega-iK)}
    {(\omega+iW_-)(\omega-iW_+)}J_{\rm q}(\omega)\quad,\label{tildeJq}\\
  \tilde{J}_{\rm t}(\omega)&=&\frac{(\omega+iK)(\omega-iK)}
    {(-\omega+iW_-)(\omega-iW_+)}J_{\rm t}(\omega)\quad.
\eeal{tildeJt}
Combining these with the transformation properties of the individual functions $f$ and $g$ derived in
Section~\ref{susy1wave}, one finds that also the mode structure is entirely conserved if $\omega^2\neq\Omega^2$,
leaving only the case $\omega^2=\Omega^2$ to be studied.

For instance, if $W_\pm=\pm K$ (an NM), then (\ref{tildeJq}) shows that in $\tilde{J}_{\rm q}$ a zero at $iK$ gets
cancelled. This is a much more meaningful statement than just noting that $A\Phi=0$ for the NM SUSY generator $\Phi$,
which in itself says nothing about the situation in the partner system. At the same time, an extra zero at $-iK$
appears. However, this latter zero must be interpreted carefully. If $-iK$ is an ordinary point for both $f$ and $g$,
it will remain so in the partner and the zero signifies that a QNM at $-iK$ is created (or its order increased). If
$-iK$ is special for either $f$ or $g$, then the extra zero merely reduces $n$ by one according to case (a2) or (a3)
in Section~\ref{susy1wave}, and the order (possibly zero) of the QNM at $-iK$ is {\em not\/} altered, consistent with
$\tilde{\Phi}$ {\em not\/} being outgoing on the `special' side. Finally, if both $f$ and $g$ are special at $-iK$
then the decrease of the associated indices by itself would reduce the pole in $J_{\rm q}$ by two orders, so that in
this case (\ref{tildeJq}) shows that the NM-generated SUSY transform {\em reduces\/} the QNM index at $-iK$ by one.
In fact this should not be surprising, for if $-iK$ is two-sided special then at least to first order the QNM agrees
with the NM, which gets cancelled in the partner; an example is the inverse of the transform considered in
Section~\ref{subsect:free}. 

Continuing with the example, one sees that $J_{\rm t}$ merely changes sign. Again, depending on the properties of $f$
and $g$, this can mean no change in the mode situation (if $-iK$ is ordinary), destruction of a TTM$_{\rm L}$ (if
$-iK$ is special for $g$), destruction of a TTM$_{\rm R}$ (if $-iK$ is special for $f$), or both of the latter. 

Generalizing from this example, one sees that the material of Section~\ref{susy1wave} together with (\ref{tildeJq})
and (\ref{tildeJt}) leads to the following statement: {\em if $-iK$ is an ordinary point for both the original and
the partner system (which in particular implies that $\Phi$ is purely incoming or outgoing on both sides), the only
effect of SUSY on the mode structure consists of the destruction of the mode $\Phi$ and the creation of the mode
$\tilde{\Phi}$, in both cases possibly by changing their order by one}. Hence, in this case the situation remains as
in Table~1 even in the presence of arbitrary potential tails. 

It may seem that (\ref{tildeJq}) holds the possibility of creating a second-order NM by starting from a system with
an NM at $iK$, and performing a SUSY transform using a generator for which $W_\pm=\mp K$. However, this can never
occur because at the eigenvalue $-K^2$, any real solution except possibly the NM itself must have a node and hence
cannot be used as a SUSY generator. For a proof which is also valid at $K=0$, assume that $\phi$ is the NM with
$\phi(x{\rightarrow}\infty)>0$ for definiteness, while $\Phi\not\propto\phi$ is a positive-definite solution at the
same eigenvalue. As $x\rightarrow\infty$, $\Phi/\phi\rightarrow\infty$ so $(\Phi/\phi)'=J(\Phi,\phi)/\phi^2>0$. Since
the Wronskian is position independent, $(\Phi/\phi)'>0$ also for $x\rightarrow-\infty$, which is only possible if
$\phi<0$ there, implying that the function has at least one node. At the rightmost node, $J(\Phi,\phi)=-\phi'\Phi<0$,
and the contradiction with the position independence of $J$ means that the function $\Phi$ of the assumption does not
exist.

\subsection{SUSY at $\Omega=0$}
\label{Omega0}

Throughout this appendix and the main text, it has been clear that there is an interplay between the outgoing waves
and possible modes at $iK$ and $-iK$. Hence, special attention should be paid to the case when these two frequencies
coincide, i.e., $\Omega=0$. For the KGE at $\omega=0$, it is still true that for $x\rightarrow\infty$, there is one
unique (up to a prefactor) small solution. This same solution is also the outgoing wave $\chi(\omega{=}0)$, as
follows by analytically continuing the small solution $g(\omega)$ from the upper half $\omega$-plane to the origin
(the same argument can {\em not\/} be used to conclude that $\chi(\omega{=}0)$ should be asymptotically large by
performing the continuation from the lower half plane, since at any frequency large functions are not unique, and the
limiting member of a sequence of them could well be asymptotically small). The issue of special points does not arise
at zero frequency, and neither does that of discontinuity indices. 

For $\Omega=0$, the six cases of Section~\ref{susy1wave} reduce to two: for $x\rightarrow\infty$, $\Phi$ can be
either (a) outgoing, i.e.\ small, or (b)~large. However, the properties of the transformation depend sensitively on
the potential tail. Here we study $V(x)\sim\kappa/x^\alpha$, with $\alpha\ge2$\cite{smalltail}; for the practically
important case $\alpha=2$ we take $\kappa=\nu(\nu+1)\ge-1/4$, with $\nu\ge-1/2$ without loss of generality. Tails
which decay faster than algebraically (including no tail at all) are easy to handle, and can be thought of as
$\alpha=\infty$. 

For $\alpha>2$, all terms in the Born approximation exist down to $\omega=0$. To first order\cite{tail},
$g(0,x)=1+\kappa x^{2-\alpha}/(\alpha{-}1)(\alpha{-}2)$, so in case (a) one has $W(x)=\kappa
x^{1-\alpha}/(\alpha{-}1)$, i.e., SUSY reverses the sign of the potential tail. In the partner system,
$\tilde{g}(0)=\lim_{\omega\rightarrow0}Ag(\omega)/i\omega=-i\partial_\omega {Ag(\omega)|}_0\propto\tilde{\Phi}$; only
if $\alpha>3$ can one write $\tilde{g}(0)=-iAg_1(0)\propto\tilde{\Phi}$ for this, since $g(\omega)$ itself is not
differentiable at $\omega=0$ for $2<\alpha\le3$. The proportionality $Ag_1(-iK)\propto\tilde{\Phi}$ has not been
encountered before in Section~\ref{susy1wave}; Eq.~(\ref{Jg-1chi}) shows why. In case (b), $\Phi\sim x+{\cal
O}(x^{3-\alpha})+{\cal O}(x^0)$, so $W(x)\sim-1/x$, $\tilde{V}(x)\sim2/x^2$, and
$\tilde{g}(\omega)=\chi(\omega)/\omega$ with $\tilde{\Phi}\propto\chi(0)=-iAg(0)\sim i/x$. Thus, for these potentials
at $\Omega=0$, {\em any\/} choice of $\Phi$ leads to a small $\tilde{\Phi}$, in contrast to the situation at finite
$\Omega$; of course, the partner potential depends on this choice. 

If $\alpha=2$, one has the exact solution $g(\omega,x)=e^{i(\nu+1)\pi/2}\sqrt{\pi\omega
x/2}\,H^{(1)}_{\nu+1/2}(\omega x)$\cite{tail} with $H^{(1)}$ a Hankel function, i.e.,
$g(\omega,x)\sim[\Gamma(\nu{+}\half)/\sqrt{\pi}](2i/\omega x)^\nu$ ($\sim(i{-}1)\ln\omega\sqrt{\omega x/\pi}$ for
$\nu=-\half$) near $\omega=0$. In case (a), the generator thus is $\Phi=x^{-\nu}$ and the transformation can be
summarized by $\tilde{\nu}=\nu-1$ ($-\nu$) for $\nu\ge\half$ ($-\half\le\nu\le\half$). In case (b), one has $\Phi\sim
x^{\nu+1}$ and $\tilde{\nu}=\nu+1$\cite{nu-half}. Thus, one sees that for $-\half\le\nu\le\half$, always
$\tilde{\Phi}\propto\tilde{\chi}(0)$ (indeed, a large $\tilde{\Phi}$ would give $\tilde{\nu}\mapsto\tilde{\nu}+1$ in
the inverse transform, which for any $\tilde{\nu}$ is incompatible with $\nu<\half$) as for $\alpha>2$. Such
potentials, which at $\Omega=0$ behave essentially the same as one with finite support, have {\em weak tails}. On the
other hand, for {\em strong tails\/} ($\alpha=2$, $\nu>\half$) $\tilde{\Phi}$ is large (small) if $\Phi$ is small
(large), as is the case for $\Omega\neq0$. 

In line with usage for other frequencies, $\omega=0$ is said to be a mode frequency if
$\chi(\omega{=}0)\propto\phi(\omega{=}0)$, where $\phi$ is the counterpart of $\chi$ for waves outgoing to the left.
Like any NM, such a mode is necessarily of first order, and there is no distinction between different types of modes.
Apart from possible overall prefactors $\phi(\omega)/f(\omega)$ and $\chi(\omega)/g(\omega)$ due to the tails, a mode
is associated with a {\em second\/}-order zero in $J_{\rm q/t}$ because the eigenvalue in the KGE is $\omega^2$. 

If the original system has a mode at $\omega=0$, the proof at the end of Section~\ref{susy2waves} implies that (if it
is nodeless) the mode itself is the only candidate for a SUSY generator, i.e., case (a) applies on both sides. If $V$
has two weak tails, $\tilde{\Phi}$ will again be a mode; combination with the above statements on the transformation
of one potential tail shows that the class of weak-tail potentials with a mode at $\omega=0$ is closed under SUSY
(since transforms at $\Omega\neq0$ obviously cannot generate strong tails), as exemplified in
Section~\ref{subsect:free} by some of the potentials SUSY-equivalent to the free field. If $V$ has a weak tail on one
side, $\tilde{\Phi}$ is outgoing on that side only so that there is no mode at $\omega=0$. If $V$ has two strong
tails, $\tilde{\Phi}$ is large on both sides. Moreover, according to the above in that case one has $\tilde{\nu}_{\rm
L/R}=\nu_{\rm L/R}-1$ for the tail amplitudes, so that
$\tilde{f}(\omega)\tilde{g}(\omega)/f(\omega)g(\omega)\sim\omega^2$; since (\ref{tildeJq}) and (\ref{tildeJt}) show
that up to a sign $J_{\rm q/t}$ are invariant under SUSY, the compensating factor $\omega^{-2}$ must come from the
disappearance of the mode. 

If the original system has no mode at $\omega=0$ then the situation is simpler, even though several types of SUSY may
be possible. If $\tilde{\Phi}$ is outgoing on both sides (which in particular will always be the case if $V$ has two
weak tails) then it is obviously a mode itself. It is not iff $\Phi$ is small on only one side and if $V$ has a
strong tail on that side. Then $\tilde{\Phi}$ is small on only the other side, i.e., if $\tilde{\Phi}$ is not a mode
then $\tilde{V}$ has no mode at all at $\omega=0$.

To conclude, let us mention that modes, special points etc.\ at $\omega^2\neq\Omega^2=0$ are not affected by SUSY, as
is the case for other generator eigenvalues. 

%==========================================================
%==========================================================

\section{Nodes in QNM wavefunctions}
\label{sect:appnode}

Nodeless eigenstates play a special role in SUSY:  they are candidates for the SUSY generator.  It is therefore
useful to highlight the differences between NMs and QNMs in this regard, especially to contrast with the well-known
property that there can be {\em only one\/} nodeless NM.

First of all, we show that for QNMs not lying on the imaginary axis (i.e., $\re\omega \ne 0$), there can be at most
one node or antinode.  This is not surprising: since the eigenvalue is complex, the wavefunction has a changing
phase, and it would be ``unlikely" that the real part and the imaginary part (or their derivatives) would vanish
together.  To prove this, we take the Schr\"odinger point of view, so that the eigenvalue is $\lambda = \omega^2$
with a nonzero imaginary part.  Now consider a time-dependent QNM and suppose that it has nodes or antinodes at two
points $x_1, x_2$.  At these two points, the current
\beq
  J = i \left[ \phi^* (\partial_x \phi) - (\partial_x \phi^*) \phi \right]
\eeql{eq:cons2}
vanishes. Then, applying the conservation law in integral form to the interval $[x_1, x_2]$, we find that the total
probablity in this interval is constant in time.  Yet the wavefunction is either growing or decaying, since
$\im\lambda \ne 0$, which is therefore a contradiction.

From the perspective of SUSY it is unfortunate that the above proof excludes the crucial imaginary axis. However, on
that axis the statement remains valid for repulsive potentials, or more generally for potentials which are so weakly
attractive that $V-\omega^2$ is positive definite. Namely, let $\phi$ be a solution with two (anti)nodes. By taking
the real or imaginary part if necessary, $\phi$ can be arranged to be real. Now between two nodes $\phi$ would have
an extremum, i.e.\ $\phi''\phi<0$ which is incompatible with the KGE. Similarly, an antinode can only be a global
maximum or minimum, precluding the presence of any other nodes or antinodes. Note that for repulsive $V$, the KGE can
be mapped to the wave equation (Appendix~\ref{sect:appkgwe}; the mapping conserves nodes but not necessarily
antinodes), for which a proof using the current $J$ (cf.\ (\ref{eq:cons2})) does not exclude the imaginary axis. 

Thus, QNMs (at least those off the imaginary axis for attractive $V$) can have at most one node or antinode.  For
symmetric potentials, in the even sector $x=0$ is already an antinode, so there can be no nodes anywhere. 

For zero modes, i.e., QNMs with $\re\omega = 0$, nodes are more ``likely": the eigenvalue is real and (if there is no
branch cut) the wavefunction has a constant phase, say purely real; it therefore requires only {\em one\/} condition,
rather than two, to have a node.  Nevertheless, in contrast to the conservative case, the proof that there is at most
one nodeless eigenstate can be bypassed. 

The interlacing nodal structure of NM eigenfunctions follows from well-established Sturm--Liouville theory.  For the
present purpose, we do not need the full apparatus.  Consider, for simplicity, a finite interval $[-a,a]$ and suppose
there are two distinct nodeless eigenfunctions $\phi_1 , \phi_2$, both of which can be chosen to be real.  Then they
can both be chosen to be positive, which violates the orthogonality condition for NMs
\beq
 \int_{-a}^{a} \phi_1(x) \phi_2(x) \, dx = 0 \quad .
\eeql{eq:orth1}

We can attempt to transplant the argument to QNMs.  For zero modes, again the wavefunctions have constant phases and
can be chosen to be real, and if they are nodeless, positive definite.  However, the analog of (\ref{eq:orth1}) for
two eigenfunctions with eigenvalues $\omega_i = -i\gamma_i$ is

\bea
&& -\left( \gamma_1 + \gamma_2 \right )
\int_{-a}^{a} \phi_1(x) \phi_2(x) \, dx
\nonumber \\
&&{} + \left[ \phi_1(-a) \phi_2(-a) + \phi_1(a) \phi_2(a) \right]
= 0
\quad .
\eeal{eq:orth2}

\nid
Note in particular the signs of the two terms.  With $\gamma_i > 0$,
this condition does {\em not\/} preclude both eigenfunctions
from being positive definite.

Thus, we have three different situations.
(a)  For NMs, there can be {\em only one\/} nodeless state.
(b)  For QNMs with $\re\omega = 0$, there
{\em could be\/} more than one state with no node.
(c)  For QNMs with $\re\omega \ne 0$, each
eigenfunction can have at most one node or antinode,
and for symmetric potentials, {\em every\/} even eigenfunction
is nodeless.

Case (b) in particular opens up the possibility of
multiple SUSY transformations, as exemplified in Sections \ref{subsect:exqnm} and ~\ref{subsect:ptfull}.

%==========================================================
%==========================================================

\section{Self-replicating potentials}
\label{sect:apprep}

In Section~\ref{subsect:ptfull} we saw that the PT potential is
self-replicating under SUSY, in the sense that
the partner is the same potential with a different strength:

\beq
{\tilde V}(x) = \alpha V(x) \quad .
\eeql{eq:selfrep1}

\nid
The condition (\ref{eq:selfrep1}) is a special case
of shape-invariant potentials described by
a family $U(a,x)$~\cite{shape}, with

\bea
V(x) &=& U(a,x) \nonumber \\
{\tilde V}(x) &=& U({\tilde a}, x) + c(a)
\quad .
\eeal{eq:shapeinv}

\nid
Here we focus on the more restrictive
sense (\ref{eq:selfrep1}) and show that the PT potential is uniquely determined by
this condition.  From (\ref{eq:selfrep1}) and (\ref{eq:vw}), and  
with $\Omega = -iK$ being purely imaginary, we find

\beq
W'(x) = \beta \left[ W(x)^2 - K^2 \right] \quad ,
\quad\beta \equiv \frac{\alpha {-} 1}{\alpha {+} 1}\quad .
\eeql{eq:selfrep2}
All nonsingular nonconstant solutions of (\ref{eq:selfrep2}) differ merely by $x$-translation. The unique
antisymmetric one reads
\bea
  W(x) &=& -K \tanh K \beta x \quad , \nonumber \\
  \Phi(x) &=& \left( \cosh K \beta x \right)^{1 / \beta} \quad .
\eeal{eq:selfrep3}
Putting this into (\ref{eq:vw}), we then find
\bea
  V(x) &=& ( \beta {-} 1 ) K^{2} \sech^2 K \beta x \quad , \nonumber \\
  {\tilde V}(x) &=& ( - \beta {-} 1 ) K^{2} \sech^2 K \beta x \quad .
\eeal{eq:selfrep4}
That is, $V$ and ${\tilde V}$ are of PT form.

%==========================================================

%==========================================================
%This is Table 1
\newpage

\vspace{1cm}

\begin{center}

\vspace{5mm}

\begin{tabular}{|c|c|c|c|c|c|c|} \hline
\mbox{ Type }& \mbox{\hspace{5mm}$\Phi$\hspace{5mm}} & 
\mbox{\hspace{5mm}${\tilde \Phi}$\hspace{5mm}} & \mbox{  NM  } & 
\mbox{  QNM  } & \mbox{ TTM${}_{\rm L}$ } & 
\mbox{ TTM${}_{\rm R}$ } \\ \hline \hline
1 & DD & II & $-1$ & +1 & 0 & 0 \\ \hline
2 & II & DD & +1 & $-1$ & 0 & 0 \\ \hline
3a & DI & ID & 0 & 0 & $-1$ & +1 \\ \hline
3b & ID & DI & 0 & 0 & +1 & $-1$ \\ \hline
4 & \multicolumn{6}{c|}{more complicated, see Appendix~\ref{sect:apptail}} \\ \hline
\end{tabular}

\vspace{5mm}
\end{center}

\nid

\vspace{1cm}
Table 1: The boundary conditions for the deleted
state $\Phi$ and the new state ${\tilde \Phi}$,
and the changes in the number of NMs, QNMs,
TTM$_{\rm L}$s and TTM$_{\rm R}$s in the SUSY transformation.

%==========================================================
%==========================================================

%Generate all the figures together with captions

\newpage

%==========================================================
%Fig 1

\newpage
%\begin{center}
\leavevmode\scalefig{0.5}\epsfbox{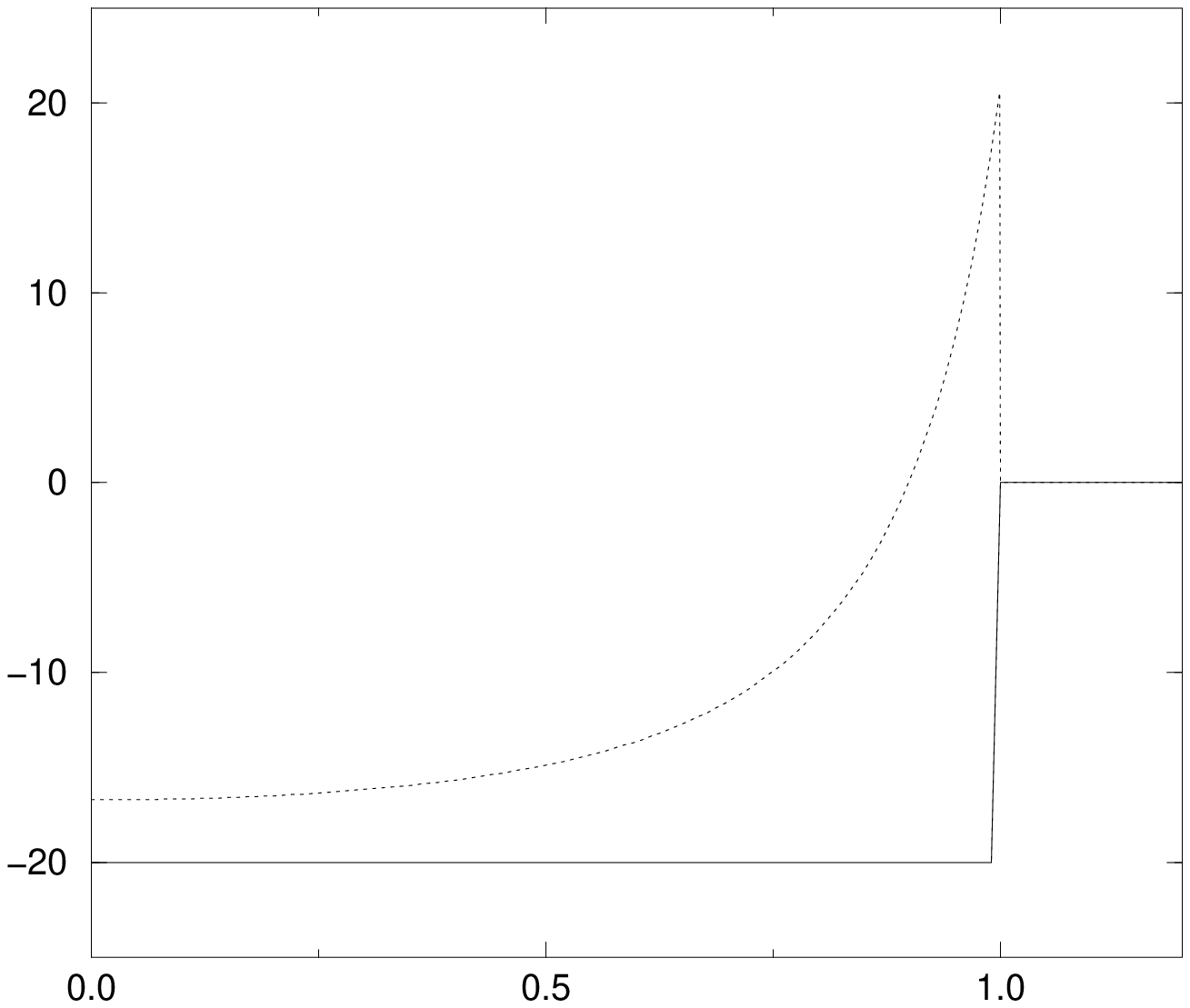}
\\
{\scriptsize Fig. 1(a): A square-well potential $V$ (solid line) defined
by (\ref{eq:exsqwell1}) with $V_0=-20$ and $a=1$, and its SUSY partner
potential ${\tilde V}$ (broken line). }
%\end{center}

\vspace{1.5cm}

%\begin{center}
\leavevmode\scalefig{0.5}\epsfbox{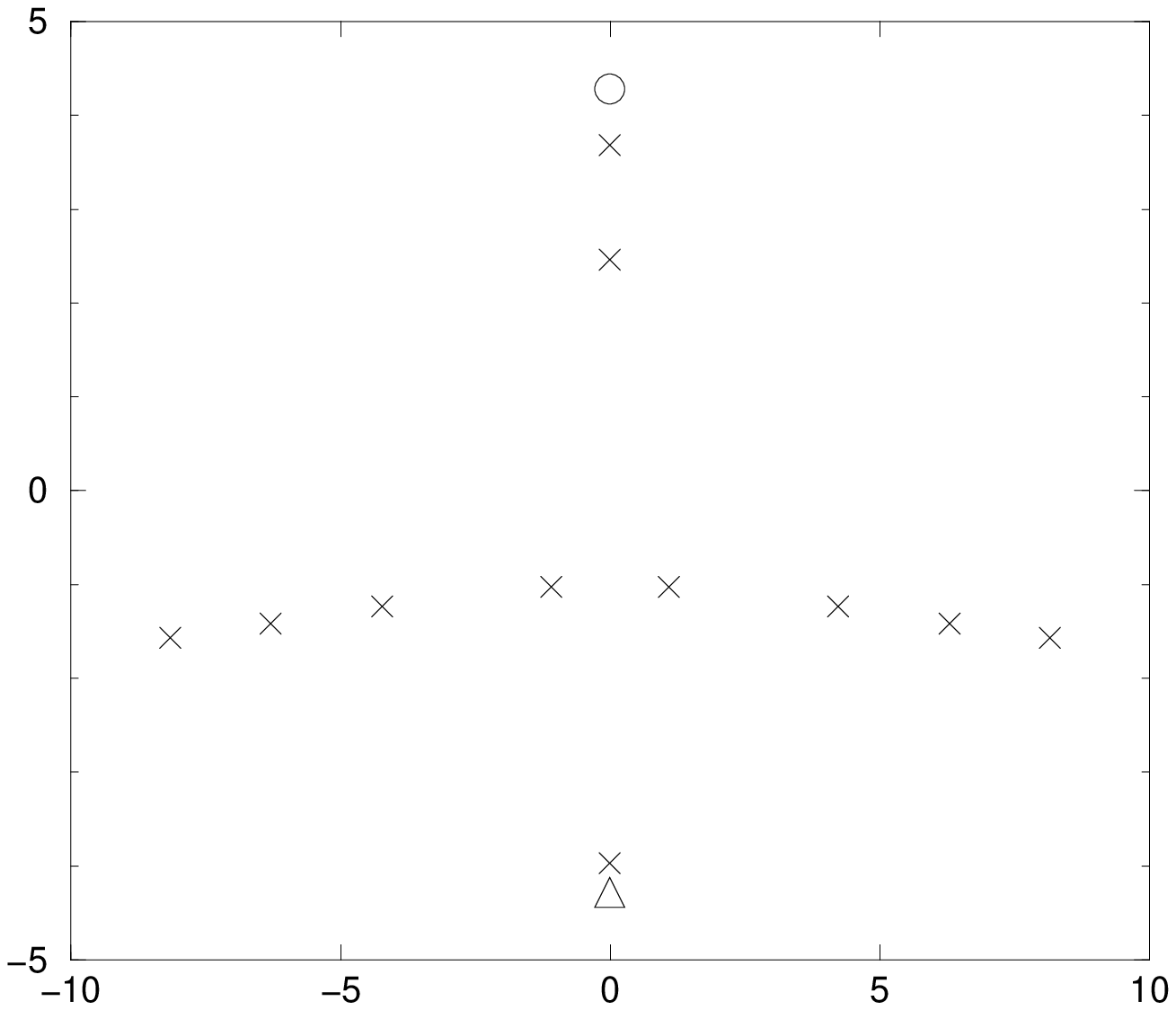}
\\
{\scriptsize Fig. 1(b): The complex $\omega$-plane showing the NMs and
QNMs common to both potentials (crosses); the mode present only in $V$
(circle), which corresponds to the generator $\Phi$; and the mode
present only in ${\tilde V}$ (triangle), which corresponds to ${\tilde
\Phi} = A \Phi$. 
}
%\end{center}

%==========================================================
%Fig 2

\newpage
%\begin{center}
\leavevmode\scalefig{0.5}\epsfbox{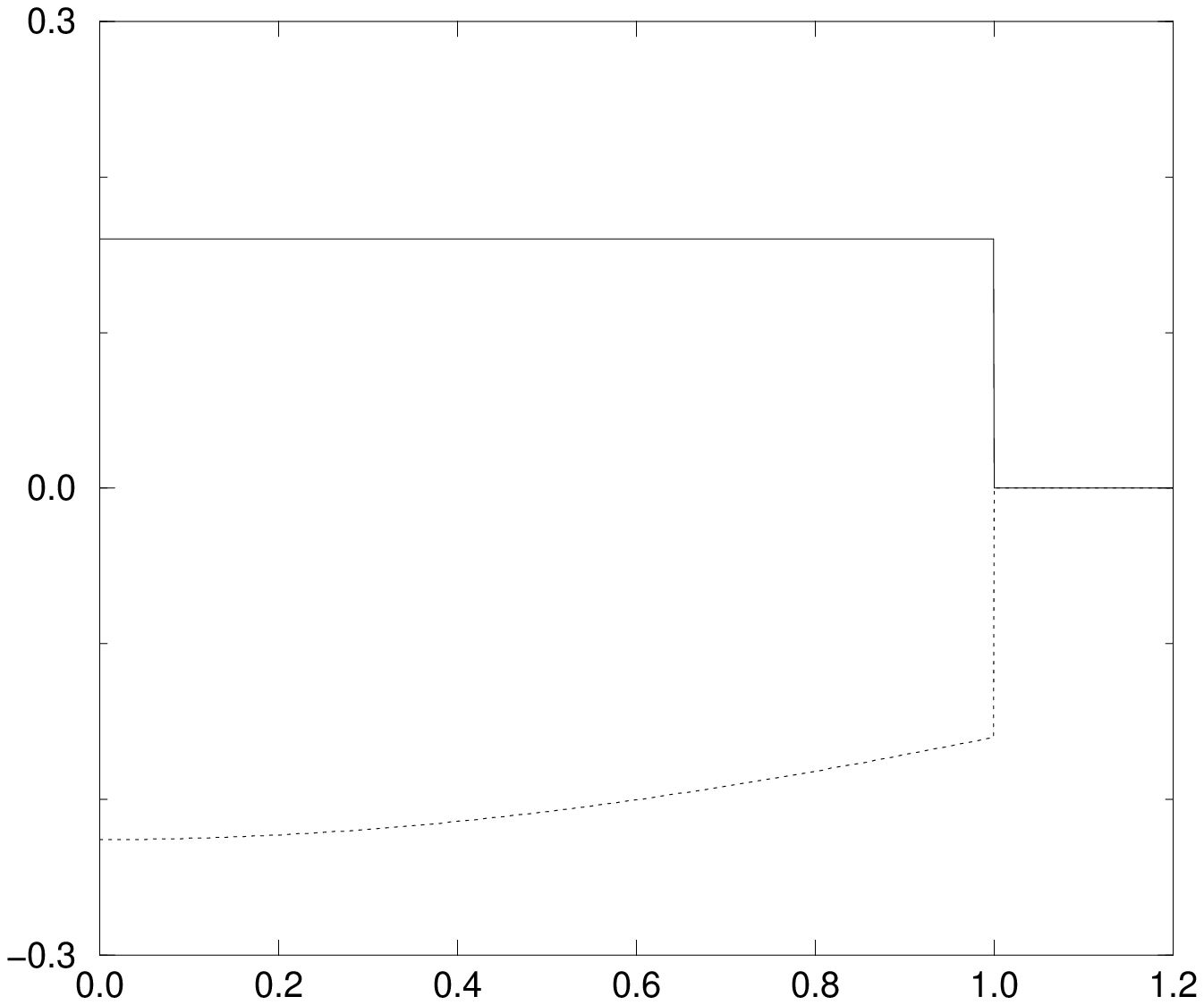}
\\
{\scriptsize Fig. 2(a): A square-barrier potential $V$ (solid line)
defined by (\ref{eq:exsqwell1}) with $V_0=0.16$ and $a=1$, and its SUSY
partner potential ${\tilde V}$ (broken line). The SUSY transformation is
constructed by using the state at $\omega_1 = -0.181i$ (circle in
Fig.~2(b)) as the generator.
}
%\end{center}

\vspace{1.5cm}

%\begin{center}
\leavevmode\scalefig{0.5}\epsfbox{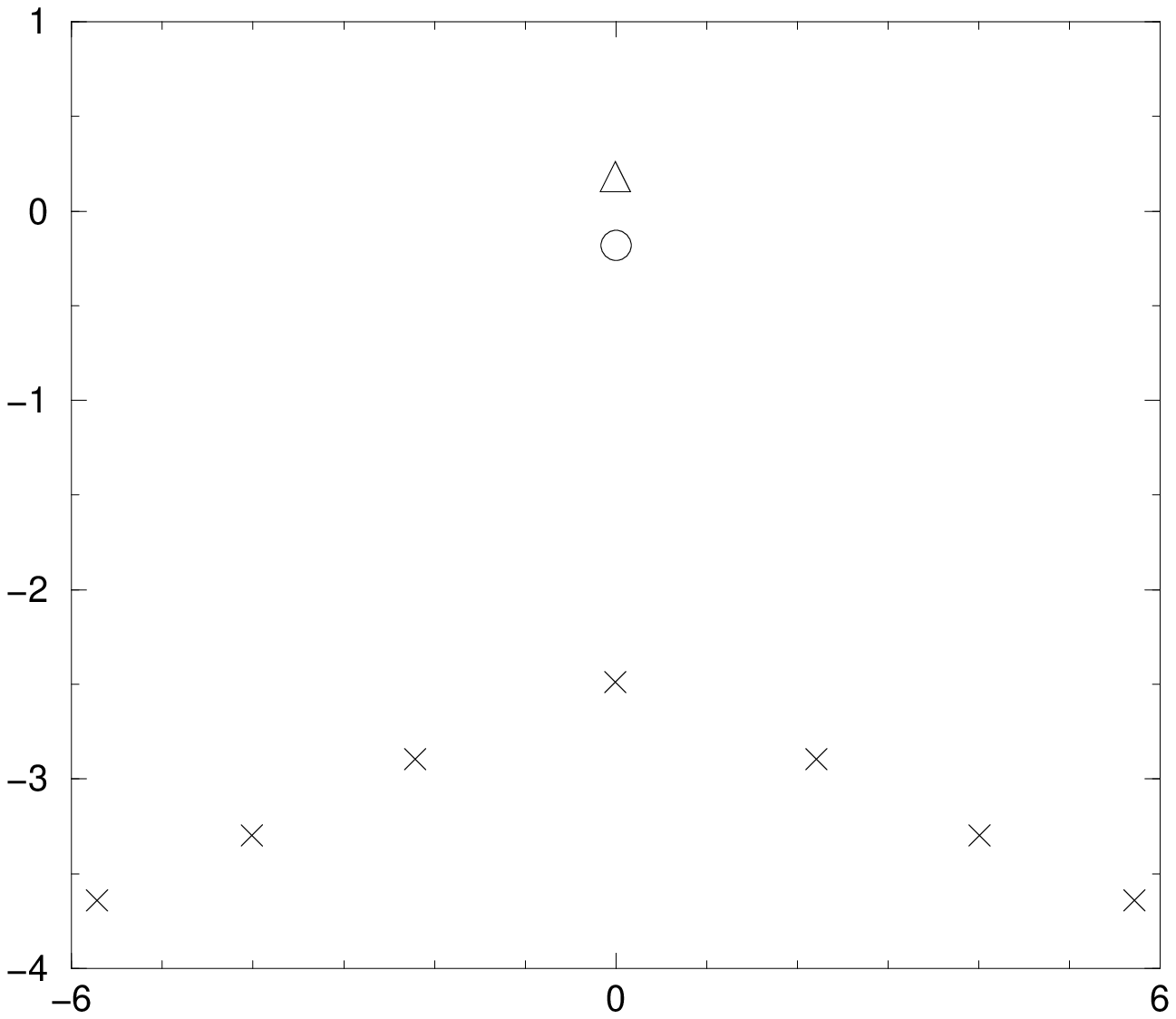}
\\
{\scriptsize Fig. 2(b): The complex $\omega$-plane showing the
QNMs common to both potentials (crosses); the mode present only in $V$
(circle), which corresponds to the generator $\Phi$; and the mode
present only in ${\tilde V}$ (triangle), which corresponds to ${\tilde
\Phi} = A \Phi$. }
%\end{center}

%==========================================================
%Fig 3

\newpage
%\begin{center}
\leavevmode\scalefig{0.5}\epsfbox{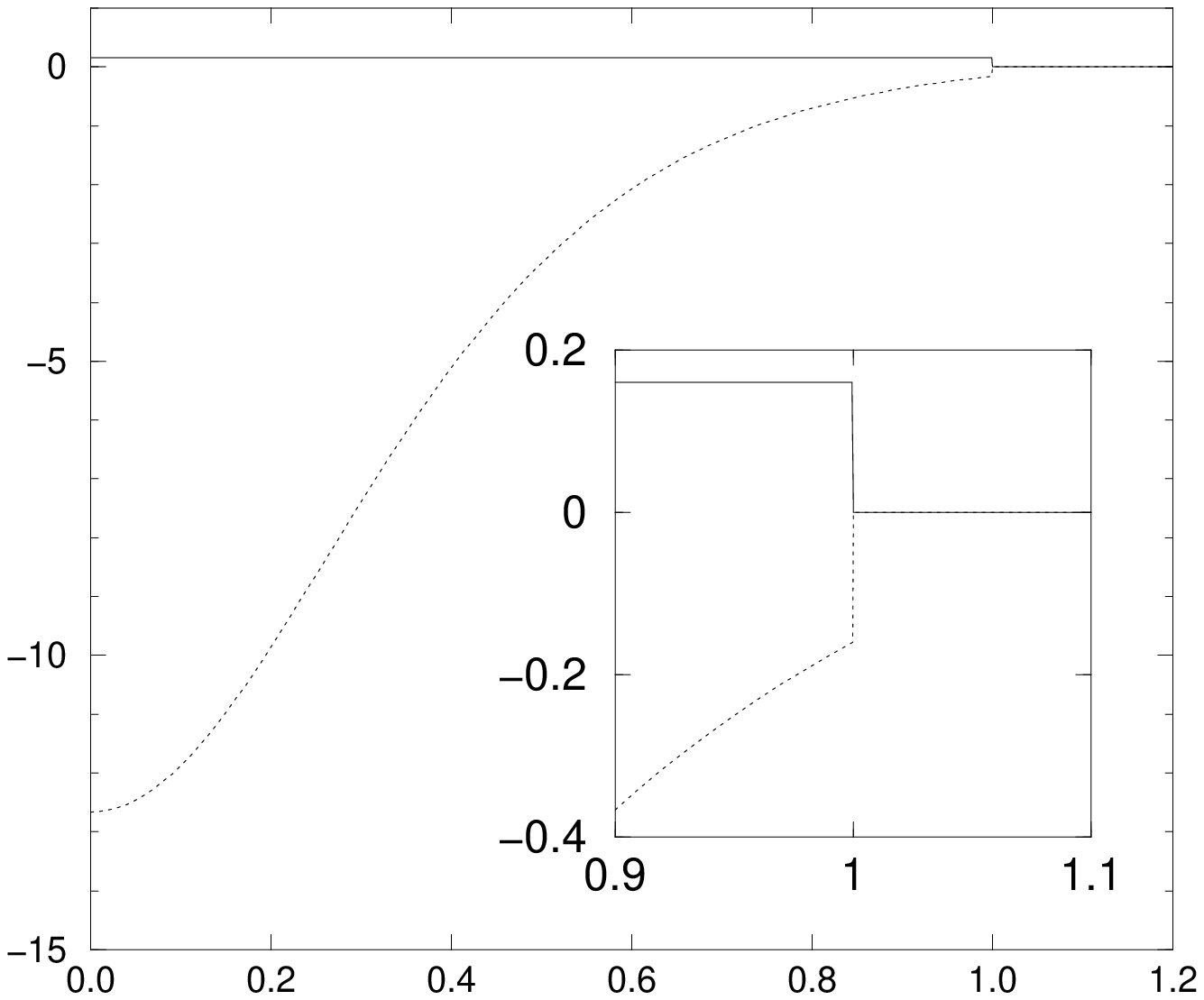}
\\
{\scriptsize Fig. 3(a):
A square-barrier potential $V$ (solid line) 
defined by (\ref{eq:exsqwell1}) with
$V_0=0.16$ and $a=1$,  
and its SUSY partner potential ${\tilde V}$ (broken line).
The SUSY transformation is constructed by using
the state at $\omega_2 = -2.500i$ (circle in Fig.~3(b))
as the generator.
Inset shows one portion enlarged.
}
%\end{center}

\vspace{1.5cm}

%\begin{center}
\leavevmode\scalefig{0.5}\epsfbox{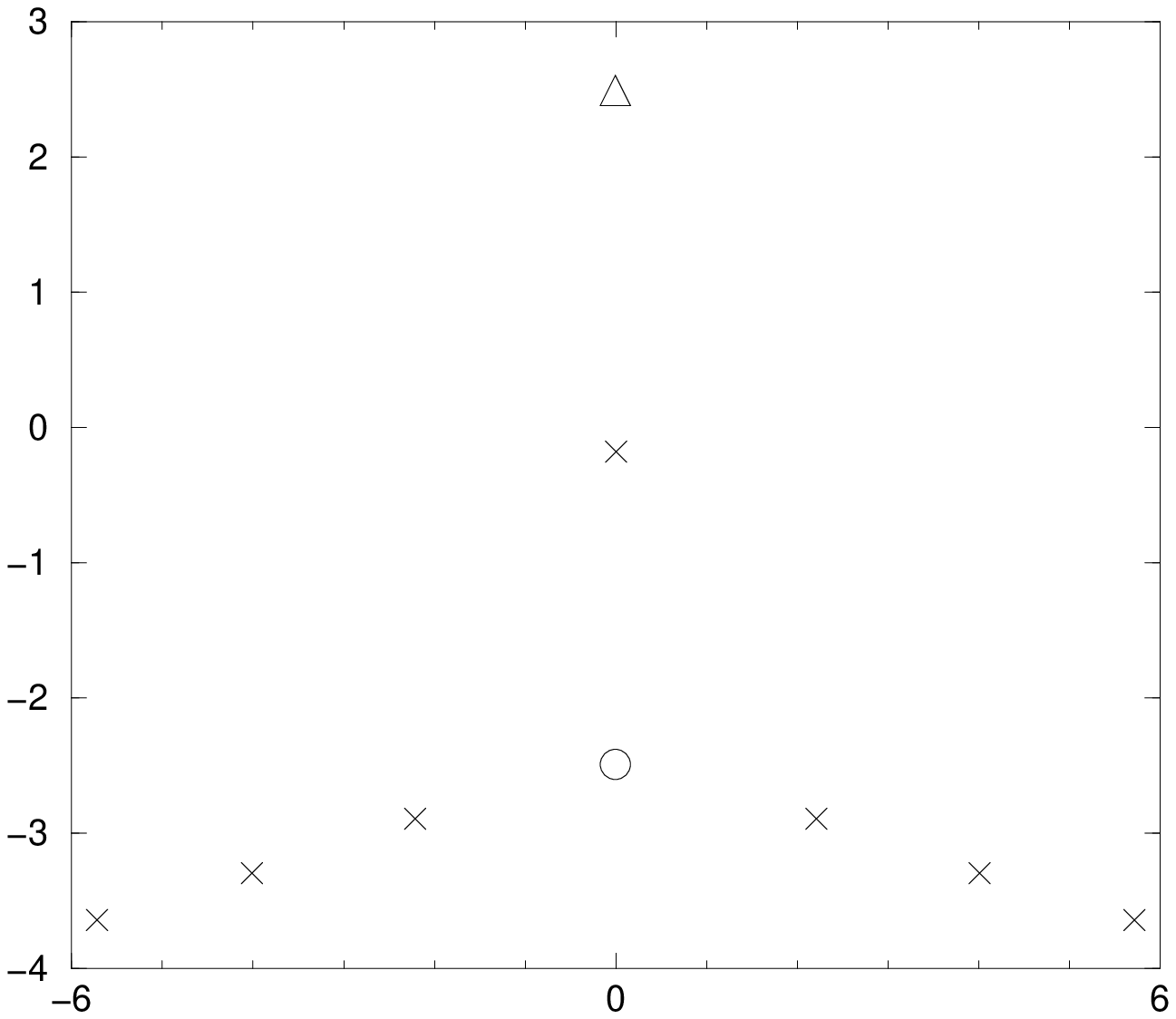}
\\
{\scriptsize Fig. 3(b):
The complex $\omega$-plane showing 
the QNMs common to both potentials
(crosses); the mode present only in $V$ (circle),
which corresponds to the generator $\Phi$; and
the mode present only in ${\tilde V}$ (triangle), which
corresponds to ${\tilde \Phi} = A \Phi$. 
}
%\end{center}

%==========================================================
%Fig 4

\newpage
%\begin{center}
\leavevmode\scalefig{0.5}\epsfbox{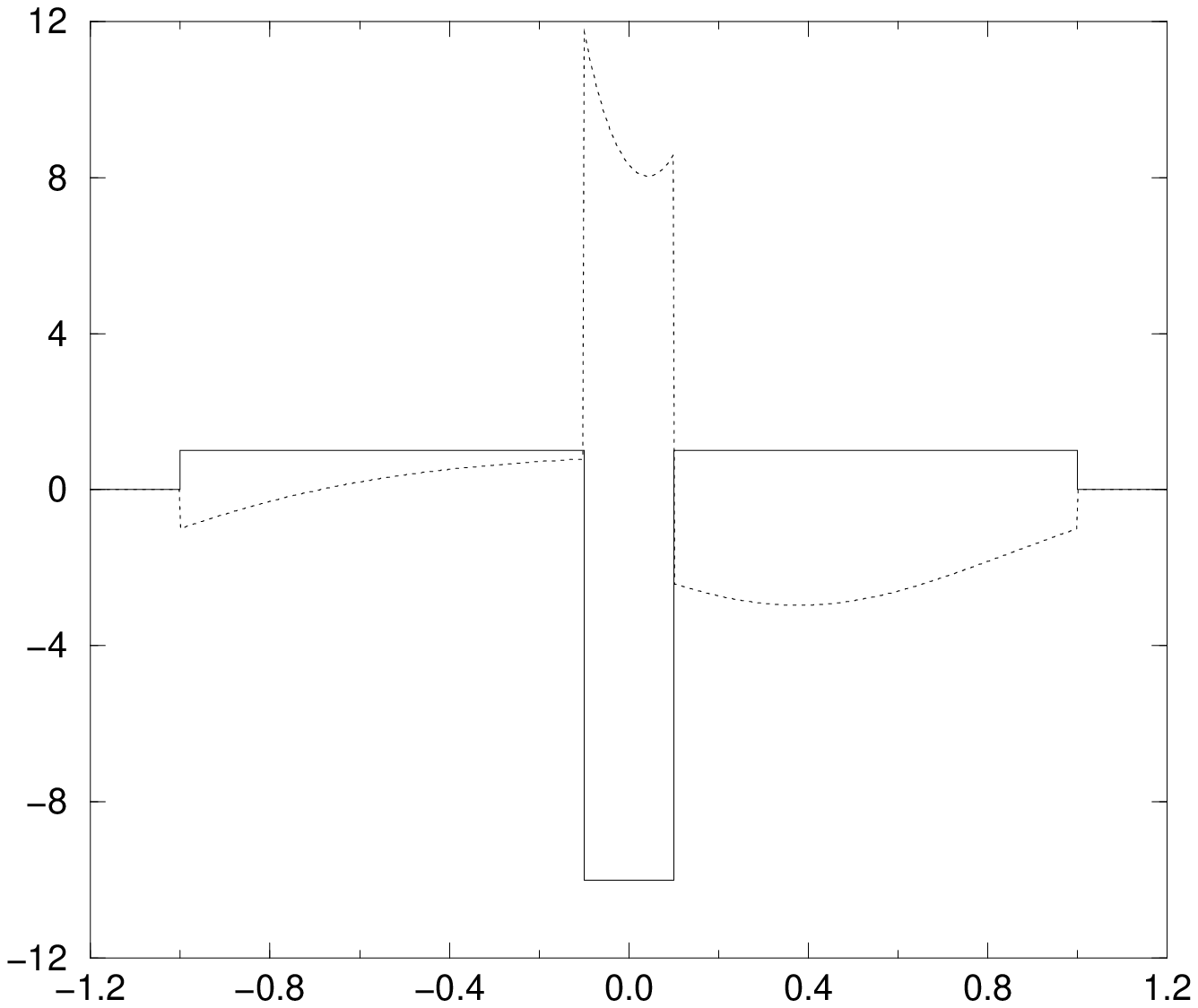}
\\
{\scriptsize Fig. 4(a):
A multi-step potential $V$ (solid line) 
defined by (\ref{eq:multistep1}) with
$V_0=1.0$, $V_1 = -10.0$,
$a=1.0$ and $b=0.1$,  
and its SUSY partner potential ${\tilde V}$ (broken line).
}
%\end{center}

\vspace{1.5cm}

%\begin{center}
\leavevmode\scalefig{0.5}\epsfbox{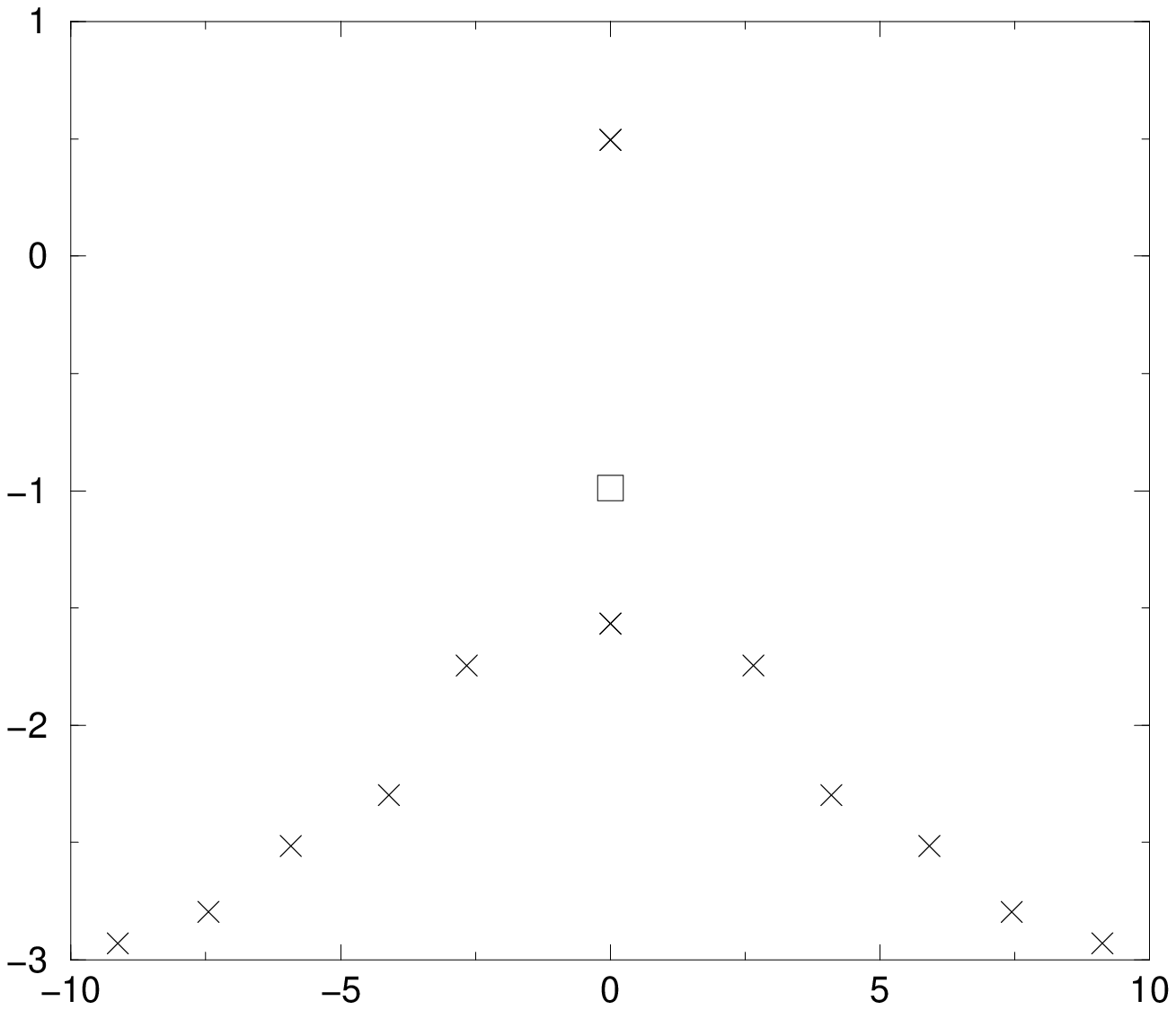}
\\
{\scriptsize Fig. 4(b):
The complex $\omega$-plane showing 
the NM and QNMs common to both potentials
(crosses). The square indicates a TTM${}_{\rm L}$
and a TTM${}_{\rm R}$ in $V$, and a doubled TTM${}_{\rm R}$ 
in ${\tilde V}$.
}
%\end{center}

%==========================================================
%Fig 5

\newpage
%\begin{center}
\leavevmode\scalefig{0.5}\epsfbox{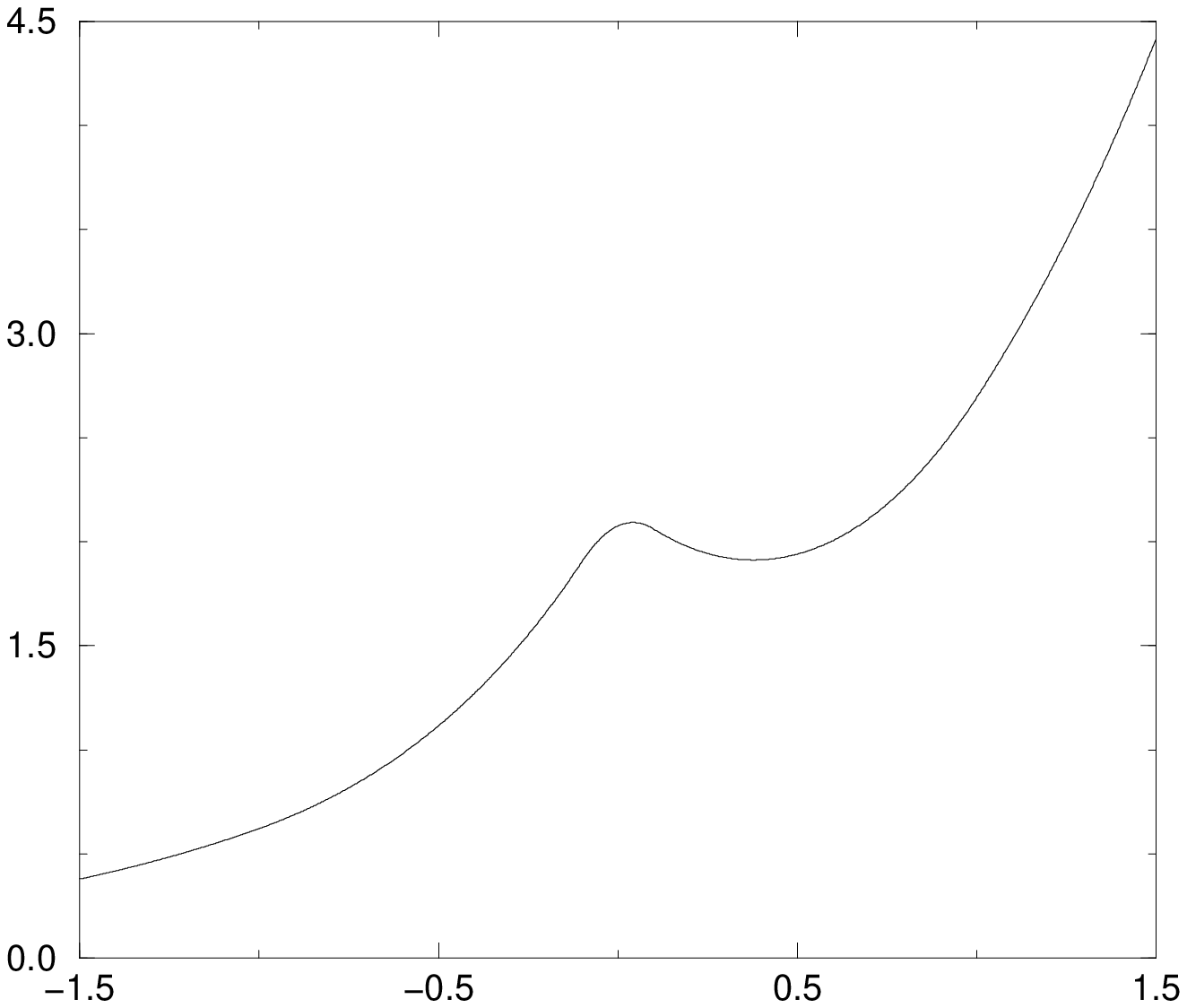}
\\
{\scriptsize Fig. 5(a):
The TTM${}_{\rm L}$ of $V$ used as the generator $\Phi$ in the SUSY %transformation in Fig.~4.
}
%\end{center}

\vspace{1.5cm}

%\begin{center}
\leavevmode\scalefig{0.5}\epsfbox{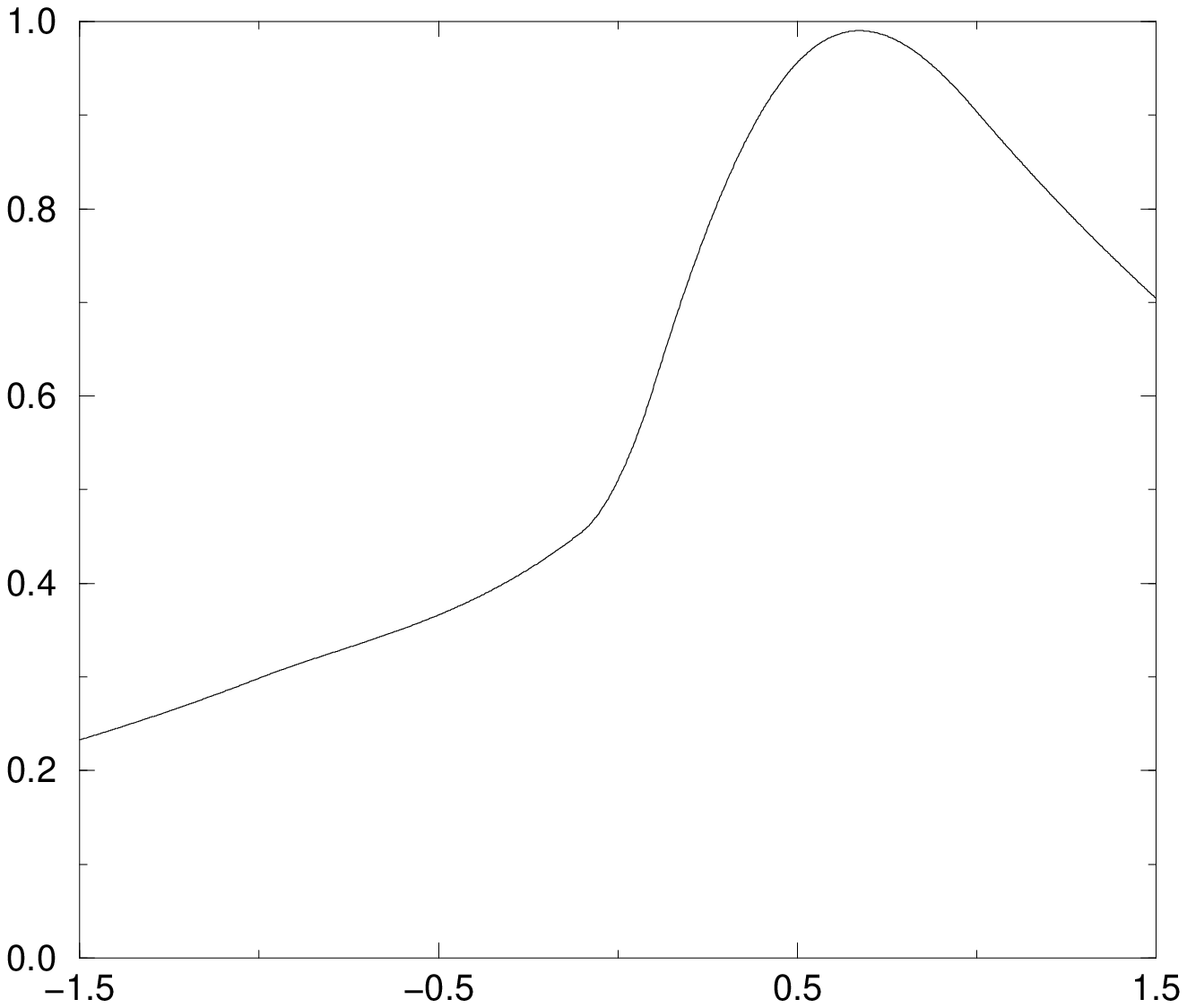}
\\
{\scriptsize Fig. 5(b):
The NM ${\tilde \phi}_1$ in ${\tilde V}$, at $\omega_{1} = 0.498i$.
}
%\end{center}

\vspace{1.5cm}

%\begin{center}
\leavevmode\scalefig{0.5}\epsfbox{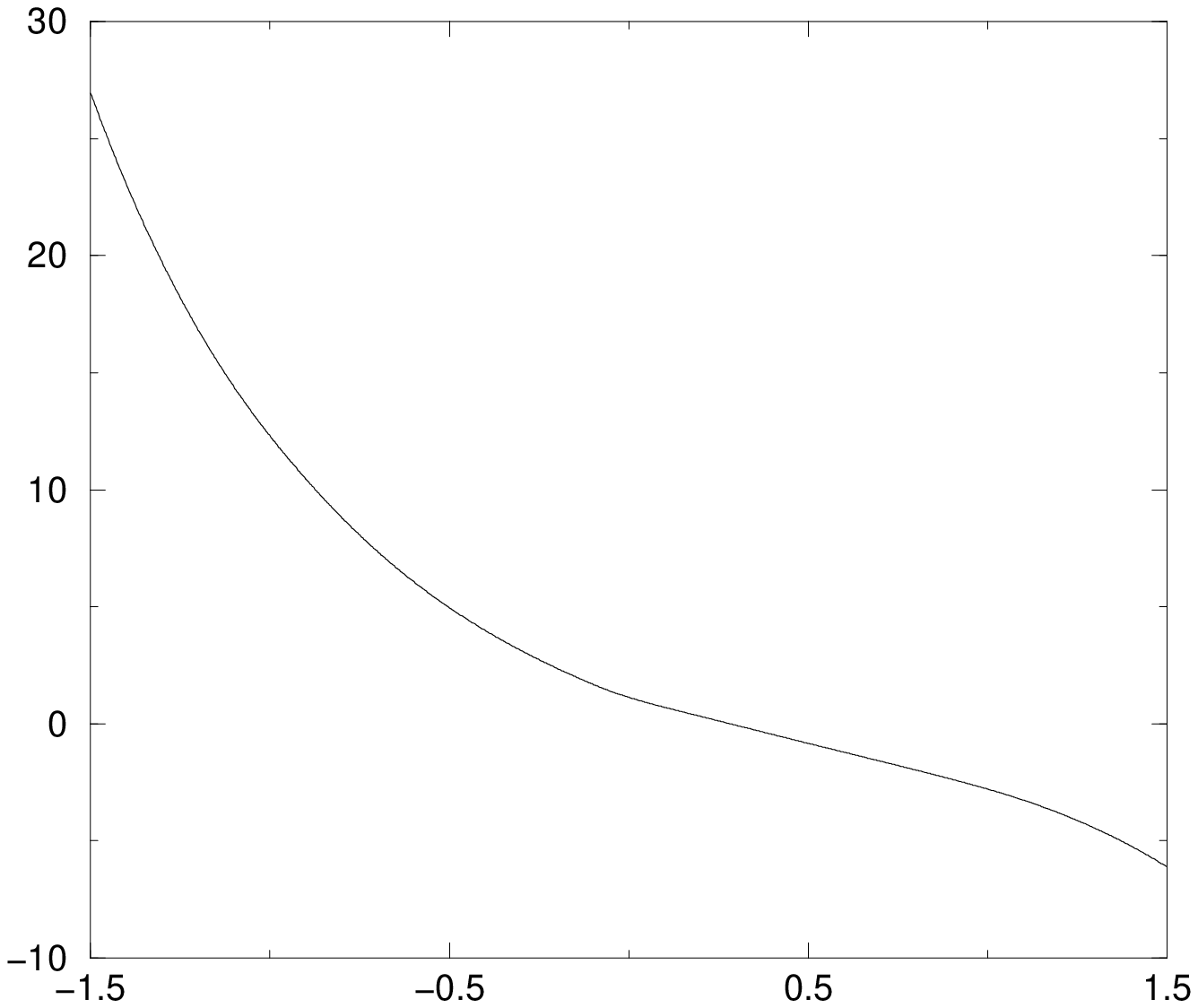}
\\
{\scriptsize Fig. 5(c):
A QNM ${\tilde \phi}_2$ in ${\tilde V}$, at $\omega_{2} = -1.570i$.
}
%\end{center}

%==========================================================
%Fig 6

\newpage
%\begin{center}
\leavevmode\scalefig{0.5}\epsfbox{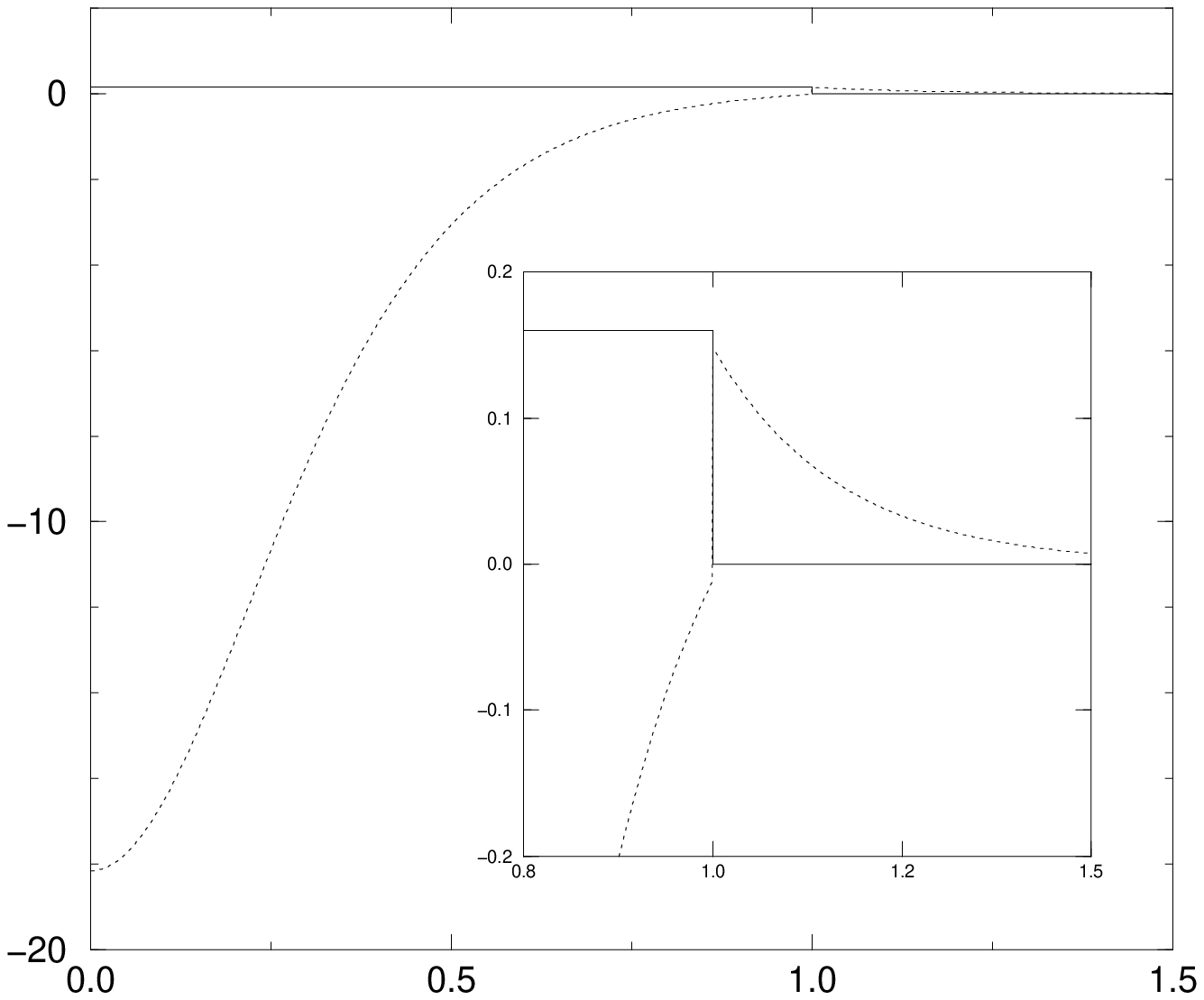}
\\
{\scriptsize Fig. 6 (a):
A square-barrier potential $V$ (solid line) 
defined by (\ref{eq:exsqwell1}) with
$V_0=0.16$ and $a=1$,  
and its SUSY partner potential ${\tilde V}$ (broken line)
generated by a Type 4 transformation, using
$K=3$, $c=1.0$ and $d=-0.829$.
Inset shows one portion enlarged.
}
%\end{center}

\vspace{1.5cm}

%\begin{center}
\leavevmode\scalefig{0.5}\epsfbox{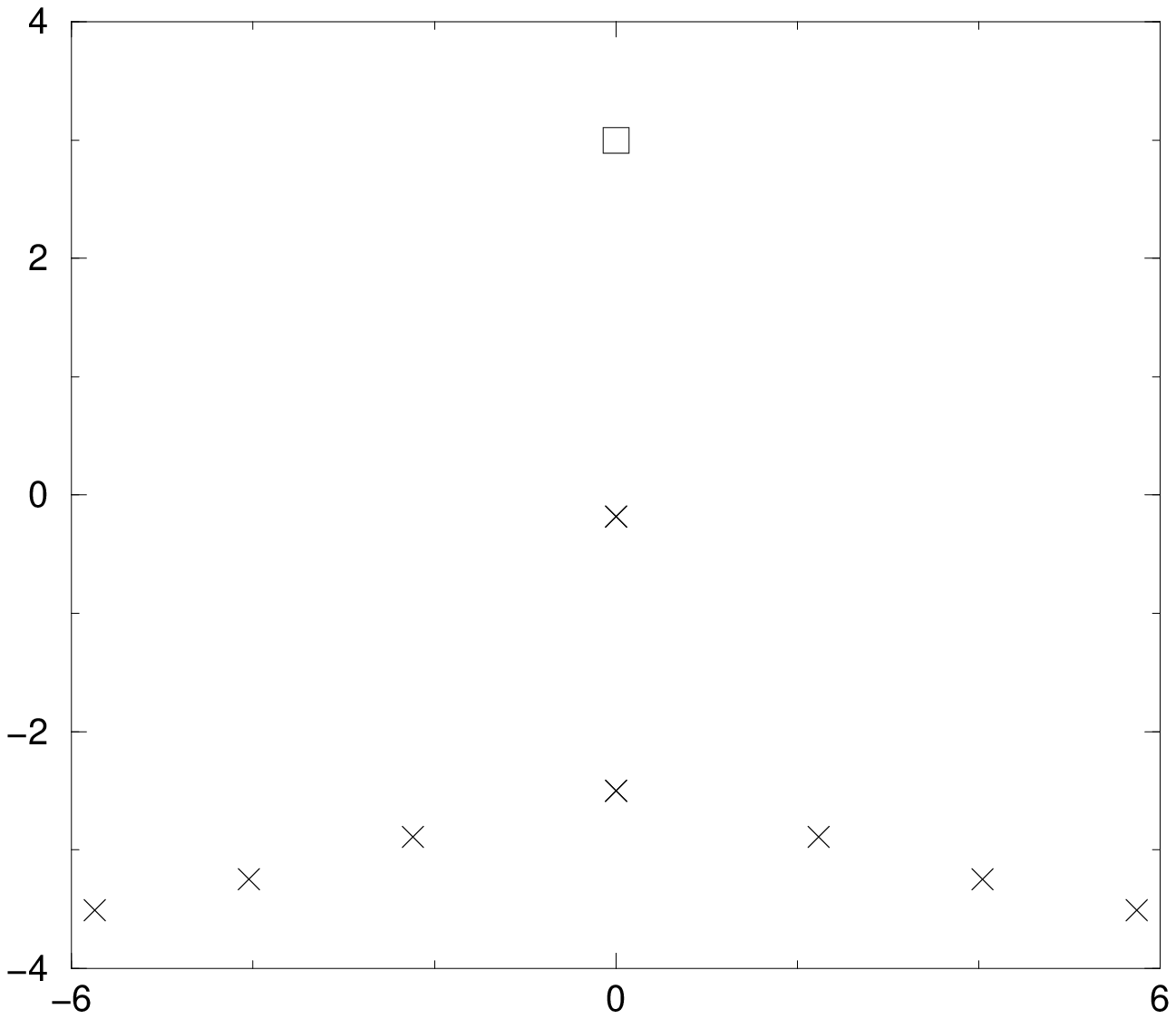}
\\
{\scriptsize Fig. 6(b):
The complex $\omega$-plane showing 
the QNMs common to both potentials
(crosses); the mixed mode $\Phi$ with boundary
condition MM in $V$, used as the generator (square).
}
%\end{center}

%==========================================================
%Fig 7

\newpage
%\begin{center}
\leavevmode\scalefig{0.5}\epsfbox{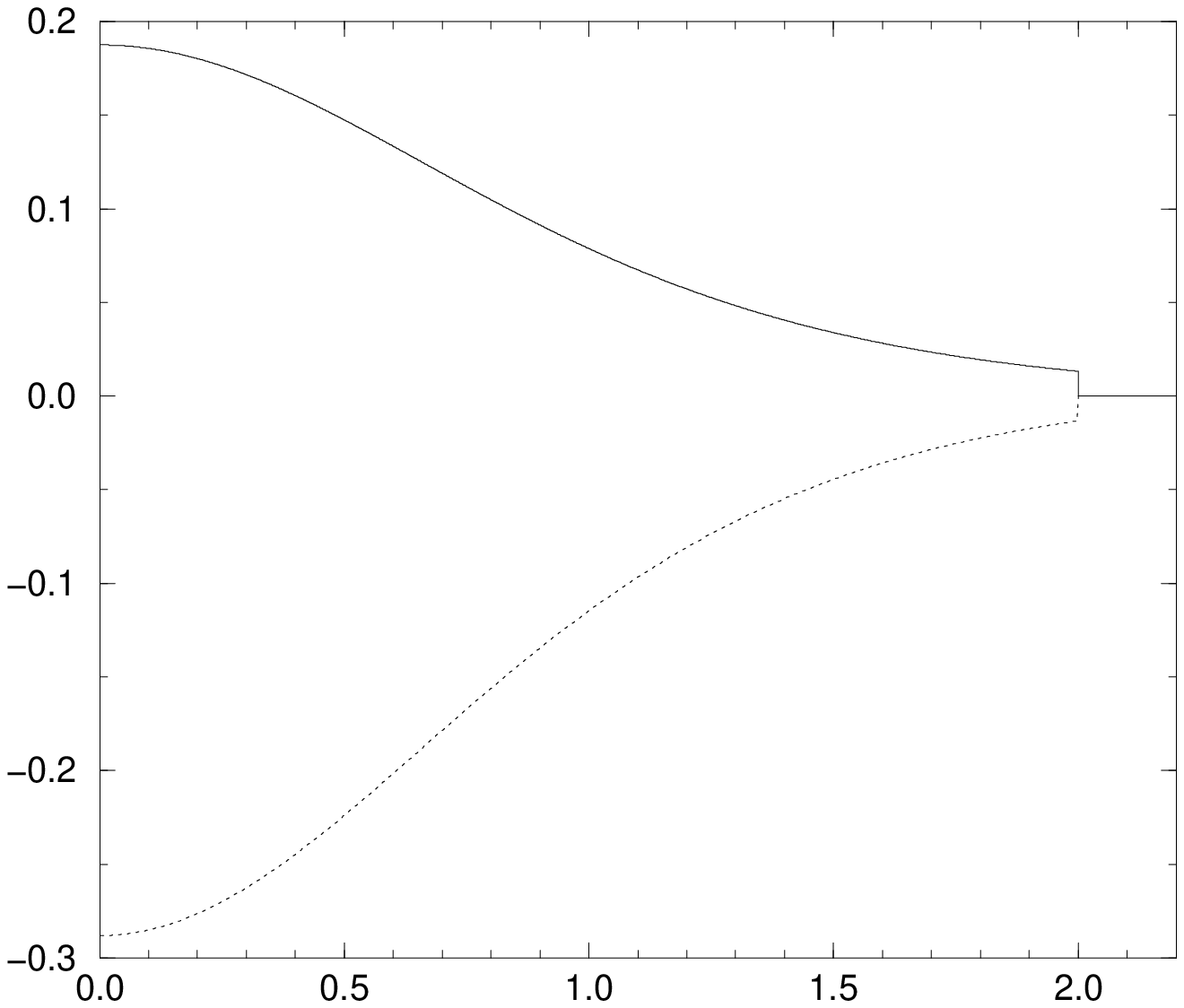}
\\
{\scriptsize Fig. 7(a): 
The truncated PT potential
(\ref{eq:expttrun1}) with
${\cal V} = 3/16$, $b=1$, and $a=2$,  
and its SUSY partner potential ${\tilde V}$ (broken line)
generated by a Type 2 transformation, using the state
at $\Omega = -0.224i$.
} 
%\end{center}

\vspace{1.5cm}

%\begin{center}
\leavevmode\scalefig{0.5}\epsfbox{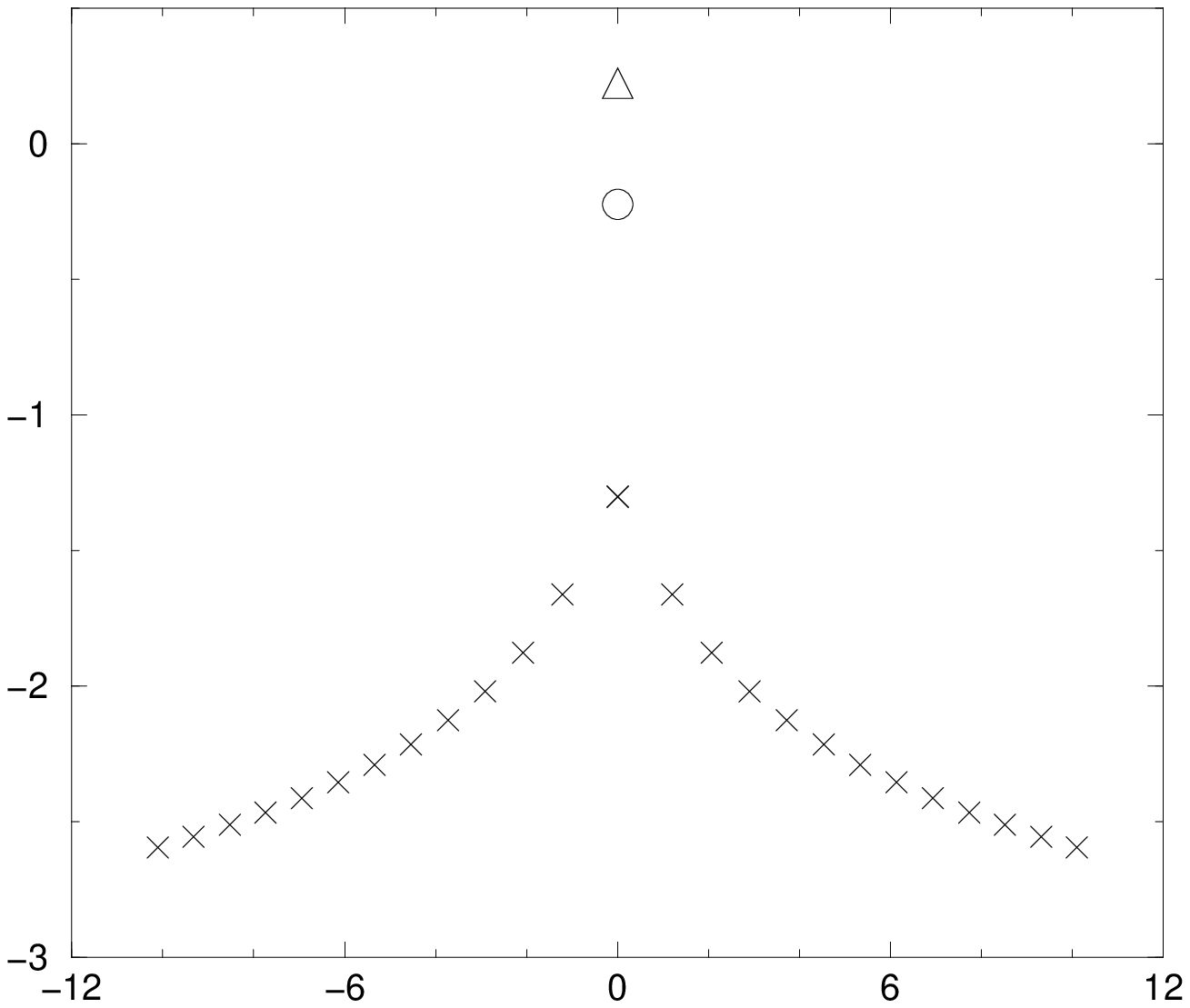}
\\
{\scriptsize Fig. 7(b):
The complex $\omega$-plane showing 
the QNMs common to both potentials
(crosses); the mode present only in $V$ (circle),
which corresponds to the generator $\Phi$; and
the mode present only in ${\tilde V}$ (triangle), which
corresponds to ${\tilde \Phi} = A \Phi$. 
}
%\end{center}

%==========================================================
%Fig 8

\newpage
%\begin{center}
\leavevmode\scalefig{0.5}\epsfbox{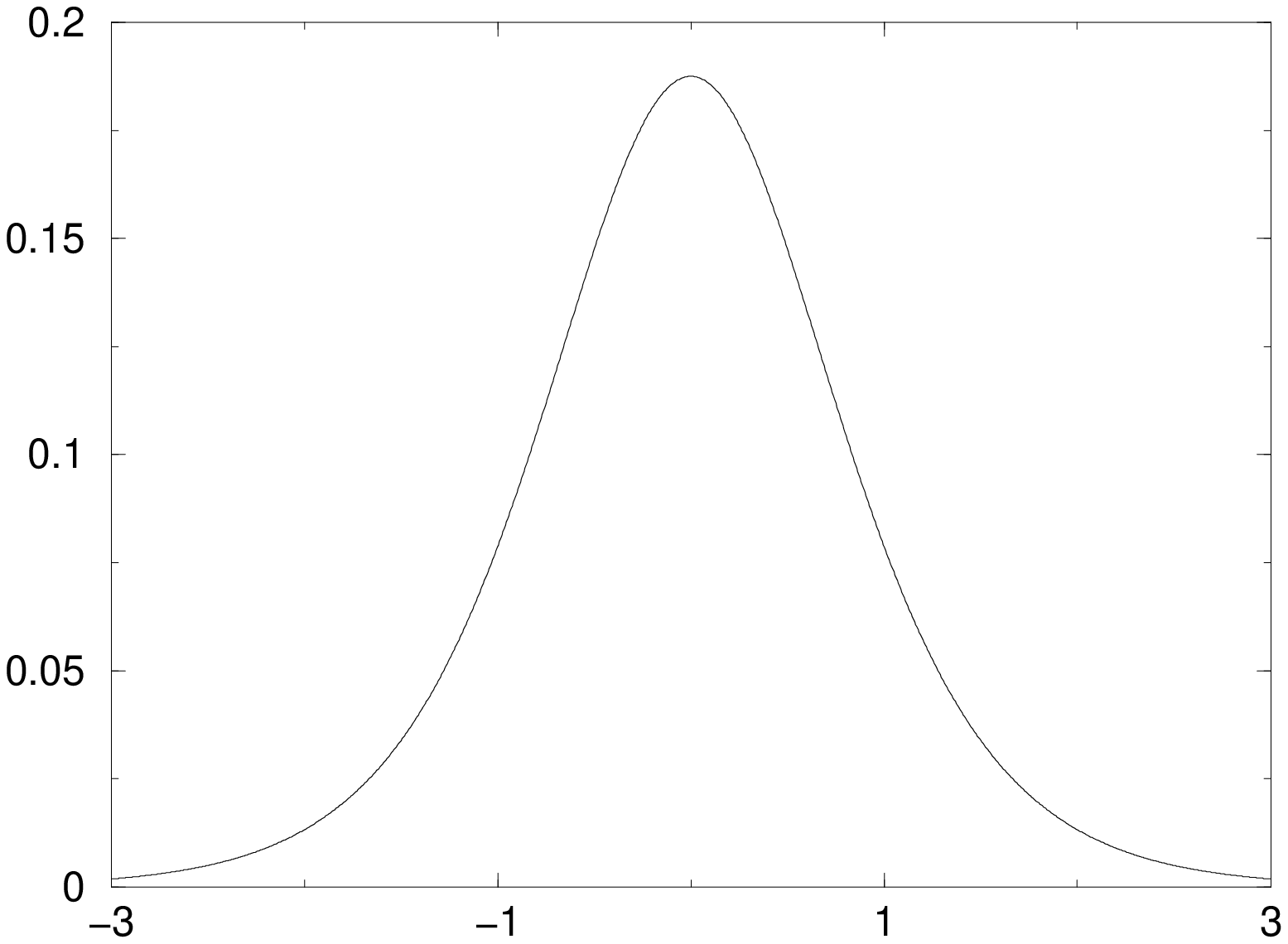}
\\
{\scriptsize Fig. 8(a):
The PT potential $V(x)$ with ${\cal V} = 3/16$, $b=1$.
} 
%\end{center}

\vspace{1.5cm}

%\begin{center}
\leavevmode\scalefig{0.5}\epsfbox{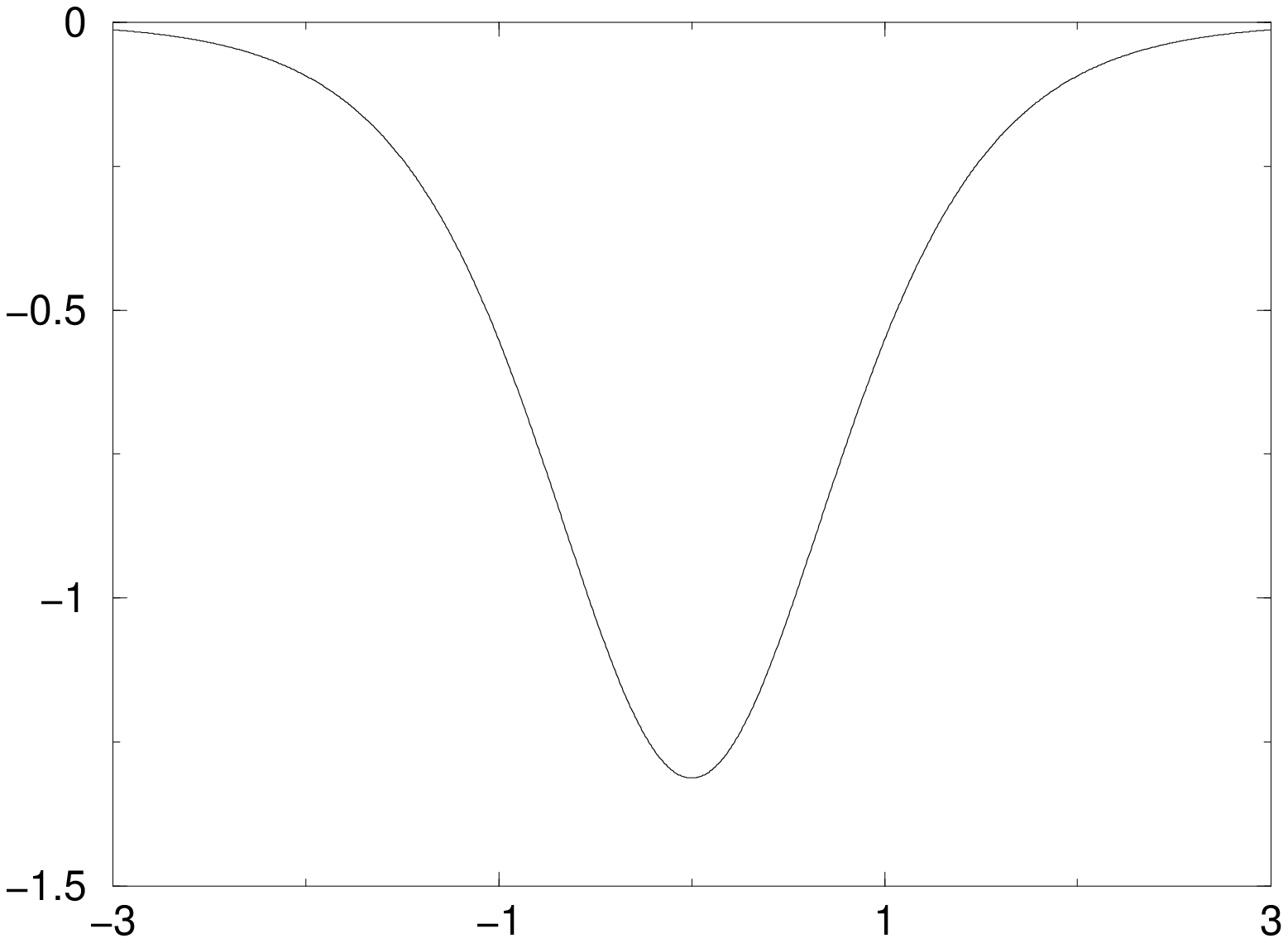}
\\
{\scriptsize Fig. 8(b):
Its SUSY partner ${\tilde V}_0^{+}(x)$.
It is also a PT potential, with ${\tilde {\cal V}} = -21/16$.
}
%\end{center}

\vspace{1.5cm}

%\begin{center}
\leavevmode\scalefig{0.5}\epsfbox{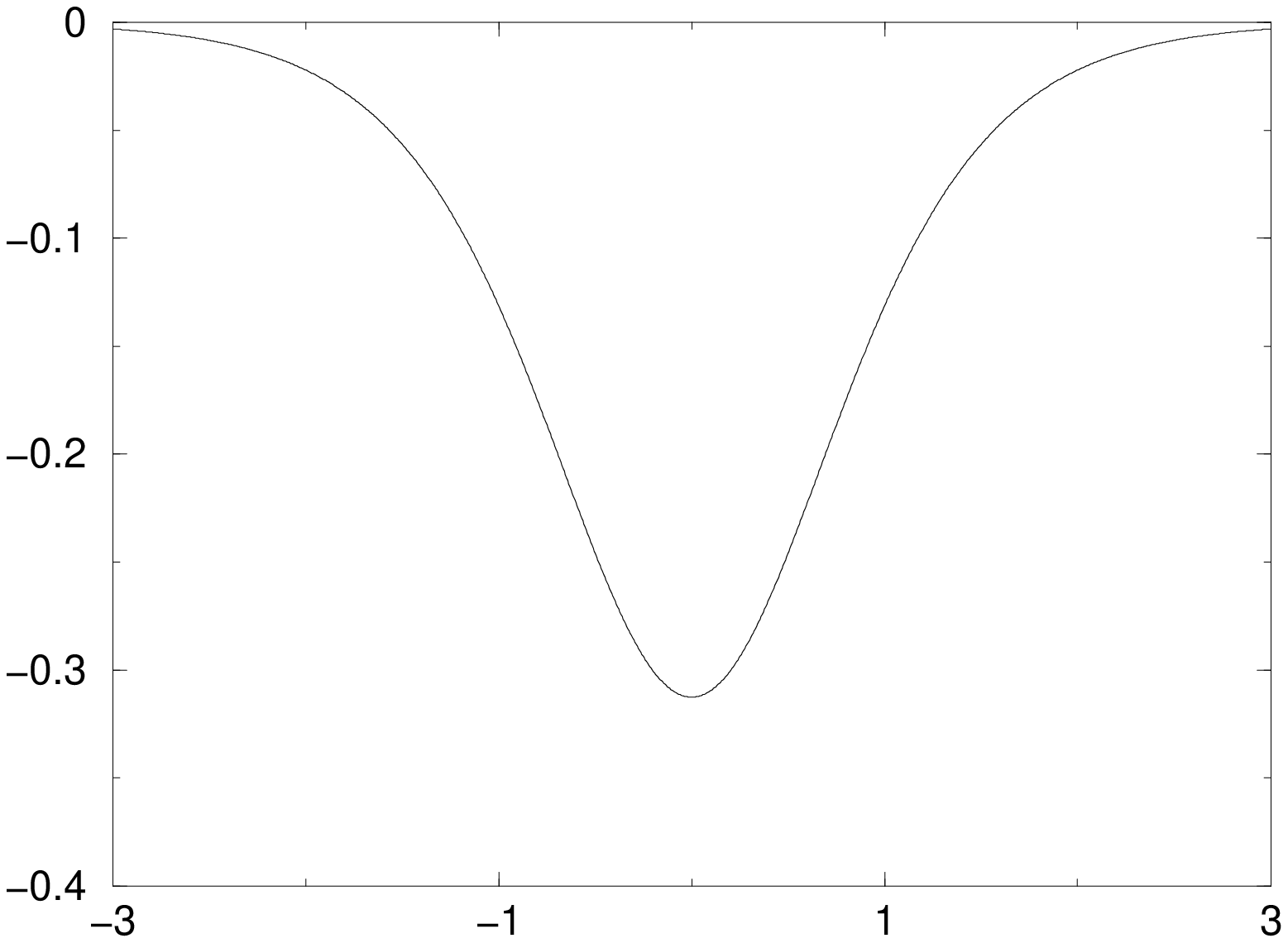}
\\
{\scriptsize Fig. 8(c):
Its SUSY partner ${\tilde V}_0^{-}(x)$.
It is also a PT potential, with ${\tilde {\cal V}} = -5/16$.
}
%\end{center}

\vspace{1.5cm}

%\begin{center}
\leavevmode\scalefig{0.5}\epsfbox{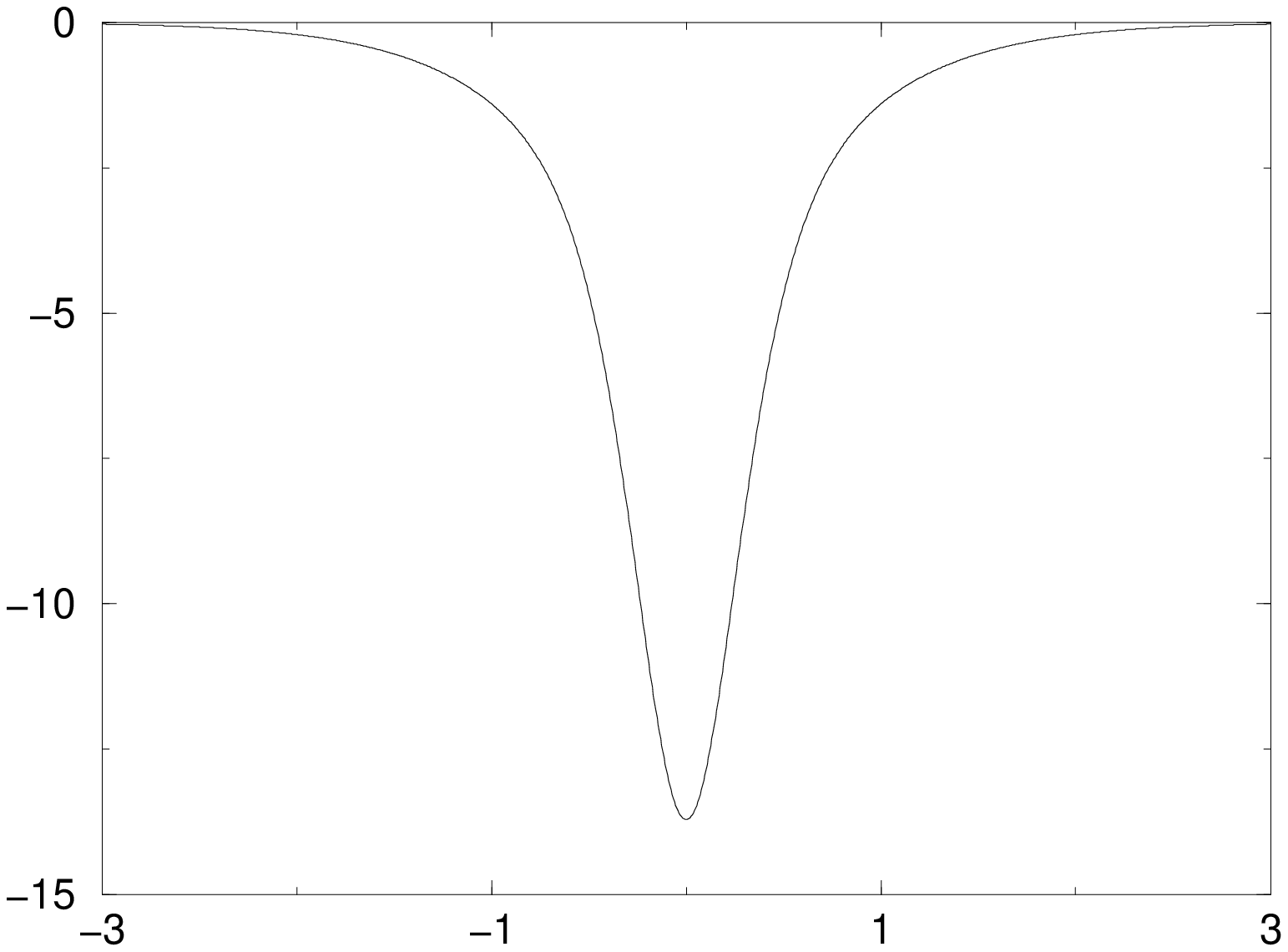}
\\
{\scriptsize Fig. 8(d):
Its SUSY partner ${\tilde V}_2^{+}(x)$.
}
%\end{center}

\vspace{1.5cm}

%\begin{center}
\leavevmode\scalefig{0.5}\epsfbox{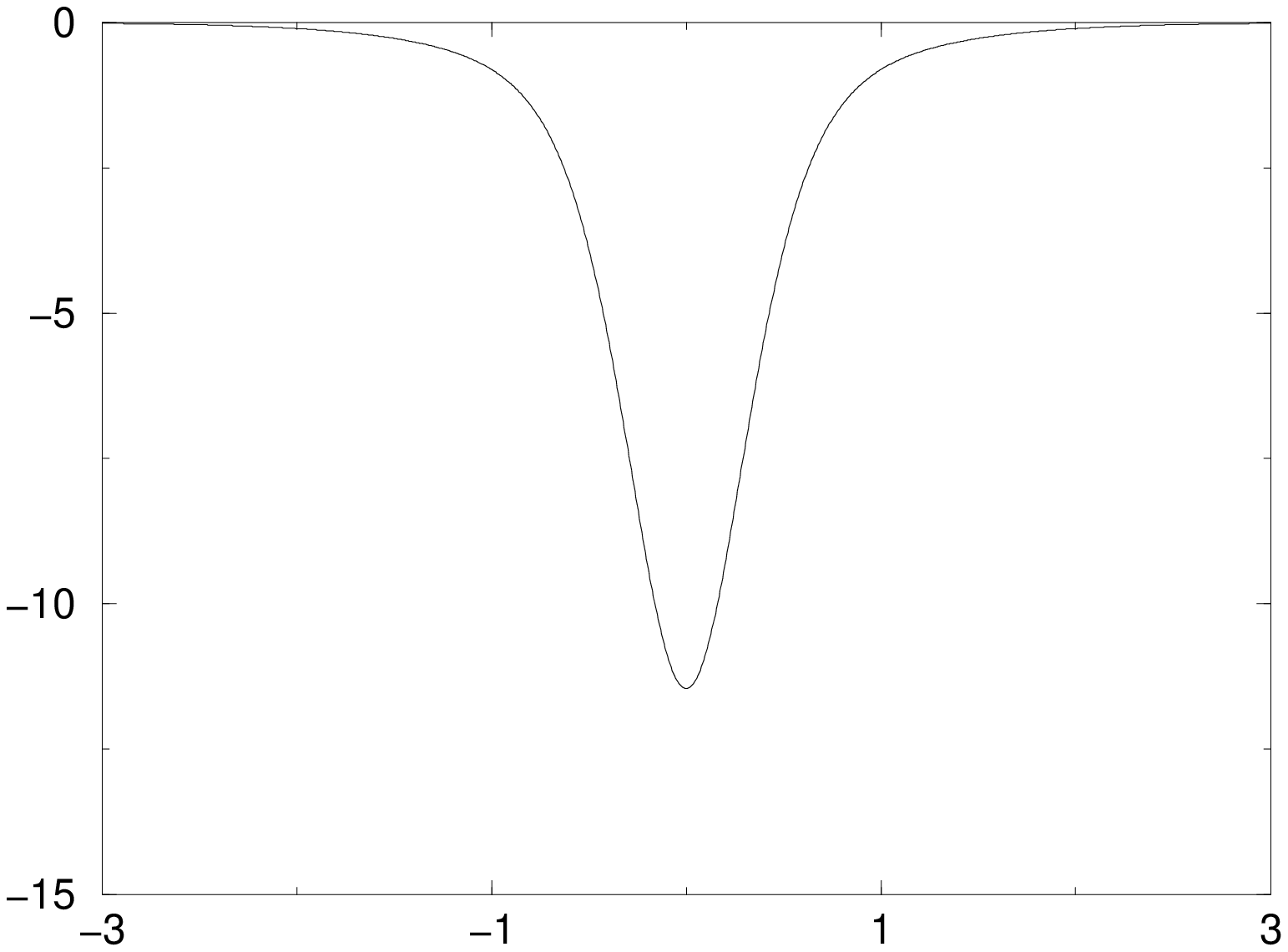}
\\
{\scriptsize Fig. 8(e):
Its SUSY partner ${\tilde V}_2^{-}(x)$.
}
%\end{center}

%==========================================================
%Fig 9

\newpage
%\begin{center}
\leavevmode\scalefig{0.6}\epsfbox{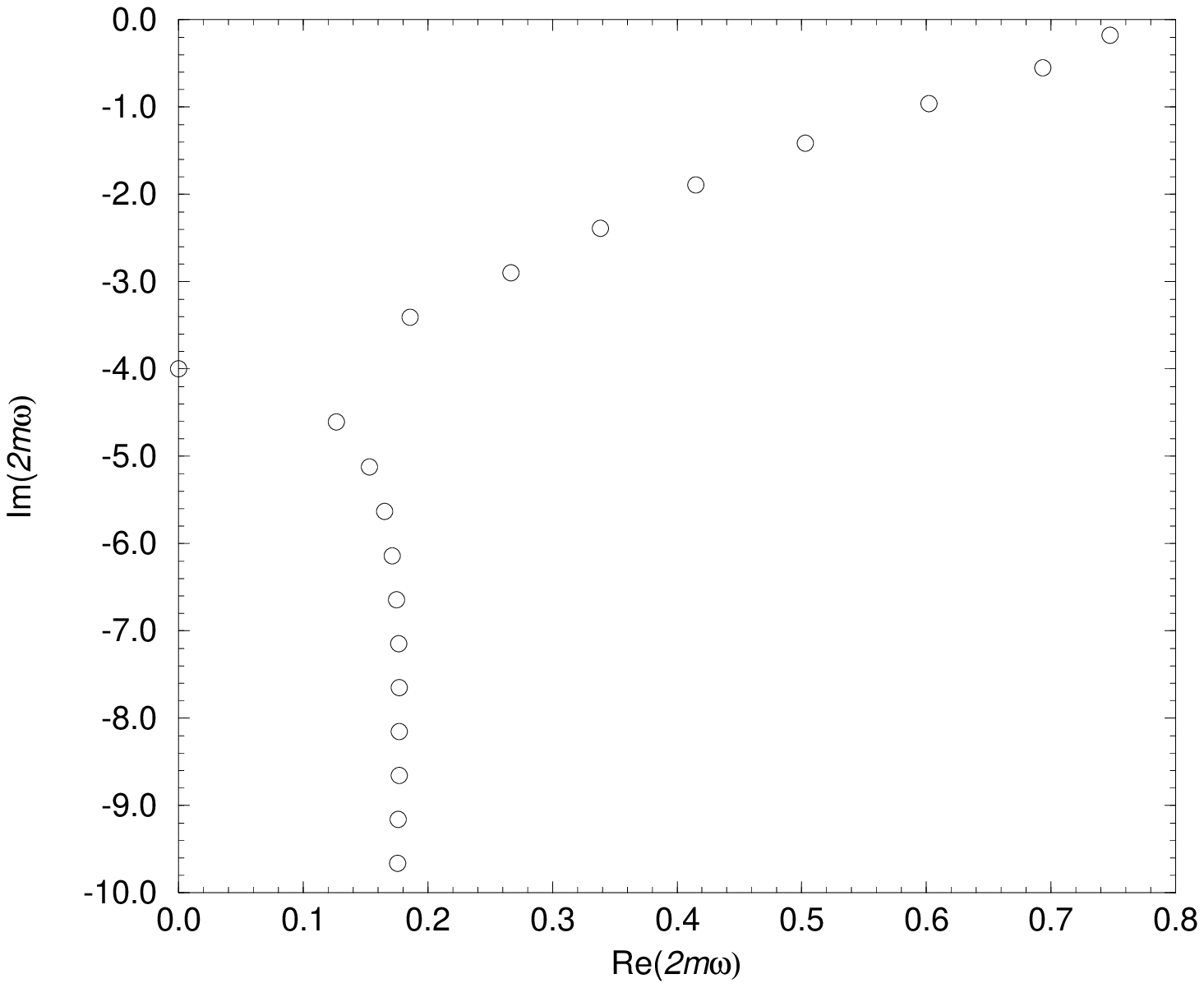}
\\
{\scriptsize Fig. 9: 
The distribution of QNMs of a Schwarzschild black hole
for $l = s = 2$.  The mode on the imaginary axis
$2m \omega = -4i$ is not a QNM, but the special mode.
} 
%\end{center}

%==========================================================


\begin{references}

\setlength{\parskip}{0mm} 

\bibitem{witten1}
E.~Witten, Nucl. Phys. B {\bf 188}, 513 (1981); {\bf 202}, 253 (1982).

\bibitem{nm}
In this paper the term normal mode refers to the spatial dependence only.
The temporal dependence is different according
to whether the Klein--Gordon or the
Schr\"odinger point of view is taken.  In the former, with $\omega^2$ 
real and negative, one has an increasing 
exponential in $t$, which is not physical.
In the latter, i.e., the real and negative eigenvalue is 
$\omega$ rather than $\omega^2$, then the temporal factor would be
$e^{-i\omega t}$, and the state would properly be a normal mode
in the time dependence as well.

\bibitem{susy1}
See, e.g., 
F.~Cooper, A.~Khare and U.~Sukhatme, Phys. Rep. {\bf 251},
267 (1995);
G.~Junker, {\em Supersymmetric Methods in Quantum
and Statistical Physics\/} (Springer, Berlin, 1996)
and references therein.

\bibitem{chand}
S.~Chandrasekhar, {\em The Mathematical Theory of Black
Holes\/} (Oxford Univ. Press, 1991).

\bibitem{lamb}
R.~Lang, M.~O. Scully and W.~E. Lamb, Phys. Lett. A {\bf 7},
1788 (1973).

\bibitem{tong1}
P.~T. Leung, S.~S. Tong and K.~Young, J. Phys. A {\bf 30},
2139 (1997).

\bibitem{nm2}
In other papers, we have also considered
NMs which are obtained as the limit of QNMs when the
escape rate of the waves is tuned to zero, e.g., by
considering potentials with a barrier
of indefinitely increasing height at $x \approx \pm a$.
These NMs are defined on finite intervals, and have
positive eigenvalues $\omega^2$; thus they are represented
on the real $\omega$ axis.  In this paper we do {\em not\/}
consider the limit where the rate of escape goes to zero,
and therefore there is no danger of confusing NMs in this
sense with the NMs on the positive imaginary axis
in the $\omega$-plane.

\bibitem{bounds} If $\im\omega>0$, the eigenfunction is square-integrable, so the Hermiticity of $H$ implies that $\omega^2\in{\bf R}$.

\bibitem{abs}
P.~T. Leung, S.~Y. Liu and K.~Young, Phys. Rev. A {\bf 49},
3982 (1994).

\bibitem{ligo}
See, e.g., A.~A. Abramovici et al., Science {\bf 256}, 325 (1992).

\bibitem{zeromode}
These are not to be confused with states with
zero eigenvalue, which are sometimes called
zero modes in the SUSY literature.  We shall not
use this latter nomenclature in this paper.

\bibitem{comp}
A.~Bachelot and A.~Motet-Bachelot, Ann. Inst. Henri Poincare,
Sect.~A, {\bf 59}, 3 (1993);
P.~T. Leung, S.~Y. Liu and K.~Young, Phys. Rev. A {\bf 49},
3057 (1994);
E.~S.~C. Ching, P.~T. Leung, W.~M. Suen and K.~Young,
Phys. Rev. Lett. {\bf 74}, 4588 (1995);
Phys. Rev. D {\bf 54}, 3778 (1996).

\bibitem{sumnm}
In calculating the Green's function $G(t)$ by inverse
Fourier transform of ${\tilde G}(\omega)$~\cite{comp}, one has
to integrate along a line in the complex $\omega$-plane that
lies above all the singularities, to ensure that $G(t{<}0)=0$;
for $t{>}0$ this then captures all the singularities,
including the NMs. 

\bibitem{pricetail}
R.~H. Price, Phys. Rev. D {\bf 5}, 2419 (1972); {\bf 5}, 2439 (1972).

\bibitem{tail}
E.~S.~C. Ching, P.~T. Leung, W.~M. Suen and K.~Young,
Phys. Rev. Lett. {\bf 74}, 2414 (1995);
Phys. Rev. D {\bf 52}, 2118 (1995).

\bibitem{tong2}
P.~T. Leung, S.~S. Tong and K.~Young, J. Phys. A {\bf 30},
2139 (1997).

\bibitem{bior}
P.~T. Leung, W.~M. Suen, C.~P. Sun and K.~Young,
Phys. Rev. E {\bf 57}, 6101 (1998).

\bibitem{quant}
K.~C. Ho, P.~T. Leung, A.~Maassen van den Brink and
K.~Young, Phys. Rev. E {\bf 58}, 2965 (1998);
A.~Maassen van den Brink,
``Exactly solvable path integral for open cavities in
terms of quasinormal modes", Preprint (1999) [quant-ph/9905082].

\bibitem{rmp}
E.~S.~C. Ching, P.~T. Leung, A.~Maassen van den Brink,
W.~M. Suen, S.~S. Tong and K.~Young,
Rev. Mod. Phys. {\bf 70}, 1545 (1998).

\bibitem{invnm}
V.~Bargmann, Phys. Rev. {\bf 75}, 301 (1949);
Rev. Mod. Phys. {\bf 21}, 488 (1949);
G.~Borg, Acta Math. {\bf 78}, 1 (1946);
N.~Levinson, Math. Tidsskr. B {\bf 25}, 24 (1949);
V.~A. Marchenko, Dokl. Akad. Nauk SSSR {\bf 72}, 457 (1950);
{\bf 104}, 695 (1955);
M.~G. Krein, Dokl. Akad. Nauk SSSR {\bf 76}, 21 (1951);
{\bf 76}, 345 (1951); {\bf 82}, 669 (1952);
R.~Jost and W.~Kohn, Phys. Rev. {\bf 87}, 979 (1952);
{\bf 88}, 382 (1952).\\
I.~M. Gelfand and B.~M. Levitan, Izv. Akad. Nauk. SSSR, Ser. Mat. {\bf 15}, 
309 (1951) 
[Am. Math. Soc. Translations (2), {\bf 1}, 253 (1955)];
B.~M. Levitan, Izv. Akad. Nauk SSSR, Ser. Mat. {\bf 28}, 63 (1964);
B.~M. Levitan and M.~G. Gasymov, UMN {\bf 19}, 3 (1964) 
[Russian Math. Survey {\bf 19}, 1 (1964)];
L.~D. Faddeev, Usp. Mat. Nauk {\bf 14}, 57 (1959) 
[English translation: B.~Seckler, J. Math. Phys. {\bf 4}, 72 (1963)];
F.~J. Dyson, in {\em Essays in Honor of Valentine Bargmann},  
E.~H. Lieb, B.~Simon and A.~S. Wightman, eds. (Princeton University Press, N. J., 1976);
V.~Barcilon,  in {\em Inverse Eigenvalue Problems (Lecture Notes in Mathematics, Vol. 1225)}, 
A.~Dold and B.~Eckmann, eds. (Springer Verlag, Berlin, 1986).

\bibitem{rundell}W.~Rundell and P.~Sacks, Math. of Comp. {\bf58}, 161 (1992); B.~D. Lowe, M.~Pilant, and W.~Rundell, SIAM J. Math. Anal. {\bf23}, 482 (1992).

\bibitem{invnmnum}
C.~P.~Sun, K.~Young and J.~Zou, J. Phys. A {\bf32}, 3833 (1999).

\bibitem{numinv} 
Consider the easier problem of {\em perturbative} inversion: given a known potential $V$
with spectrum $\{ \omega_n \}$, can an unknown perturbation $\Delta V$ be determined from the corresponding
shifts $\{ \Delta \omega_n^2 \}$? If this problem is posed for NMs with 
only {\em one\/} spectrum, e.g., that
for the nodal problem, unique inversion is definitely {\em im}possible.  This is most easily seen by
recognizing (e.g., Ref.~\cite{invnm}) that the antinodal spectrum is required as well.  In other words,
we are free to assume any antinodal spectrum, and upon continuous
distortions of the latter, one can obtain a
continuous family of $\Delta V$ that would correspond to the same nodal-spectrum shifts $\{ \Delta \omega_n^2
\}$.  However, numerical experiments \cite{numinv2} seem to indicate that perturbative inversion is possible
and unique for many examples of open systems, with only {\em one\/} set of complex frequency shifts given. 
This then suggests that if there are remaining ambiguities in inversion, at least they do not form a
continuous family, and are limited only to discrete solutions. 

\bibitem{numinv2}
W.~S. Lee, Thesis, The Chinese University of Hong Kong (1998).

\bibitem{freq} 
Suppose one has a signal $\phi(t) = \sum_n a_n \exp(-i\omega_n t)$ in terms of a set of
(possibly complex) eigenfrequencies $\omega_n$.  The coefficients $a_n$ depend on the initial conditions,
whereas the frequencies $\omega_n$ are determined by the system (i.e., the potential $V$).   Thus, going
through the frequencies is the natural way of filtering the information to retain only those properties
intrinsic to the system. 

\bibitem{bhsign}
B.~Schutz and C.~Will, Astrophys. J. {\bf 291},
L33 (1985);
S.~Detweiler, in {\em Sources of Gravitational
Radiation}, L.~Smarr, ed. (Cambridge Univ. Press,
Cambridge, 1979), p.~211 and references therein.

\bibitem{dirt}
P.~T. Leung, Y.~T. Liu, W.~M. Suen, C.~Y. Tam
and K.~Young, Phys. Rev. Lett. {\bf 78}, 2894 (1997);
Phys. Rev. D {\bf 59}, 044034 (1999).

\bibitem{suk}
C.~V. Sukumar, J. Phys. A {\bf 18}, 2917 (1985).

\bibitem{complexW}In A.~A. Andrianov, M.~V. Ioffe, F.~Cannata, and J.-P. Dedonder, Int. J. Mod. Phys. A {\bf14}, 2675
(1999), SUSY transformations between complex potentials using a complex $W$ are considered, with an emphasis on cases
where one of the partners is real/hermitian, giving also the complex partner an (essentially) real spectrum. However,
for complex potentials in general one has neither completeness nor causality. In contrast, we study complex spectra
of real potentials. 

\bibitem{fn2} 
The nomenclature follows the NM case.  It may seem paradoxical that the case where the spectra
are imperfectly matched is said to be good, while the case of perfect matching is said to be broken; the
reason stems from boson--fermion symmetry in field theory. 

\bibitem{jordan} 
A.~Maassen van den Brink and K.~Young, 
``Jordan blocks and generalized biorthogonal bases: realizations 
in open wave systems", Preprint (1998) 
[math-ph/9905019].

\bibitem{pt}
G.~P\"oschl and E.~Teller, Z. Phys. {\bf 83},
143 (1933);
V.~Ferrari and B.~Mashhoon, Phys. Rev. D {\bf 30}, 295 (1984).

\bibitem{pottail}
P.~T. Leung, Y.~T. Liu, C.~Y. Tam and K.~Young,
Phys. Lett. A {\bf 247}, 253 (1998).

\bibitem{bytong}
B.~Y. Tong, Solid State Comm. {\bf 104}, 679 (1999).

\bibitem{chanddet}
S.~Chandrasekhar and S.~Detweiler,
Proc. Roy. Soc. Lond. A {\bf 344}, 441 (1975).

\bibitem{price}
A.~Anderson and R.~H. Price,
Phys. Rev. D {\bf 43}, 3147 (1991).

\bibitem{spmode}
H.~Lin and B.~Mashhoon,
Class. Quantum Grav. {\bf 13}, 233 (1996); see also Ref.~\cite{chand}.

\bibitem{onozawa}H. Onozawa, Phys. Rev. D {\bf55}, 3593 (1997).

\bibitem{and}
N.~Andersson and S.~Linnaeus, Phys. Rev. D {\bf 46},
4179 (1992);
N.~Andersson, Class. Quantum Grav. {\bf 11}, L39 (1994).

\bibitem{leaver}E. Leaver, Proc. Roy. Soc. London {\bf A402}, 285 (1985); J. Math. Phys. {\bf27}, 1238 (1986).

\bibitem{zel}
Ya.~B.~Zeldovich, Zh. Eksp. Teor. Fiz. {\bf 39},
776 (1960) 
[Sov. Phys. JETP {\bf 12}, 542 (1961)].

\bibitem{zelgen}
H.~M. Lai, P.~T. Leung, K.~Young, P.~Barber and S.~Hill,
Phys. Rev. A {\bf 41}, 5187 (1990);
H.~M. Lai, C.~C. Lam, P.~T. Leung and K.~Young,
J. Opt. Soc. Am. B {\bf 8}, 1962 (1991);
P.~T. Leung and K.~Young, Phys. Rev. A {\bf 44}, 3152 (1991);
P.~T. Leung, S.~Y. Liu, S.~S. Tong and K.~Young,
Phys. Rev. A {\bf 49}, 3068 (1994).

\bibitem{factor}
A.~Maassen van den Brink and K.~Young,
``Factor-space construction of outgoing waves", Preprint (1999).

\bibitem{missing}
If the double pole is split along the imaginary axis, 
there are two distinct SUSY transforms generated by 
$\tilde{\Phi}$ and $\tilde{\Psi}_j$ respectively, which become 
identical when the poles merge. 
Thus, in the spirit of the JB approach to the field expansion, 
one might try to find the ``missing" SUSY map generated 
by $\tilde{\Psi}_{j,1}$. However, no meaningful transformation emerges. 
This difference with the case of the field expansion can probably 
be understood by realizing that, when the poles split perpendicular 
to the imaginary axis upon changing the sign of the splitting 
perturbation, the SUSY transforms cease to exist altogether.

\bibitem{miracle2}It is remarkable that the values of $q$ for which these ``miracles" occur are the same as those for
which the PT potential has total transmission (Section~\ref{subsect:free}), for the former property is completely
determined by the asymptotics of the potential on one side, whereas the latter is global.

\bibitem{Jzero}With the replacement $g(\Omega)\mapsto\chi(\Omega)$, the calculation leading to (\ref{Jg-1g}) remains valid in cases a2 and a3. However, it does not convey information since on the l.h.s.\ $A\chi(\Omega)=0$.

\bibitem{smalltail}For $\alpha<2$, $V(x{\rightarrow}\infty)$ does not behave as a small tail anymore. That $\alpha=2$
is the marginal case is caused by considering $\omega=0$\cite{tail}. For finite $\omega$, the marginal case would be
$\alpha=1$ (cf.\ the asymptotics of Coulomb wave functions), as can readily be seen from the WKB approximation. 

\bibitem{nu-half} At $\nu=-\half$, the large generator is $\Phi\sim\sqrt{x}\ln x$, and there are significant
sub-leading corrections in the $\tilde{\nu}=\half$ partner potential so that its small solution
$\tilde{\chi}(\omega{=}0)$ is $\tilde{\Phi}=\Phi^{-1}$ instead of $x^{-1/2}$.

\bibitem{shape}
L.~E.~Gendenshtein, JETP Lett. {\bf 38}, 356 (1983);
F.~Cooper and J.~N.~Ginocchio, Phys. Rev. D {\bf 36}, 2458 (1987);
A.~Khare and U.~P.~Sukhatme, J. Phys. A {\bf 21}, L501 (1988);
see also Ref.~\cite{susy1}.

\end{references}
\end{document}